\documentclass[fleqn,usenatbib]{mnras}

\usepackage{newtxtext,newtxmath}

\usepackage[T1]{fontenc}

\DeclareRobustCommand{\VAN}[3]{#2}
\let\VANthebibliography\thebibliography
\def\thebibliography{\DeclareRobustCommand{\VAN}[3]{##3}\VANthebibliography}

\DeclareRobustCommand{\DE}[3]{#2}
\let\DEthebibliography\thebibliography
\def\thebibliography{\DeclareRobustCommand{\DE}[3]{##3}\DEthebibliography}

\usepackage{graphicx}	
\usepackage{amsmath}	

\usepackage{booktabs}
\usepackage{float}
\usepackage{threeparttable}
\usepackage{longtable}
\usepackage{pdflscape}
\usepackage{breqn}
\usepackage{tikz,xcolor,hyperref}

\definecolor{lime}{HTML}{A6CE39}
\DeclareRobustCommand{\orcidicon}{%
    \begin{tikzpicture}
    \draw[lime, fill=lime] (0,0) 
    circle [radius=0.16] 
    node[white] {{\fontfamily{qag}\selectfont \tiny ID}};
    \draw[white, fill=white] (-0.0625,0.095) 
    circle [radius=0.007];
    \end{tikzpicture}
    \hspace{-2mm}
}

\newcommand{\orcidmpr}{\href{https://orcid.org/0000-0001-9164-2882}{\orcidicon}}
\newcommand{\orcidnad}{\href{https://orcid.org/0000-0003-4842-8834}{\orcidicon}}

\title[Study on four new chemically peculiar stars]{High-resolution spectroscopic analysis of four new chemically peculiar stars}

\author[M. P. Roriz et al.]{
M. P. Roriz\orcidmpr$^{1}$\thanks{E-mail: michelle@on.br}, 
C. B. Pereira$^{1}$\thanks{E-mail: claudio@on.br}, 
S. Junqueira$^{1}$, 
M. Lugaro$^{2,3,4}$, 
N. A. Drake\orcidnad$^{1,5,6}$ 
and C. Sneden$^{7}$ 
\\
$^{1}$ Observat\'orio Nacional/MCTI, Rua Gen. Jos\'e Cristino, 77, 20921-400, Rio de Janeiro, Brazil\\
$^{2}$ Konkoly Observatory, Research Centre for Astronomy and Earth Sciences, Eötvös Loránd Research Network (ELKH), H-1121 Budapest, \\Konkoly Thege M. \'ut 15-17, Hungary\\
$^{3}$ ELTE E\"{o}tv\"{o}s Lor\'and University, Institute of Physics, Budapest 1117, P\'azm\'any P\'eter s\'et\'any 1/A, Hungary\\
$^{4}$ School of Physics and Astronomy, Monash University, VIC 3800, Australia\\
$^{5}$ Laboratory of Observational Astrophysics, Saint Petersburg State University, Universitetski pr. 28, 198504, Saint Petersburg, Russia\\
$^{6}$ Laborat\'orio Nacional de Astrof\'{\i}sica/MCTI, Rua dos Estados Unidos 154, Bairro das Na\c c\~oes, 37504-364, Itajub\'a, Brazil\\
$^{7}$ Department of Astronomy and McDonald Observatory, The University of Texas, Austin, TX 78712, USA\\
}

\date{Accepted XXX. Received YYY; in original form ZZZ}

\pubyear{2022}

\begin{document}
\label{firstpage}
\pagerange{\pageref{firstpage}--\pageref{lastpage}}
\maketitle

\begin{abstract}
We present detailed chemical compositions of four stars on the first-ascent red giant branch that are classified as chemically peculiar, but lack comprehensive analyses at high spectral resolution. For BD$+03\degr$2688, HE~0457-1805, HE~1255-2324, and HE~2207-1746, we derived metallicities [Fe/H] $=-1.21$, $-0.19$, $-0.31$, and $-0.55$, respectively, indicating a range in Galactic population membership. In addition to atmospheric parameters, we extracted elemental abundances for 28 elements, including the evolutionary-sensitive CNO group and $^{12}$C/$^{13}$C ratios. Novel results are also presented for the heavy elements tungsten and thallium. All four stars have very large enhancements of neutron-capture elements, with high [La/Eu] ratios indicating enrichments from the \textit{slow} neutron capture ($s$-process). To interpret these abundances, all indicative of [$s$/Fe] $> 1.0$, we compared our results with data from literature, as well as with predictions from the Monash and {\sc fruity} $s$-process nucleosynthesis models. BD$+03\degr$2688, HE~1255-2324, and HE~2207-1746 show C/O $>1$, while HE~0457-1805 has C/O $<1$. Since HE~0457-1805 and HE~1255-2324 are binary stars, their peculiarities are attributable to mass transfer. We identified HE~0457-1805 as a new barium giant star, and HE~1255-2324 as a new CH star, in fact a higher metallicity analogue CEMP-$r/s$ star; the single object reported in literature so far with similar characteristics is the barium star HD~100503 ([Fe/H] $= -0.72$). A systematic monitoring is needed to confirm the binary nature of BD$+03\degr$2688 and HE~2207-1746, which are probably CH stars.
\end{abstract}




\section{Introduction}\label{sec:intro}

The chemical abundance profiles of stellar atmospheres, revealed from their observed spectra, give us strong constraints on the environments where the stars were formed as well as their subsequent evolution. Chemically peculiar stars, in particular, are objects whose abundance distributions  differ from those observed in normal stars of the same age and Galactic field population. Many of these chemical peculiarities can be interpreted in the light of the mass transfer hypothesis between the components of a binary systems or even as consequence of a previous enrichment of the natal cloud. Such characteristics make these stars important targets for investigation. To disentangle the possible events able to produce these peculiarities, detailed chemical abundance studies are needed as they provide valuable data set for testing and constraining theoretical nucleosynthesis models, as well as the feasible astrophysical sources. Barium stars \citep{bidelman1951}, CH stars \citep{keenan1942}, and the Carbon-Enhanced Metal-Poor (CEMP) stars \citep{beers2005}, among others, are examples of different classes of chemically peculiar objects, each displaying its own characteristics.

Barium stars and CH stars exhibit atmospheres rich in both carbon and elements created by the \textit{slow} neutron-capture process \citep[the $s$-process;][]{kappeler2011}. The $s$-process takes place within the deep layers of evolved stars, when they reach the asymptotic giant branch \citep[AGB;][]{karakas2014} stage of the stellar evolution. During their thermally-pulsing (TP-AGB) phase, the stars self-enrich their atmospheres through the convective mixing episodes known as third dredge-up (TDU), which bring the by-products of nucleosynthesis to their external layers. However, since barium stars and CH stars are not luminous enough objects to be considered TP-AGB stars, they were not able to internally synthesize heavy elements observed on their atmospheres.

In fact, the characteristic overabundance of $s$-elements, as easily seen in the very strong $n$-capture transitions such as the Ba\,{\sc ii} and Sr\,{\sc ii} lines in barium and CH stars, as well as their anomalously deep molecular CH, CN and C$_{2}$ bands, are understood in the light of the mass transfer mechanism between the components of a binary system; their chemical peculiarities are from extrinsic origin. In this picture, the primary evolved star, now an invisible white dwarf, transferred the material previously processed in its AGB phase to the secondary star, now observed as a barium star or CH star. Indeed, since the 1980s, this astrophysical scenario has been confirmed \citep{mcclure1980, mcclure1984, mcclure1990} and extended \citep[e.g.][]{jorissen1998, jorissen2019} from data of the radial velocity monitoring programs that show orbital motion for these objects. Thus, the chemical abundance pattern recorded in the atmospheres of CH stars and barium stars provide us with clues from which we can probe and infer about the physical conditions inside the former AGB companion star.

Although the same astrophysical scenario is invoked to explain the origins of CH stars and barium stars, these objects have individual spectroscopic characteristics that warrant closer attention. CH stars are the Population II analogues of barium stars. While CH stars are metal-poor objects, belonging to the halo population, showing high radial velocities, barium stars are objects commonly associated with the disk of the Galaxy. CH stars exhibit C/O $>1$, whereas barium stars exhibit C/O $<1$. Several studies have provided elemental abundances of different chemical elements for barium stars \citep[e.g.][]{barbuy1992, allen2006, pereira2011, decastro2016, karinkuzhi2018, shejeelammal2020, roriz2021a, roriz2021b} and CH stars \citep[e.g.][]{vanture1992b, vanture1992c, pereira2009, pereira2012, karinkuzhi2014, goswami2005, goswami2016, purandardas2019}. As far as the enrichment levels of the $s$-process elements are concerned, the atmospheres of CH stars are more $s$-rich than the atmospheres of barium stars; in fact, this is a typical feature of the $s$-process in low-metallicities, where seed nuclei experience higher neutron exposures. In general, barium stars have orbital periods longer than those commonly observed in CH stars, and display more eccentric orbits than CH stars. Taken together, the abundance patterns and the orbital elements observed in these systems are used to probe the binary evolution of barium stars and CH stars, in order to constrain the mass transfer mechanism. This is a debated topic in literature, with Roche-lobe overflow and wind accretion as the main channels of mass transfer to account the observations \citep[e.g.][]{han1995, jorissen1998, jorissen2016}. Therefore, studies focused on these peculiar systems are important to improve our understanding of both nucleosynthesis and the physics of mass transfer.

CEMP stars, in turn, are objects found in lower metallicity regimes ([Fe/H] $<-2.0$). They exhibit high carbon excess ([C/Fe] $>+1.0$) in their atmospheres, although the limit for the carbon-to-iron ratio is not strictly defined in literature. In addition to their carbon-rich nature, CEMP stars can also exhibit a high content of heavy elements ($Z>30$), which are mainly synthesized by the $s$-process and the \textit{rapid} neutron-capture mechanism (the $r$-process). To emphasize these characteristics, CEMP stars are labeled as CEMP-$s$ and CEMP-$r$, respectively. In contrast, the CEMP-no stars do not exhibit $n$-enhancements on their surface. The so called CEMP-$r/s$ stars are more puzzling objects, since their atmospheres are simultaneously enriched in $r$- and $s$-elements. However, as demonstrated by \citet{hampel2016}, the \textit{intermediate} neutron-capture process \citep[the $i$-process;][]{cowan1977} is able to reproduce the chemical patterns observed in these systems. Nonetheless, it is worth noting that there are also interesting objects displaying features of CEMP-$r/s$, but without carbon enhancement \citep{roederer2016}, or with abundance profiles that are not interpreted in the light of a single/simple $i$-process \citep{koch2019}. In fact, these classifications are based from abundance data of different chemical elements, such as Ba, La, Sr, and Eu \citep[see, e.g.][]{beers2005, aoki2007, masseron2010, spite2013, hansen2019}. Some astrophysical scenarios are proposed and debated in literature to explain the origins of the peculiarities observed in CEMP stars \citep[see, for example,][and their references]{goswami2021}. However, the same formation scenario of the barium stars and CH stars (i.e. mass transfer) is also invoked to explain the origin of the CEMP-$s$ stars, considered as the metal-poor counterparts of the CH stars. Indeed, \citet{lucatello2005} reported a binary frequency of $\sim 68\%$ in their sample of CEMP-$s$. The high binary frequency among the CEMP-$s$ stars is also confirmed in further studies \citep[e.g.][]{starkenburg2014, hansen2016} which, from extensive radial velocity monitoring, corroborate that most, if not all, CEMP-$s$ stars are members of binary systems.

In the present study, we report a high-resolution optical spectroscopy analysis for four red giants that have been classified as either barium stars or CH stars, but have not yet been subjected to comprehensive chemical composition analyses: BD$+03\degr$2688, HE~0457-1805, HE~1255-2324, and HE~2207-1746. The star BD$+03\degr$2688 was first recognized as a chemically peculiar star by \citet{luck1991}, who included this object into a small group of four stars as broadly labeled as "metal-deficient barium stars" (BD$+03\degr$2688, BD$+04\degr$2466, HD~55496, and HD~104340). These authors noticed that the stars belonging to this group exhibit strong CH and C$_{2}$ molecular bands, as well as weak metallic lines; therefore, they could be the Population II analogs of the classical barium stars and probably CH candidate stars. Although these classifications were interesting and deserved further investigation at that time, the relatively low resolution ($\sim 0.2$~\AA) of \citeauthor{luck1991}'s spectra precluded deeper insight into the true nature of these stars. Given this, we decided to re-investigate and revise the meaning of "metal-deficient barium stars", using a new high-resolution optical CCD spectra.

In fact, the stars HD~55496 and HD~104340 could not be classified as CH candidate stars due to weakness of C$_{2}$ bands. As seen by \citet{junqueira2001}, and later by \citet{jorissen2005} and \citet{drake2008}, HD~104340 is probably a metal-poor AGB candidate star. Another example is the star HD~55496, which was considered as a field star that may have escaped from a globular cluster \citep{pereira2019b}. The other two stars of the \citet{luck1991}'s group, BD$+04\degr$2466 \citep{pereira2009} and BD$+03\degr$2688 (this work), in fact, exhibit strong CH and C$_{2}$ molecular bands, along with weak metallic lines, and could be strong CH candidate stars. BD$+04\degr$2466 is a CH star, since it displays the typical characteristics of the stars belonging to this class, $s$-process and carbon enrichment \citep{pereira2009}, and it is a binary system \citep{jorissen2005}. BD$+03\degr$2688 is the last object of the \citeauthor{luck1991}'s group not yet investigated with better spectroscopy data; we included it also in our investigation.

Along with BD$+03\degr$2688, we have also added as targets the stars HE~0457-1805, HE~1255-2324, and HE~2207-1746, from the objective prism plates of the Hamburg/ESO Survey \citep[][]{christlieb2001}. The stars HE~0457-1805 and HE~2207-1746 came to our attention because they were previously classified as chemically peculiar stars by \citet{goswami2005}, i.e. stars with prominent C$_{2}$ bands having respectively a 'HD~26' and a 'HD~209061' type spectra, both well known CH stars. HE~1255-2324 was discovered in the objective prism plates of the Hamburg/ESO Survey carbon stars \citep[][]{christlieb2001}. In their list, HE~1255-2324 presents low values for the band indexes $I$(C$_{2}~\lambda$4\,737~\AA) and $I$(C$_{2}~\lambda$5\,165~\AA), thus probably indicating not being a cool carbon star, where the spectrum is crowded due to the strong molecular line opacities.

The paper is organized in the following way. In Section~\ref{sec:observations}, we briefly describe the setup instrumentation for the data acquisition; in Section~\ref{sec:atm_par}, we present the adopted methodology to derive the atmospheric parameters of the target stars; in Section~\ref{sec:abd_anal}, we describe the techniques used to obtain the chemical abundances. The results and discussion are presented in Section~\ref{sec:results}. Finally, in Section~\ref{sec:conclusions}, we drawn the main conclusions of this study.

\section{Observations}\label{sec:observations}

The high-resolution spectra of the stars analyzed in this work were obtained with the Fiberfed Extended Range Optical Spectrograph (FEROS; \citealt{kaufer1999}) at the 2.2\,m ESO telescope in La Silla (Chile). The wavelength range obtained with FEROS covers the spectral region between 3\,900~\AA\ and 9\,200~\AA. The resolving power of FEROS is approximately 48\,000 with a degradation less than 10\% over the whole wavelength range \citep{kaufer1998}. The spectra were reduced following the standard procedure, including bias subtraction, flat-fielding, order extraction and wavelength calibration following the FEROS Data Reduction Software pipeline based on ESO-MIDAS environment. The spectra were later normalised and shifted to rest wavelengths in the regions of interest for spectral synthesis. Table~\ref{tab:spec_data} provides an observing log of the spectroscopic observations.

\begin{table*}
\caption{Log of the observations for the target stars analyzed in this work. Column 1 gives the name of the star, column 2 gives the dates of observation, column 3 gives the S/N ratio around 6\,000~\AA, and the last column gives the observed heliocentric radial velocity (RV) derived from Doppler shifts based on ten absorption lines.}
\label{tab:spec_data}
    \begin{tabular}{lcccc}\toprule
    Star               & Date of          &   S/N            & Exp.              &      RV            \\
                       & observation      & (at 6\,000~\AA)  & (secs)            & (km\,s$^{-1}$)     \\
    \midrule
    BD$+03\degr$2688   & Apr. 4, 2007     &    180           & 2\,400            &  32.64$\pm$0.29    \\
                       & Mar. 18, 2016    &    120           & 1\,500            &  34.36$\pm$0.19    \\ 
    \midrule
    HE~0457-1805       & Oct. 15, 2007    &    100           & 2 $\times$ 3\,600 &  61.22$\pm$0.34    \\
    \midrule
    HE~1255-2324       & May 4, 2010      &    100           & 1\,800            &  37.25$\pm$0.26    \\
                       & Aug. 1, 2010     &    110           & 1\,800            &  18.21$\pm$0.21    \\
    \midrule
    HE~2207-1746       & Sep. 7, 2007     &    100           & 2 $\times$ 3\,600 & $-$8.01$\pm$0.21   \\
    \bottomrule
    \end{tabular}
\end{table*}

\section{Atmospheric parameters}\label{sec:atm_par}

The atmospheric parameters of BD$+03\degr$2688, HE~0457-1805, HE~1255-2324 and HE~2207-1746 were determined in the same way as in the previous studies done for the stars HD~104340 and HD~206983 \citep{drake2008}, the CEMP-$s$ star CD$-50^\circ$776 \citep{roriz2017}, and the star HD~55496 \citep{pereira2019a}. We first measured the equivalent widths $W/\lambda$ of the Fe\,{\sc i} and Fe\,{\sc ii} absorption lines by fitting Gaussian profiles to the observed spectral lines. The atomic parameters, excitation potential ($\chi$), and $\log$ \textit{gf} values for the Fe\,{\sc i} and Fe\,{\sc ii} lines were taken from \citet{lambert1996}. The equivalent widths of the measured Fe\,{\sc i} and Fe\,{\sc ii} lines are shown in Table~\ref{tab:iron_lines}, presented in Appendix~\ref{app:obs_lines}; we used the task {\sl splot} of {\sc iraf} to measure the equivalent widths.

\begin{table*} 
\caption{Atmospheric parameters obtained in this work for the program stars, as well as values from other studies available in literature.} \label{tab:atm_par}
    \begin{threeparttable}
        \begin{tabular}{@{\extracolsep{4pt}}lcccccccc}\toprule
        & \multicolumn{2}{c}{BD$+03\degr$2688} & \multicolumn{2}{c}{HE~0457-1805} &
        \multicolumn{2}{c}{HE~1255-2324} & \multicolumn{2}{c}{HE~2207-1746}\\
        \cline{2-3}
        \cline{4-5}
        \cline{6-7}
        \cline{8-9}
        
        Parameter        &      Value       &  Ref. & Value           & Ref. & Value            & Ref.& Value            & Ref.\\
        \midrule
        
        $T_{\rm eff}$(K) & 4\,600$\pm$40    & 1     & 4\,720$\pm$100  & 1    & 4\,960$\pm$110   & 1   & 4\,780$\pm$80    & 1   \\
                         & 4\,747$\pm$46    & 2     & 4\,484          & 4    & 4\,405           & 4   & 4\,737           & 4   \\
                         & 4\,300           & 3     & 4\,580          & 5    & 4\,760           & 5   &                  &     \\
                         &                  &       & 4\,437          & 6    & 4\,486           & 6   &                  &     \\
                         &                  &       &                 &      & 4\,487           & 7   & 4\,577           & 7   \\
                         &                  &       &                 &      & 4\,530           & 8   &                  &     \\
        \midrule
      
        $\log g$ (dex)   & 1.6$\pm$0.3      & 1    & 2.2$\pm$0.4      & 1    & 3.0$\pm$0.3      & 1   &  2.6$\pm$0.3     & 1   \\
                         & 1.43$\pm$0.14    & 2    & 0.77             & 4    & 0.69             & 4   &  1.23            & 4   \\
                         & 0.00             & 3    & 1.58             & 6    & 1.73             & 6   &                  &     \\
                         &                  &      &                  &      & 1.73             & 7   &                  &     \\
                         &                  &      &                  &      & 2.63             & 8   &                  &     \\
        \midrule

        [Fe/H] (dex)     & $-$1.21$\pm$0.13 & 1    & $-$0.19$\pm$0.17 & 1    & $-$0.31$\pm$0.16 & 1   & $-$0.55$\pm$0.15 & 1  \\
                         & $-$1.27$\pm$0.07 & 2    & $-$1.46          & 4    & $-$1.47          & 4   & $-$1.87          & 4  \\
                         & $-$1.42          & 3    & $+$0.33          & 6    & $+$0.32          & 6   &                  &    \\
                         &                  &      &                  &      & $+$0.20          & 7   &                  &    \\
        \midrule

        $\xi$ (km\,s$^{-1}$) & 1.5$\pm$0.2  & 1    &  1.3$\pm$0.2    & 1     & 1.5$\pm$0.2      & 1   & 1.4$\pm$0.2      & 1  \\
                             & 2.7          & 3    &                 &       &                  &     &                  &    \\
        \bottomrule

        \end{tabular}
        \textbf{References:} (1) This work; (2) \citet{arentsen2019}; (3) \cite{luck1991}; (4) \citet{kennedy2011}; (5) \citet{munari2014}; (6) \citet{kunder2017}; (7) \citet[][]{stevens2017}; (8) \citet[][]{mcdonald2017}.
        
    \end{threeparttable}
\end{table*}

These absorption lines were used to obtain the stellar atmospheric parameters, which are the effective temperature, $T_{\rm eff}$, the surface gravity, $\log g$, the microturbulent velocity, $\xi$, and the metallicity, [Fe/H]\footnote{Throughout this study, we have used the standard spectroscopy notation, [A/B] $= \log(N_{\rm A}/N_{\rm B})_{\star} -\log(N_{\rm A}/N_{\rm B})_{\odot}$, for two generic elements A and B.}. In order to determine the best atmospheric model that describes the physical conditions of the stellar atmosphere of BD$+03\degr$2688, HE~0457-1805, HE~1255-2324, and HE~2207-1746, we have used the version 2013 of the spectral line analysis code {\sc moog}\footnote{{\sc moog} is available at: \url{https://www.as.utexas.edu/~chris/moog.html}.} \citep{sneden1973}, which assumes the conditions of local thermodynamic equilibrium (LTE), and the 1D plane-parallel model atmospheres provided by \citet{kurucz1993}.

The effective temperature and the microturbulent velocity are determined when, for a given atmopsheric model, there are no trends for Fe\,{\sc i} lines between iron abundance and both excitation potential and reduced equivalent width ($W_{\lambda}/{\lambda}$). The surface gravity is obtained by means of the ionization equilibrium, that is, when the abundances of Fe\,{\sc i} and Fe\,{\sc ii} are equal at the selected effective temperature. The final Fe\,{\sc i} abundance defines the metallicity of the star. Our derived adopted atmospheric parameters for each star are given in Table~\ref{tab:atm_par}, along with values from previous studies. The final metallicity was normalized to the iron abundance given by \cite{grevesse1998} which is 7.50 dex.

Comparing the $\log$ \textit{gf} values of the Fe\,{\sc ii} lines between the ones we used from \citet{lambert1996} with those given by \citet{melendez2009}, we found a mean difference of $-0.025$ ($\sigma=0.03$) for eleven lines in common. Therefore, there will not be a significant impact in our derived stellar parameters. In Table 7 of \citeauthor{melendez2009}, it can be seen that the mean difference by using the $\log$~\textit{gf} values from \citeauthor{lambert1996} and those of \citeauthor{melendez2009} is $-0.01$ ($\sigma=0.07$).

A very low value of the surface gravity ($\log g=0.0$) for BD$+03\degr$2688 was determined by \cite{luck1991}, while we have derived $\log g=1.6$. For this star, \citeauthor{luck1991} used a photographic image-tube spectrogram with a dispersion of 4.6~\AA/mm and a relatively low resolution of $\sim 0.2$~\AA\ ($3\,800-5\,200$~\AA). This may explain the low gravity and the high microturbulent velocity (2.7~km\,s$^{-1}$) found by these authors. This high value for the microturbulent velocity, derived by means of neutral lines, might in turn result in the low value for $\log g$, derived by forcing the Fe\,{\sc ii} and Fe\,{\sc i} lines to yield the same iron abundance.

For the stars HE~0457-1805, HE~1255-2324, and HE~2207-1746, previous atmospheric parameters (Table~\ref{tab:atm_par}) were obtained by \citet{kennedy2011}, who used a different technique from that used in this work. These authors first obtained photometric temperatures, and by considering that all the stars analyzed in their work were low metallicity stars (down to [Fe/H]\,=\,$-$1.7), they obtained the surface gravities using evolutionary tracks for that metallicities. Our derived metallicities for these stars are more than an order of magnitude higher.

The errors in temperature and microturbulent velocity were estimated considering the uncertainty in the value of the slope of the null trend, respectively between the iron abundance given by the Fe\,{\sc i} lines and the excitation potential, and between the iron abundance and the reduced equivalent width. For the gravity, the error is estimated as the difference between the nominal value given by the ionization equilibrium and the value of $\log g$ that produces a difference between the mean abundances of Fe\,{\sc i} and Fe\,{\sc ii}, that is equal to (or greater than) the standard deviation of the mean [Fe\,{\sc i}/H]. The uncertainties of the obtained atmospheric parameters are also provided in Table~\ref{tab:atm_par}.

\section{Abundance analysis}\label{sec:abd_anal}

We have determined abundances for several chemical elements, including light elements (C, N, O, Na, Mg, Al, Si, Ca, and Ti), iron-group elements (Cr and Ni), and neutron-capture elements (Rb, Sr, Y, Zr, Nb, Mo, Ru, La, Ce, Nd, Sm, Eu, Er, W, Tl, and Pb), either by matching predicted and measured equivalent widths or computing synthetic spectra and comparing them with the observed spectra. The elemental abundances derived in this work were also obtained from the code {\sc moog} \citep{sneden1973}.

The abundances of carbon, nitrogen, oxygen, and the $^{12}$C/$^{13}$C isotopic ratio were determined exclusively with spectral syntheses. The carbon abundance was obtained using the C$_{2}$ (0,1) band head of the system $A^3\Pi_{g} - X^3\Pi_{u}$ at 5\,635~\AA, while the nitrogen abundance was obtained based on the $^{12}$CN lines $A^2\Pi-X^2\Sigma$ in the 7\,994 -- 8\,020~\AA\ wavelength range. To obtain the $^{12}$C/$^{13}$C isotopic ratio, we used the $^{13}$CN lines around the region at 8\,004 -- 8\,005~\AA. Details about the molecular constants and dissociation energies of the molecules involved in the determination of the carbon and nitrogen abundances are given in \cite{drake2008}. The observed and synthetic spectra for the spectral regions of the C$_{2}$ molecular band, the CN molecular lines used for nitrogen abundance and the $^{13}$C lines used for the determination of the $^{12}$C/$^{13}$C isotopic ratio are shown in Figures~\ref{fig:syn_c2}, \ref{fig:syn_cn} and \ref{fig:syn_c12c13}, provided in Appendix~\ref{app:extra_figures}. For oxygen, we determined its abundance based on the forbidden line at 6\,300.3~\AA.

\begin{figure}
    \centering
    \includegraphics[width=\columnwidth]{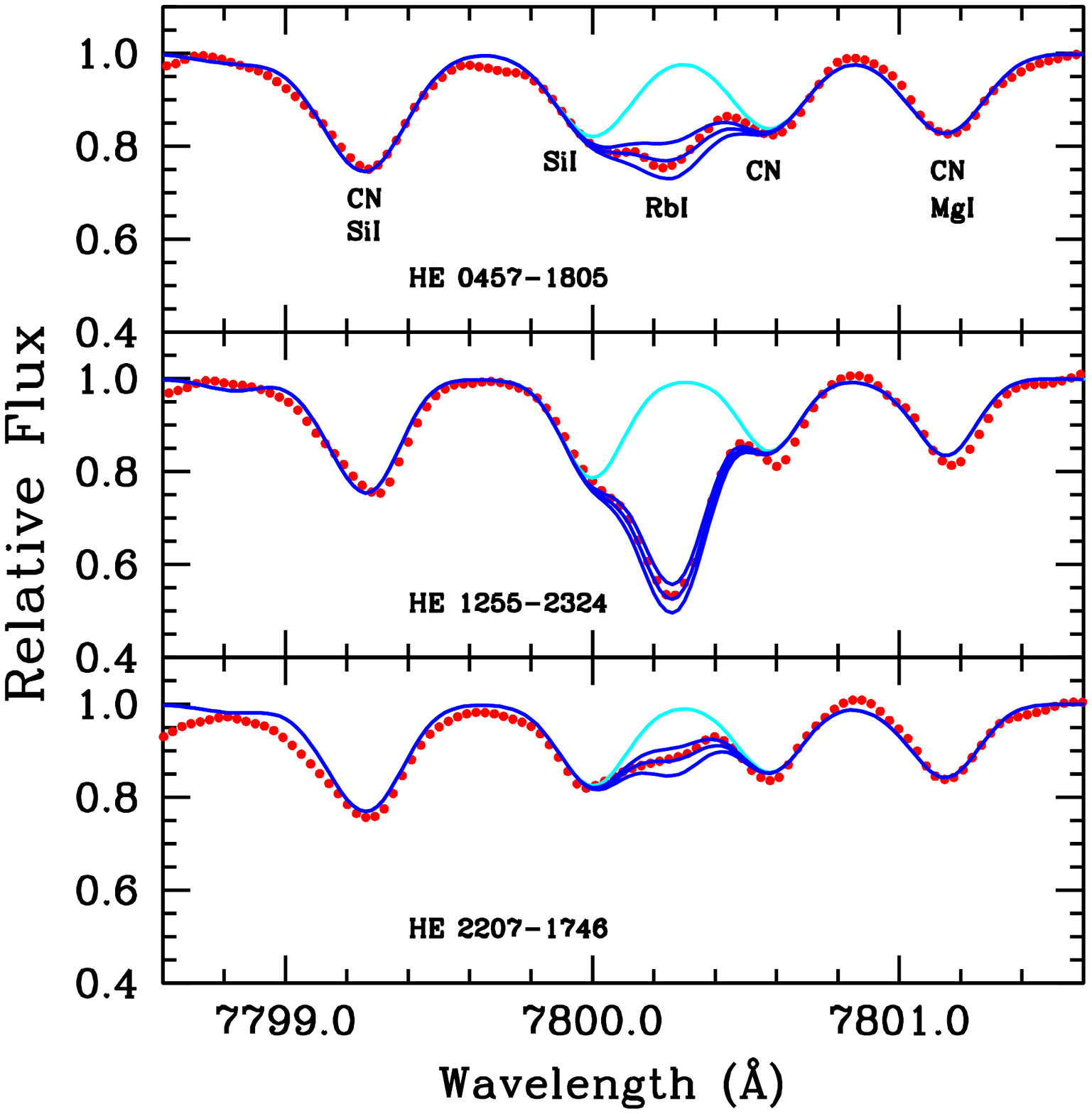}
    \caption{Observed (red dots) and synthetic (curves) spectra close to the region of the Rb\,{\sc i} line at 7\,800.20~\AA\ for the stars HE~0457-1805, HE~1255-2324, and HE~2207-1746. The upper cyan curves shown in the panels are synthetic spectra calculated without contribution from the rubidium line. The middle blue curves are synthetic spectra calculated for the adopted solutions, $\log \epsilon$(Rb)\,=\,2.85, 3.90, and 2.45 dex, that provide the best fits for HE~0457-1805, HE~1255-2324, and HE~2207-1746, respectively; the upper and lower blue curves show the spectral synthesis for $\Delta \log \epsilon (\textrm{Rb})=\pm 0.14$ dex around the adopted solutions.}
    \label{fig:syn_rb}
\end{figure}

\begin{figure}
    \centering
    \includegraphics[width=\columnwidth]{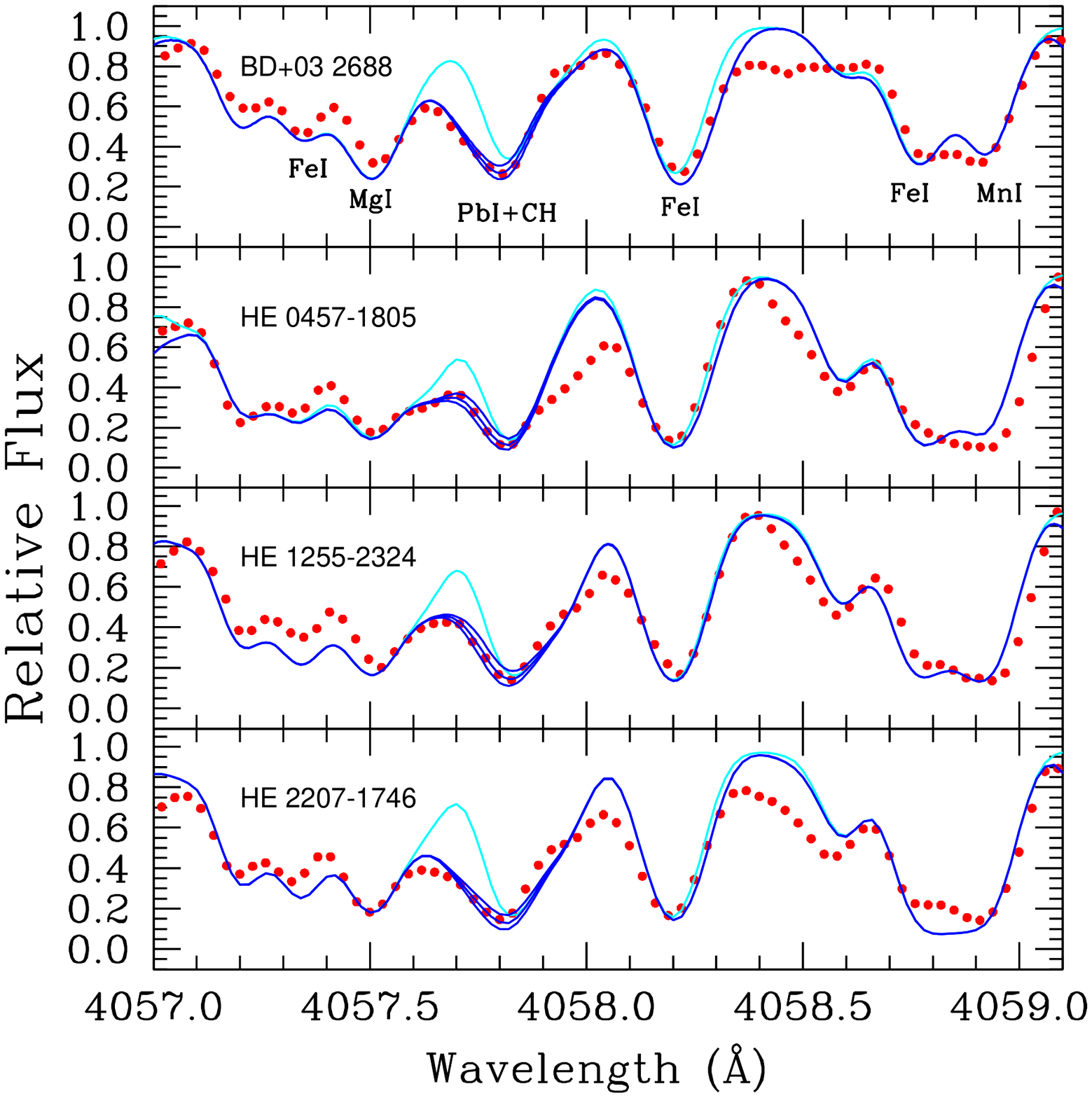}
    \caption{Observed (red dots) and synthetic (curves) spectra close to the region of the Pb\,{\sc i} line at 4\,057.8~\AA\ for the stars BD$+03\degr$2688, HE~0457-1805, HE~1255-2324, and HE~2207-1746. The upper cyan curves shown in the panels are synthetic spectra calculated without contribution from the CH lines. The middle blue curves are synthetic spectra calculated for the adopted solutions, $\log \epsilon$(Pb)\,=\,2.25, 3.25, 3.05, and 2.95 dex, that provide the best fits for BD$+03\degr$2688, HE~0457-1805, HE~1255-2324, and HE~2207-1746, respectively; the upper and lower blue curves show the spectral synthesis for $\Delta \log \epsilon (\textrm{Pb})=\pm 0.20$ dex around the adopted solutions.}
    \label{fig:syn_pb}
\end{figure}

The abundances of rubidium, niobium, and europium were also determined by means of the spectral synthesis technique. Details of the spectral regions and lines used for the abundance determination are given in \cite{roriz2021a, roriz2021b}. Figure~\ref{fig:syn_rb} shows spectral synthesis for the element rubidium around the resonant Rb\,{\sc i} line at 7\,800.2~\AA. However, we were not able to derive the rubidium content for the star BD$+03\degr$2688; for this star, the Rb line was too weak to be detected in its spectrum. The lead abundance was derived from the Pb\,{\sc i} line at $\lambda$ 4\,057.81~\AA, where the isotopic shifts and hyperfine splitting (HFS) were taken from \cite{vaneck2003}. In Figure~\ref{fig:syn_pb}, we show the observed and synthetic spectra for the region around the lead line at 4\,057.81~\AA. As one can see in Figure~\ref{fig:syn_pb}, the CH transitions may contribute to the region where the lead line was used for the spectral synthesis. We considered the previous carbon abundance determination when we synthesized the profile at 4\,057.81~\AA, as it was done in \citet[][]{roriz2017}.

For the other chemical elements considered in this study, we list in Table~\ref{tab:other_lines} the lines used to obtain their abundances along with their excitation potentials, $\log$ \textit{gf} values, their respective references, and the equivalent width measurements. For lanthanum, in particular, we used the same technique as employed by \cite{roriz2021b}; we measured the equivalent widths and run the driver {\sl blends} of {\sc moog}, which provides abundances from blended spectral lines, since the atomic terms of lanthanum are strongly affected by HFS. 

\begin{figure}
    \centering
    \includegraphics[width=\columnwidth]{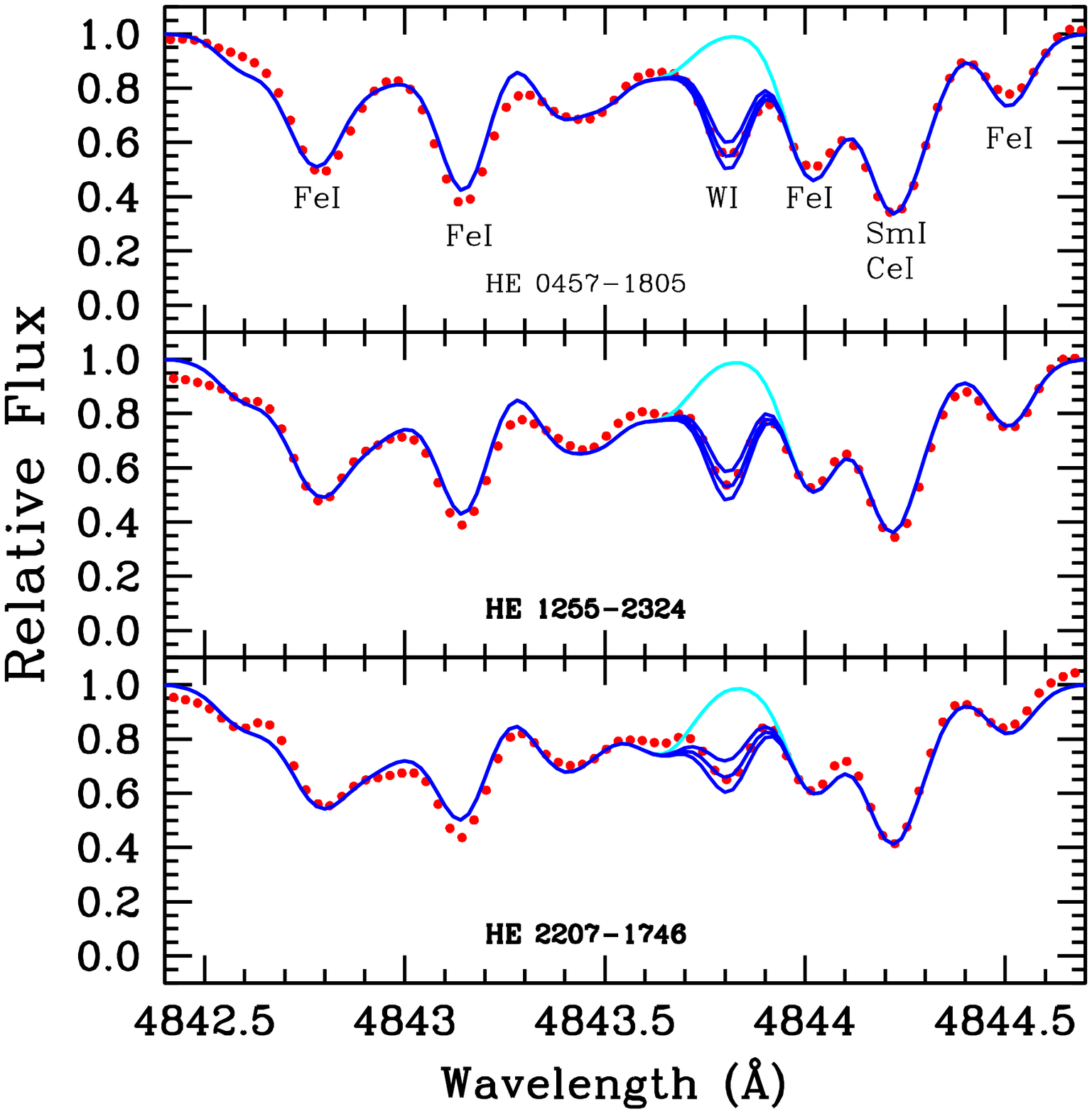}
    \caption{Observed (red dots) and synthetic (curves) spectra close to the region of the W\,{\sc i} line at 4\,843.81~\AA\ for the stars HE~0457-1805, HE~1255-2324, and HE~2207-1746. The upper cyan curves shown in the panels are synthetic spectra calculated without contribution from the tungsten line. The middle blue curves are synthetic spectra calculated for the adopted solutions, $\log \epsilon$(W)\,=\,2.31, 2.71, and 2.01 dex, that provide the best fits for HE~0457-1805, HE~1255-2324, and HE~2207-1746, respectively; the upper and lower blue curves show the spectral synthesis for $\Delta \log \epsilon (\textrm{W})=\pm 0.18$ dex around the adopted solutions.}
    \label{fig:syn_w}
\end{figure}

We have also determined abundances of three other heavy elements not investigated before in the spectra of barium stars: erbium (Z\,=\,68), tungsten (Z\,=\,74), and thallium (Z\,=\,81). The erbium absorption lines used in the abundance analysis are given in Table~\ref{tab:other_lines}. The tungsten abundances were obtained using spectral synthesis technique of two W\,{\sc i} lines, one at 4\,843.81~\AA, with $\log$ \textit{gf} = $-$1.50, and the other line at 5\,224.66~\AA, with $\log$ \textit{gf} = $-$1.70. The $\log$ \textit{gf} values were taken from Vienna Atomic Line Database \citep[{\sc vald};][]{ryabchikova2015}. Figure~\ref{fig:syn_w} shows the spectral synthesis for tungsten around the line at 4\,843.81~\AA.

To determine thallium abundances, we also used the spectral synthesis technique. We have taken into account the HFS of the atomic terms 6 $^{2}P_{3/2}$ and 7$^{2}S_{1/2}$ involved in the transition at 5\,350.46~\AA, for its two stable isotopes, $^{203}$Tl (29.5\%) and $^{205}$Tl (70.5\%), which have the same nuclear spin, $I=1/2$. The adopted HFS constants, as well as the isotopic shift, were taken from \citet{lambert1972}, and the energy levels are provided by the National Institute of Standards and Technology \citep[{\sc nist};][]{kramida2020} database. Thus, we were able to compute the $\log$ \textit{gf} of each hyperfine component by distributing the total $\log$ \textit{gf} according to the relative intensities. In this procedure, we have adopted $\log$ \textit{gf}\,=\,$-$0.21, also provided by the {\sc nist} database. In fact, our results are similar to that given by \citet{wahlgren2000}. We summarize in Table~\ref{tab:hfs_tl} the list of the atomic lines used to derive abundances via spectral synthesis technique, where we provide the references for $\log$ \textit{gf} and HFS. In particular, for thallium, we have shown in Table~\ref{tab:hfs_tl} the HFS components of the 5\,350.46~\AA\ line along with the respective $\log$ \textit{gf} values also derived in this work.

\begin{table}
\centering
\caption{Atomic lines used to obtain the abundance by spectral synthesis. In the last column, we provide the references for $\log$ \textit{gf} and HFS data.}\label{tab:hfs_tl}
    \begin{threeparttable}
        \begin{tabular}{lcccl}
        \toprule
        Element     & Wavelength (\AA) & $\chi$\,(eV) & $\log$ \textit{gf} & Ref. \\
        \midrule
        O\,{\sc i}  & 6\,300.3         &  0.00        & $-$9.72            & 1   \\
        \midrule
        Rb\,{\sc i} & 7\,800.268       &  0.000       & $+$0.137           & 2,3 \\
        \midrule
        Nb\,{\sc i} & 4\,606.756       &  0.348       & $-$0.377           & 2,4 \\
        Nb\,{\sc i} & 5\,344.158       &  0.348       & $-$0.730           & 2,4 \\
        Nb\,{\sc i} & 5\,350.722       &  0.267       & $-$0.910           & 2,4 \\
        \midrule
        Eu\,{\sc i} & 6\,645.100       &  1.379       & $+$0.120           & 5   \\
        \midrule
        W\,{\sc i}  & 4\,843.810       &  0.412       & $-$1.500           & 2   \\
        W\,{\sc i}  & 5\,224.660       &  0.599       & $-$1.700           & 2   \\
        \midrule
        Tl\,{\sc i} & 5\,350.423       &  0.965       & $-$1.651           & 6,7 \\
                    & 5\,350.428       &              & $-$0.952           &     \\
                    & 5\,350.541       &              & $-$1.350           &     \\
                    & 5\,350.406       &              & $-$1.273           &     \\
                    & 5\,350.411       &              & $-$0.574           &     \\
                    & 5\,350.523       &              & $-$0.972           &     \\
        \midrule
        Pb\,{\sc i} & 4\,057.810       &  1.320       & $-$0.220           & 8   \\
        \bottomrule
    \end{tabular}
    \textbf{References:} (1) \citet{allende2001} (2) {\sc vald}; (3) \citet{roriz2021a}; (4) \citet{roriz2021b}; (5) \citet{lawler2001b}; (6) {\sc nist}; (7) This work; (8) \citet{vaneck2003}.
    \end{threeparttable}
\end{table}

\begin{figure}
    \centering
    \includegraphics[width=\columnwidth]{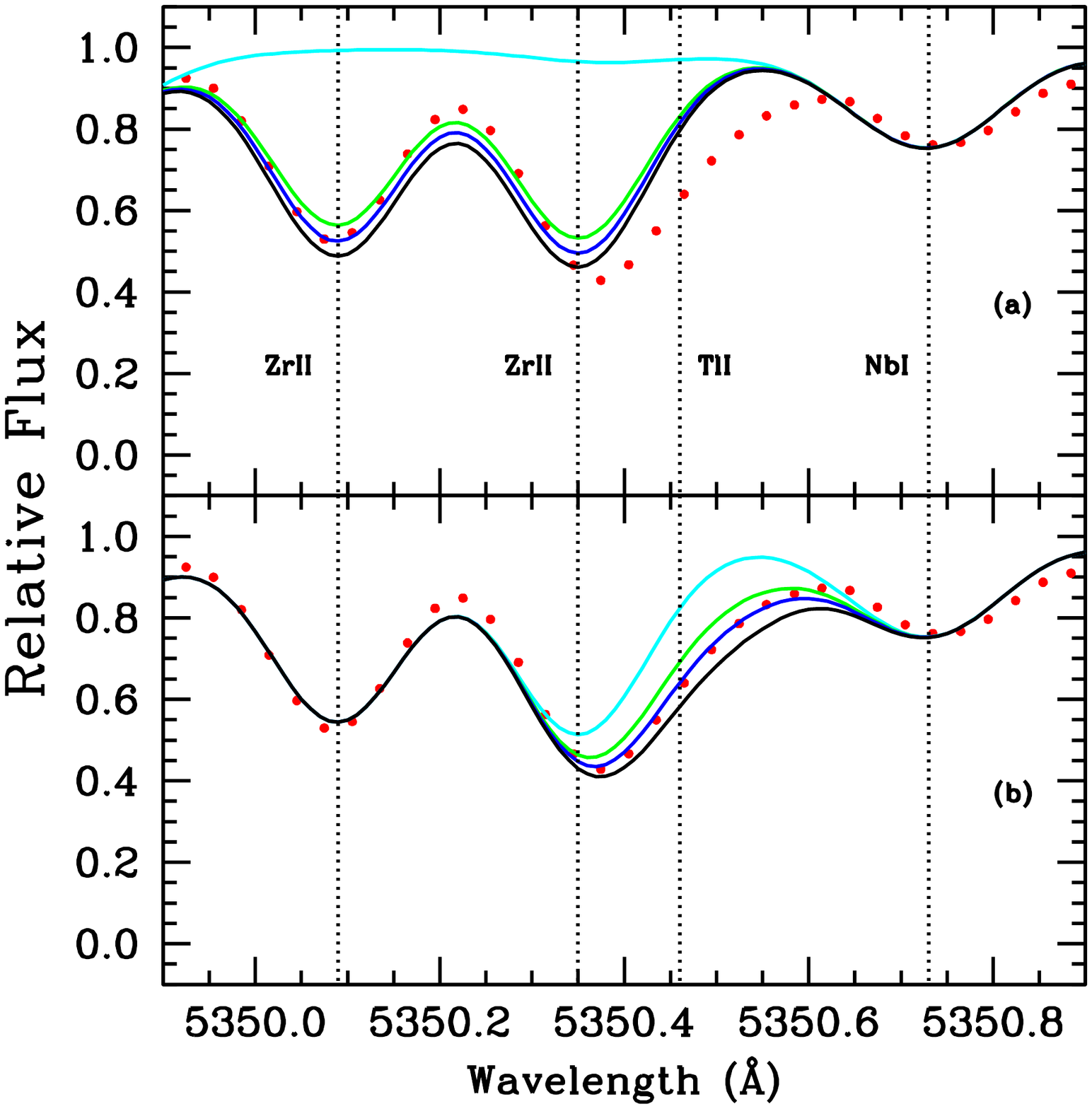}
    \caption{Observed (red dots) and synthetic (curves) spectra close to the region of the Tl\,{\sc i} line at 5\,350.46~\AA\ for the stars HE~1255-2324. Vertical dotted lines represent the observed transitions due to Zr\,{\sc ii} at 5\,350.09~\AA, Zr\,{\sc ii} at 5\,350.35~\AA, Tl\,{\sc i} at 5\,350.46~\AA, and Nb\,{\sc ii} at 5\,350.72~\AA. In panel (a), we show synthetic spectra calculated for different zirconium abundances, and without contribution from the thallium line. The upper cyan curve shows the synthesis without contribution from the zirconium line. In blue, we show the best fit based on the Zr\,{\sc ii} line at 5\,350.09~\AA, obtained for $\log \epsilon$(Zr)\,=\,4.33 dex ([Zr/Fe]\,=$+2.04$). Green and black curves are synthetic spectra calculated for $\Delta \log \epsilon (\textrm{Zr}) \pm 0.20$ dex. In panel (b), we show synthetic spectra calculated for different thallium abundances, considering the best fit for the zirconium abundance seen in panel (a). The upper cyan curve shows the synthesis without contribution from the thallium line. In blue, we show the best fit based on the Tl\,{\sc i} line at 5\,350.46~\AA, obtained for $\log \epsilon$(Tl)\,=\,2.35 dex ([Tl/Fe]\,=$+1.76$). In order to fit the whole line profile of the Zr\,{\sc ii} at 5\,350.35~\AA, we need to add the thallium contribution in the red wing of the zirconium line. Green and black curves are synthetic spectra calculated for $\Delta \log \epsilon (\textrm{Tl}) \pm 0.20$ dex.}
    \label{fig:syn_tl}
\end{figure}

In Figure~\ref{fig:syn_tl}, we show the thallium line at 5\,350.46~\AA, for HE~1255-2324, where red dots represent the observed spectrum. The thallium transition lies at the red wing of the Zr\,{\sc ii} at 5\,350.35~\AA. In panel (a) of Figure~\ref{fig:syn_tl}, we show that only considering the Zr\,{\sc ii} transition, assuming the same abundance of its neighbour Zr\,{\sc ii} line at 5\,350.09~\AA, it is not possible to reproduce the observed line profile (blue curve synthesis). Higher zirconium abundance does not also fit the profile (black line synthesis). In panel (b) of Figure~\ref{fig:syn_tl}, we show that we were able to synthesize and reproduce to whole profile assuming a thallium abundance of $\log \epsilon$(Tl)\,=\,2.35 dex (blue curve synthesis). By doing the same analysis for HE~0457-1805 and HE~2207-1746, we obtained a thallium abundance, respectively of $\log \epsilon$(Tl)\,=\,1.80 and 1.90 dex.

Finally, Tables~\ref{tab:bd_and_he} and \ref{tab:he_and_he} provide the obtained abundances, their standard deviation, and the number of lines used for each species. The final abundances were normalized to the solar photospheric abundances given by \cite{grevesse1998}. 

\begin{table*}
\centering
\caption{Elemental abundances derived for BD$+03\degr$2688 and HE~0457-1805. The second column provides the solar photospheric abundances as recommended by \citet[][]{grevesse1998}. The third and eighth columns give the abundances in the scale $\log\epsilon$({\rm H}) = 12, and their respective dispersion among the lines of the elements with more than three available lines are shown in the fourth and ninth columns. The fifth and tenth columns give the information whether abundances were determined from spectrum synthesis technique (syn) or based on the equivalent width measurements; in this latter case, we provide the number of lines ($n$) used for the abundance determination. Abundances in the notations [X/H] and [X/Fe] are also provided. The $^{12}\mathrm{C}/\,^{13}\mathrm{C}$ isotopic ratio is given in the last line of the Table.}\label{tab:bd_and_he}
    \begin{tabular}{@{\extracolsep{4pt}}lccccccccccccc}
    \toprule
    & & & \multicolumn{5}{c}{BD$+03\degr$2688} & & \multicolumn{5}{c}{HE~0457-1805}\\
    \cline{4-8}
    \cline{10-14}

    Species & $\log\epsilon_{\odot}$ & & $\log\epsilon$ & $\sigma_{\rm obs}$ & $n$(\#) & [X/H] & [X/Fe]  & &  $\log\epsilon$ & $\sigma_{\rm obs}$ & $n$(\#) & [X/H] & [X/Fe] \\
    \midrule
        
    C\,(C$_{2}$)  & 8.52 & & 7.67 & ---  & syn & $-$0.85 & $+$0.36 & &  8.36 & ---  & syn & $-$0.16 & $+$0.03 \\
    N\,(CN)       & 7.92 & & 6.90 & ---  & syn & $-$1.02 & $+$0.19 & &  8.65 & ---  & syn & $+$0.73 & $+$0.92 \\      
    O             & 8.83 & & 7.58 & ---  & syn & $-$1.25 & $-$0.04 & &  8.48 & ---  & syn & $-$0.35 & $-$0.16 \\
    Na\,{\sc i}   & 6.33 & & 4.89 & 0.09 & 04  & $-$1.44 & $-$0.23 & &  6.70 & 0.09 & 03  & $+$0.37 & $+$0.56 \\      
    Mg\,{\sc i}   & 7.58 & & 6.86 & ---  & 02  & $-$0.72 & $+$0.49 & &  7.53 & ---  & 02  & $-$0.05 & $+$0.14 \\
    Al\,{\sc i}   & 6.47 & & ---  & ---  & --- &     --- &     --- & &  6.38 & 0.10 & 04  & $-$0.09 & $+$0.10 \\
    Si\,{\sc i}   & 7.55 & & 6.72 & 0.13 & 04  & $-$0.83 & $+$0.38 & &  7.61 & 0.19 & 04  & $+$0.06 & $+$0.25 \\      
    Ca\,{\sc i}   & 6.36 & & 5.55 & 0.10 & 11  & $-$0.81 & $+$0.40 & &  6.43 & 0.08 & 06  & $+$0.07 & $+$0.26 \\
    Ti\,{\sc i}   & 5.02 & & 3.95 & 0.10 & 20  & $-$1.07 & $+$0.14 & &  4.92 & 0.08 & 11  & $-$0.10 & $+$0.09 \\
    Fe\,{\sc i}   & 7.50 & & 6.29 & 0.13 & 65  & $-$1.21 &    ---  & &  7.31 & 0.17 & 22  & $-$0.19 &    ---  \\
    Fe\,{\sc ii}  & 7.50 & & 6.29 & 0.09 & 12  & $-$1.21 &    ---  & &  7.32 & 0.15 & 06  & $-$0.18 &    ---  \\
    Cr\,{\sc i}   & 5.67 & & 4.35 & 0.14 & 10  & $-$1.32 & $-$0.11 & &  5.50 & 0.09 & 07  & $-$0.17 & $+$0.02 \\         
    Ni\,{\sc i}   & 6.25 & & 4.91 & 0.11 & 22  & $-$1.34 & $-$0.13 & &  6.26 & 0.13 & 20  & $+$0.01 & $+$0.20 \\
    Rb\,{\sc i}   & 2.60 & & ---  & ---  & --- &    ---  &    ---  & &  2.85 & ---  & syn & $+$0.25 & $+$0.44 \\
    Sr\,{\sc i}   & 2.97 & & 3.03 & 0.21 & 03  & $+$0.06 & $+$1.27 & &  3.80 & ---  & 02  & $+$0.83 & $+$1.02 \\
    Y\,{\sc ii}   & 2.24 & & 1.86 & 0.17 & 06  & $-$0.38 & $+$0.83 & &  3.35 & 0.02 & 03  & $+$1.11 & $+$1.30 \\      
    Zr\,{\sc i}   & 2.60 & & 2.73 & ---  & 02  & $+$0.13 & $+$1.34 & &  3.54 & 0.19 & 14  & $+$0.94 & $+$1.13 \\
    Nb\,{\sc i}   & 1.42 & & ---  & ---  & --- &    ---  &    ---  & &  2.62 & 0.10 & syn & $+$1.20 & $+$1.39 \\
    Mo\,{\sc i}   & 1.92 & & ---  & ---  & --- &    ---  &    ---  & &  3.47 & 0.18 & 04  & $+$1.55 & $+$1.74 \\
    Ru\,{\sc i}   & 1.84 & & ---  & ---  & --- &    ---  &    ---  & &  2.86 & 0.16 & 03  & $+$1.02 & $+$1.21 \\
    La\,{\sc ii}  & 1.17 & & 1.28 & 0.10 & 05  & $+$0.11 & $+$1.32 & &  2.75 & 0.14 & 04  & $+$1.58 & $+$1.77 \\
    Ce\,{\sc ii}  & 1.58 & & 2.05 & 0.22 & 08  & $+$0.47 & $+$1.68 & &  3.28 & 0.15 & 08  & $+$1.70 & $+$1.89 \\
    Nd\,{\sc ii}  & 1.50 & & 1.88 & 0.14 & 16  & $+$0.38 & $+$1.59 & &  3.20 & 0.12 & 14  & $+$1.70 & $+$1.89 \\
    Sm\,{\sc ii}  & 1.01 & & 1.14 & 0.21 & 07  & $+$0.13 & $+$1.34 & &  2.29 & 0.07 & 05  & $+$1.28 & $+$1.47 \\     
    Eu\,{\sc ii}  & 0.51 & & 0.00 & ---  & syn & $-$0.51 & $+$0.70 & &  1.11 & ---  & syn & $+$0.60 & $+$0.79 \\
    Er\,{\sc ii}  & 0.93 & & ---  & ---  & --- &    ---  &    ---  & &  2.70 & 0.26 & 03  & $+$1.77 & $+$1.96 \\
    W\,{\sc i}    & 1.11 & & ---  & ---  & --- &    ---  &    ---  & &  2.31 & ---  & syn & $+$1.20 & $+$1.39 \\
    Tl\,{\sc i}   & 0.90 & & ---  & ---  & --- &    ---  &    ---  & &  1.80 & ---  & syn & $+$0.90 & $+$1.09 \\ 
    Pb\,{\sc i}   & 1.95 & & 2.25 & ---  & syn & $+$0.30 & $+$1.51 & &  3.25 & ---  & syn & $+$1.30 & $+$1.49 \\
    \midrule

    $^{12}\mathrm{C}/\,^{13}\mathrm{C}$ & & & 18 & & syn & &       &  &  20  &      & syn &         &        \\
    \bottomrule

    \end{tabular}
\end{table*}

\begin{table*}
\centering
\caption{Same as in Table~\ref{tab:bd_and_he}, for the stars HE~1255-2324 and HE~2207-1746.}\label{tab:he_and_he}
    \begin{tabular}{@{\extracolsep{4pt}}lccccccccccccc}
    \toprule
    & & & \multicolumn{5}{c}{HE~1255-2324} & & \multicolumn{5}{c}{HE~2207-1746}\\
    \cline{4-8}
    \cline{10-14}
    
    Species & $\log\epsilon_{\odot}$ & & $\log\epsilon$ & $\sigma_{\rm obs}$ & $n$(\#) & [X/H] & [X/Fe] & & $\log\epsilon$ & $\sigma_{\rm obs}$ & $n$(\#) & [X/H] & [X/Fe] \\
    \midrule

    C\,(C$_{2}$) & 8.52 & &  8.61 & ---  & syn &  $+$0.09 & $+$0.40 & &  8.49 & ---  & syn &  $-$0.03 & $+$0.52 \\
    N\,(CN)      & 7.92 & &  8.44 & ---  & syn &  $+$0.52 & $+$0.83 & &  8.25 & ---  & syn &  $+$0.33 & $+$0.88 \\
    O            & 8.83 & &  8.42 & ---  & syn &  $-$0.41 & $-$0.10 & &  8.33 & ---  & syn &  $-$0.50 & $+$0.05 \\
    Na\,{\sc i}  & 6.33 & &  6.35 & 0.21 & 04  &  $+$0.02 & $+$0.33 & &  6.02 & 0.11 & 03  &  $-$0.31 & $+$0.24 \\
    Mg\,{\sc i}  & 7.58 & &  7.49 & 0.10 & 04  &  $-$0.09 & $+$0.22 & &  7.36 & 0.04 & 04  &  $-$0.22 & $+$0.33 \\
    Al\,{\sc i}  & 6.47 & &  6.37 & 0.02 & 04  &  $-$0.10 & $+$0.21 & &  6.08 & 0.08 & 03  &  $-$0.39 & $+$0.16 \\
    Si\,{\sc i}  & 7.55 & &  7.30 & 0.07 & 03  &  $-$0.25 & $+$0.06 & &  7.29 & 0.12 & 04  &  $-$0.26 & $+$0.29 \\
    Ca\,{\sc i}  & 6.36 & &  6.11 & 0.05 & 05  &  $-$0.25 & $+$0.06 & &  5.99 & 0.12 & 06  &  $-$0.37 & $+$0.18 \\
    Ti\,{\sc i}  & 5.02 & &  4.87 & 0.07 & 10  &  $-$0.15 & $+$0.16 & &  4.67 & 0.12 & 15  &  $-$0.35 & $+$0.20 \\
    Fe\,{\sc i}  & 7.50 & &  7.19 & 0.16 & 31  &  $-$0.31 &    ---  & &  6.95 & 0.15 & 37  &  $-$0.55 &    ---  \\
    Fe\,{\sc ii} & 7.50 & &  7.20 & 0.15 & 05  &  $-$0.30 &    ---  & &  6.94 & 0.13 & 05  &  $-$0.56 &    ---  \\
    Cr\,{\sc i}  & 5.67 & &  5.49 & 0.12 & 08  &  $-$0.18 & $+$0.13 & &  5.13 & 0.17 & 09  &  $-$0.54 & $+$0.01 \\
    Ni\,{\sc i}  & 6.25 & &  6.16 & 0.07 & 13  &  $-$0.09 & $+$0.22 & &  5.73 & 0.03 & 13  &  $-$0.52 & $+$0.03 \\
    Rb\,{\sc i}  & 2.60 & &  3.90 & ---  & syn &  $+$1.30 & $+$1.61 & &  2.45 & ---  & syn &  $-$0.15 & $+$0.40 \\
    Sr\,{\sc i}  & 2.97 & &  4.11 & 0.24 & 03  &  $+$1.14 & $+$1.45 & &  3.35 & ---  & 02  &  $+$0.38 & $+$0.93 \\
    Y\,{\sc ii}  & 2.24 & &  3.54 & 0.14 & 04  &  $+$1.30 & $+$1.61 & &  2.93 & 0.10 & 04  &  $+$0.69 & $+$1.24 \\
    Zr\,{\sc i}  & 2.60 & &  3.99 & 0.10 & 10  &  $+$1.39 & $+$1.70 & &  3.19 & 0.13 & 12  &  $+$0.59 & $+$1.14 \\
    Nb\,{\sc i}  & 1.42 & &  3.22 & 0.15 & syn &  $+$1.80 & $+$2.11 & &  2.62 & 0.20 & syn &  $+$1.20 & $+$1.75 \\
    Mo\,{\sc i}  & 1.92 & &  3.92 & 0.02 & 03  &  $+$2.00 & $+$2.31 & &  ---  & ---  & --- &  ---     & ---     \\
    Ru\,{\sc i}  & 1.84 & &  3.03 & 0.03 & 03  &  $+$1.19 & $+$1.50 & &  2.38 & 0.10 & 03  &  $+$0.54 & $+$1.09 \\
    La\,{\sc ii} & 1.17 & &  2.77 & 0.12 & 04  &  $+$1.60 & $+$1.91 & &  2.40 & 0.15 & 04  &  $+$1.23 & $+$1.78 \\
    Ce\,{\sc ii} & 1.58 & &  3.06 & 0.16 & 07  &  $+$1.48 & $+$1.79 & &  2.73 & 0.15 & 08  &  $+$1.15 & $+$1.70 \\
    Nd\,{\sc ii} & 1.50 & &  3.14 & 0.15 & 20  &  $+$1.64 & $+$1.95 & &  2.88 & 0.16 & 17  &  $+$1.38 & $+$1.93 \\
    Sm\,{\sc ii} & 1.01 & &  2.31 & 0.11 & 04  &  $+$1.30 & $+$1.61 & &  1.78 & 0.12 & 04  &  $+$0.77 & $+$1.32 \\
    Eu\,{\sc ii} & 0.51 & &  1.40 & ---  & syn &  $+$0.89 & $+$1.20 & &  0.91 & ---  & syn &  $+$0.40 & $+$0.95 \\
    Er\,{\sc ii} & 0.93 & &  3.37 & 0.16 & 04  &  $+$2.44 & $+$2.75 & &  2.16 & ---  & 01  &  $+$1.23 & $+$1.78 \\
    W\,{\sc i}   & 1.11 & &  2.74 & ---  & syn &  $+$1.63 & $+$1.94 & &  2.06 & ---  & syn &  $+$0.95 & $+$1.50 \\
    Tl\,{\sc i}  & 0.90 & &  2.35 & ---  & syn &  $+$1.45 & $+$1.76 & &  1.90 & ---  & syn &  $+$1.00 & $+$1.55 \\
    Pb\,{\sc i}  & 1.95 & &  3.05 & ---  & syn &  $+$1.10 & $+$1.41 & &  2.95 & ---  & syn &  $+$1.00 & $+$1.55 \\
    \midrule

    $^{12}\mathrm{C}/\,^{13}\mathrm{C}$ & & & 18 & & syn &  &  &  & $\geq$ 40.0 & & syn &  & \\
    \bottomrule

    \end{tabular}
\end{table*}

\subsection{Abundance uncertainties}

The uncertainties in abundances come from errors associated with the (i) stellar atmospheric parameters ($T _{\rm eff}$, $\log g$, $\xi$, and [Fe/H]) and (ii) line-to-line scatter, associated with line parameters, as equivalent width ($W_{\lambda}$) measurements, line blending, continuum normalisation, and oscillator strength. The uncertainties due to errors in the equivalent width measurements were calculated following the expression given in \citet{cayrel1988} which takes into account the S/N ratio and the resolution of the spectra. For a resolution of $R=48\,000$, and the typical S/N ratios mentioned in Section~\ref{sec:observations}, the expected uncertainties in the equivalent widths are about $2-3$\,m{\AA}. To take into account the other components that contribute to the line-to-line scatter, we adopt the standard random errors, given by $\sigma_{\textrm{ran}}=\sigma_{\textrm{obs}}/\sqrt{n}$, where $n$ is the number of lines considered to derive the abundance of a given chemical element X, and $\sigma_{\textrm{obs}}$ is the standard deviation. In order to simplify the evaluation of the systematic errors associated with the stellar atmospheric parameters, we have assumed that their uncertainties are independent, following the methodology adopted in previous studies \citep[e.g.][]{decastro2016, karinkuzhi2018, cseh2018, roriz2021b}.

The final abundance uncertainties, $\sigma_{\log \epsilon (\rm{X})}$, were calculated according to the the following equation:

\begin{dmath}\label{eq:eq3}
     \sigma_{\log \epsilon(\textrm{X})_{\star}}^{2}=\sigma_{\textrm{ran}}^{2}+\left(\frac{\partial \log \epsilon}{\partial T_{\textrm{eff}}}\right)^{2}\sigma_{T_{\textrm{eff}}}^{2}+ \left(\frac{\partial \log \epsilon}{\partial \log g}\right)^{2}\sigma_{\log g}^{2}+ \left(\frac{\partial \log \epsilon}{\partial \xi}\right)^{2}\sigma_{\xi}^{2}+ \left(\frac{\partial \log \epsilon}{\partial \textrm{[Fe/H]}}\right)^{2}\sigma_{\textrm{[Fe/H]}}^{2}+ \left(\frac{\partial \log \epsilon}{\partial W_{\lambda}}\right)^{2}\sigma_{W_{\lambda}}^{2}, 
\end{dmath}

\noindent similar to the formalism presented by \citet[][]{mcwilliam1995}, where the partial derivatives correspond to variations in abundance when we change one parameter, keeping the others constant. Finally, the uncertainty in the [X/Fe] ratio is given by:

\begin{equation}
    \sigma_{\textrm{[X/Fe]}}^{2}=\sigma_{\textrm{X}}^{2}+\sigma_{\textrm{Fe}}^{2}.
\end{equation}

\noindent The partial derivatives in Equation~(\ref{eq:eq3}) were evaluated for the stars BD$+03\degr$2688 and HE~1255-2324. The results for HE~1255-2324 were also assumed for the other two targets, HE~0457-1805 and HE~2207-1746; this assumption is reasonable, since these two objects have uncertainties similar to HE~1255-2324. However, the $\sigma_{\textrm{ran}}$ term was computed for each stars, when three or more lines were used in the abundance derivation. 

The changes in $\log \epsilon (\rm X)$ for the elements from sodium to lead are presented in Tables~\ref{tab:uncertainties_bd03} and \ref{tab:uncertainties_he1255} of Appendix~\ref{app:uncertainties}, respectively for BD$+03\degr$2688 and HE~1255-2324. From these tables, one can see that the abundances based on the neutral elements are more sensitive to the temperature variation that those based on singly-ionized elements, which are more sensitive to surface gravity variation, such as seen in the variations of the abundances of Fe\,{\sc i} and Fe\,{\sc ii}. 

For the light elements, carbon, nitrogen, and oxygen, we followed the same procedure by varying the atmospheric parameters and then computing independently the abundance changes introduced by them, as shown in Tables~\ref{tab:uncertainties_cno_bd03} and \ref{tab:uncertainties_cno_he1255}, respectively for BD$+03\degr$2688 and HE~1255-2324. The derived CNO abundances are not very sensitive to the variations of the microturbulent velocity, since weak lines were used for their determination. In addition, the uncertainties in one of the light elements may also affect the abundance of another light element, such as an uncertainty of the oxygen abundance affects the carbon abundance and vice versa. In the same way, uncertainties in the carbon abundance result in variation of nitrogen abundances, since the CN molecule lines are used for the N abundance determination. These changes are also presented in Tables~\ref{tab:uncertainties_cno_bd03} and \ref{tab:uncertainties_cno_he1255}. The uncertainty in the $^{12}$C/$^{13}$C ratio was estimated from changes in the atmospheric parameters of HE~1255-2324, as shown in the last line of Table~\ref{tab:uncertainties_cno_he1255}.

\section{Results and discussion}\label{sec:results}

\subsection{The position of the stars in the HR diagram and their luminosities}\label{sub:luminosities}

\begin{figure}
    \centering
    \includegraphics{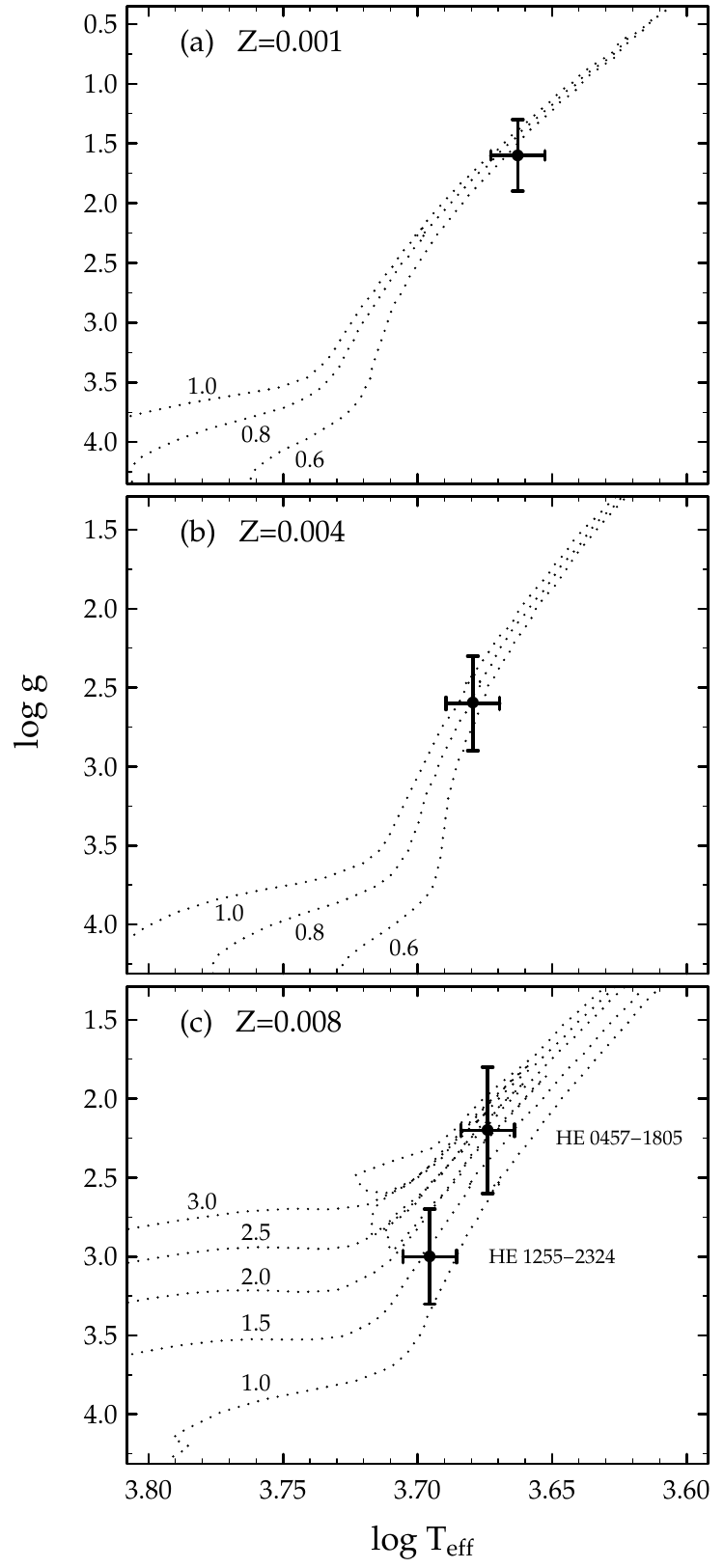}
    \caption{Position of the stars BD$+03\degr$2688 (panel a), HE~2207-1746 (panel b), HE~0457-1805, and HE~1255-2324 (panel c) in the $\log g$\,-\,$\log T_{\rm eff}$  diagram. The dotted curves are evolutionary tracks taken from \citet{girardi2000}, where the numbers correspond to stellar masses in units of solar mass (M$_{\odot}$).}
    \label{fig:tracks}
\end{figure}

We estimate the stellar masses of the stars analyzed in this work from their position in the $\log g$\,-\,log\,$T_{\rm eff}$ diagram which are seen in Figure~\ref{fig:tracks}. We used the evolutionary tracks of \citet{girardi2000} for metallicities $Z$\,=\,0.001 for BD$+03\degr$2688, $Z$\,=\,0.004 for HE~2207-1746, and $Z$\,=\,0.008 for HE~0457-1805 and HE~1255-2324, taking into account the results for $T_{\rm eff}$ and $\log g$ given in Table~\ref{tab:atm_par}. We obtained a mass of $0.6-1$\,M$_{\odot}$ for BD$+03\degr$2688, $\sim 1.5-3$\,M$_{\odot}$ for HE~0457-1805, $1-2$\,M$_{\odot}$ for HE~1255-2324, and $0.6-1$\,M$_{\odot}$ for HE~2207-1746.

Their luminosities, along with uncertainties associated with the temperatures and surface gravities given in Table~\ref{tab:atm_par}, were then obtained from the following relation:

\begin{equation}\label{eq:luminosity}
\log \left (\frac{L_{\star}}{L_{\odot}} \right )=4\log T_{\rm eff \star} - \log g_{\star} + \log \left (\frac{M_{\star}}{M_{\odot}} \right )-10.61.
\end{equation}

\noindent For BD$+03\degr$2688, HE~0457-1805, HE~1255-2324, and HE~2207-1746, the resulting luminosities according to the above equation are log\,($L_{\star}/L_{\odot})$\,=\,2.22$_{-0.29}^{+0.31}$, 2.28$_{-0.49}^{+0.44}$, 1.35$_{-0.34}^{+0.34}$, and 1.41$_{-0.33}^{+0.33}$, respectively. To compute these luminosities based on distance, we adopted the bolometric correction (BC) relation of \citet{alosno1999}, using the effective temperature and metallicity derived here, and $M_{{\rm bol} \odot} = +4.74$ \citep{bessell1998}. The extinction was determined by using calibrations among $A_{\rm V}$, Galactic coordinates, and distances given by \citet{chen1998}. For the stars BD$+03\degr$2688 and HE~2207-1746, which are objects with high Galactic latitudes, the extinctions are very small; we have adopted the values $A_{\rm V}=$ 0.06 and 0.07, respectively, according to the reddening map provided by the Infrared Science Archive \citep[][]{irsa}. From the obtained distances, based in the parallaxes given by GAIA DR2 \citep[][]{bailer-jones2018}, we derived log\,($L_{\star}/L_{\odot})$\,=\,2.15, 2.38, 1.44, and 1.67, respectively for BD$+03\degr$2688, HE~0457-1805, HE~1255-2324, and HE~2207-1746, in good agreement with the luminosities derived from Equation~\ref{eq:luminosity}.

The luminosities of the target stars are all log\,($L_{\star}/L_{\odot})$\,$\leq$\,2.1, far below the threshold needed to be considered AGB stars, and therefore to become self enriched by the elements of the $s$-process (see Section~\ref{sub:sub:heavy_elements}). According to \citet{lattanzio1986} and \citet{vassiliadis1993}, the minimal luminosity for a star to be considered an AGB star is 1\,800\,$L_{\odot}$ and 1\,400\,$L_{\odot}$, or log\,($L_{\star}/L_{\odot})$\,=\,3.26 and 3.15, respectively.

An additional evidence that these stars are not currently experiencing internal $s$-process nucleosynthesis is the absence of technetium (Tc) in their atmospheres. Tc is an element with no stable isotope; the radioactive $s$-nuclide $^{99}$Tc has a half-live $T_{1/2} \sim 2.1 \times 10^{5}$ years. Therefore, its detection in the stellar spectra is a remarkable signature of the $s$-process currently taking place inside the stars. We searched for Tc lines in the spectra of the program stars. In Figure~\ref{fig:technetium}, we show the regions close to the Tc\,{\sc i} lines at 4\,238.19~\AA, 4\,262.27~\AA, and 5\,924.47~\AA, commonly used to its detection. The Tc\,{\sc i} transitions around 4\,238.19~\AA\ and 5\,924.47~\AA\ are clearly absent; the feature at 4\,262.27~\AA, as already reported by \citet{vaneck1999}, has a Nd\,{\sc ii} contribution at 4\,262.228~\AA. Then, as an exercise, we computed synthetic spectra close to this region (see middle panels of Figure~\ref{fig:technetium}), where we note that the absorption in 4\,262.2~\AA\ is due to the high amount of Nd present in these stars. Therefore, the non-detection of Tc in our targets is a further clue that $s$-process is not currently taking place inside these stars, and supports the hypothesis of the mass transfer mechanism to explain the overabundance of the $s$-process observed in these systems.

\subsection{Radial velocity}\label{sub:radial_velocity}

In this section, we will discuss the binary status of the program stars, a fundamental requirement to support the mass transfer scenario as origin of their chemical peculiarities. 

BD$+03\degr$2688 displays properties of a halo star; although its radial velocity is low (+32.8~km\,s$^{-1}$), it is a metal-poor star and has a high Galactic latitude. In addition, BD$+03\degr$2688 satisfies several pre-requisities to be considered a CH star: it is a $s$-process enriched star (as we will see in Section~\ref{sub:sub:heavy_elements}, BD$+03\degr$2688 has [$s$/Fe] = 1.36, where [$s$/Fe] is the averaged $s$-process abundance), with C/O = 1.23 (see Section~\ref{sub:sub:cno}), and absolute magnitude $M_V\!=\!-0.07\pm0.01$, which is slightly lower than the typical values among the CH stars \citep[according to Table IV of][CH stars have $M_V$ values between $-$0.25 and $-$2.2]{hartwick1985}. Like the CH stars, BD$+03\degr$2688 should be a binary system.

\begin{table}
\centering
\caption{Date of observation, Modified Julian Date (MJD), and the known values for the observed heliocentric radial velocity (Vr) of the star BD$+03\degr$2688.}\label{tab:vr_bd032688}
    \begin{threeparttable}
        \begin{tabular}{ccccc}
        \toprule
        
        Date of            & MJD       &  Vr            &  Error  & Ref.\\
        observation        &           & (km\,s$^{-1}$) &         &     \\
        \midrule
        
        Apr. 8, 1993  & 49086.481 & 33.16          &  0.43   & 1  \\
        Mar. 5, 1994  & 49417.506 & 32.35          &  0.47   & 1  \\
        Apr. 3, 1994  & 49446.510 & 32.33          &  0.44   & 1  \\
        May 14, 1994  & 49487.388 & 33.00          &  0.47   & 1  \\
        Feb. 3, 1995  & 49752.658 & 33.52          &  0.52   & 1  \\
        Mar. 30, 1995 & 49807.506 & 32.64          &  0.71   & 1  \\
        Dec. 20, 1995 & 50072.729 & 33.07          &  0.46   & 1  \\
        Apr. 4, 2007  & 54194.146 & 32.64          &  0.29   & 2  \\
        Mar. 18, 2016 & 57465.155 & 34.36          &  0.19   & 2  \\
        \bottomrule
        
        \end{tabular}
        \textbf{References:} (1) \cite{jorissen2005}; (2) This work.
    \end{threeparttable}
\end{table}

\begin{figure}
    \centering
    \includegraphics[width=\columnwidth]{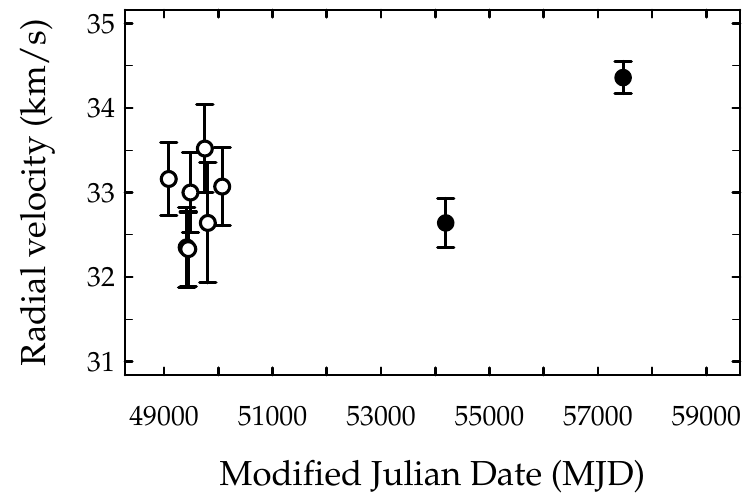}
    \caption{Radial velocity of BD$+03\degr$2688 as a function of the Modified Julian Date (MJD). Open points are data taken from \citet{jorissen2005}; the filled points (this work) at MJD\,=\,54194.146 and 57465.155 are data observed with the FEROS spectrograph at April 4, 2007 and March 18, 2016, respectively.}
    \label{fig:vr_bd032688}
\end{figure}

As seen in Section~\ref{sub:luminosities}, BD$+03\degr$2688 is not luminous enough to be considered an AGB star, in order to account for the observed overabundance of carbon and the elements created by the $s$-process. Therefore, the observed heavy-element overabundance could be due to a mass transfer happened in the past. In Table~\ref{tab:vr_bd032688} we list all
known heliocentric radial velocity measurements for BD$+03\degr$2688 available in the literature and determined in this work. Their mean value is $33.01 \pm 0.21$ km/s ($\sigma = 0.64$ km/s), essentially constant to within the stated measurement uncertainties. A similar conclusion was also found by \cite{jorissen2005}. The absence of the radial velocity variations may also be due to the predominance of the transverse velocity in the space velocity of this star. In Figure~\ref{fig:vr_bd032688}, we show the radial velocity measurements obtained by \cite{jorissen2005} and the two our values of 32.64$\pm$0.29 km\,s$^{-1}$ and 34.36$\pm$0.19 km\,s$^{-1}$ obtained in this study. Therefore, a systematic radial velocity monitoring is necessary to confirm the possible binary nature of BD$+03\degr$2688.

For the stars HE~0457-1805, HE~1255-2324, and HE~2207-1746, the heliocentric radial velocities were also determined based on Doppler shifts of ten absorption lines. For HE~0457-1805, we obtained 61.22$\pm$0.24 km\,s$^{-1}$, which is in the range between 60.92 km\,s$^{-1}$ to 73.99 km\,s$^{-1}$ due the variation of its binary period of 2724 days (\citealt{pourbaix2004}; see also \citealt{jorissen2016}). For HE~1255-2324, we obtained two different radial velocities: 37.25$\pm$0.26 km\,s$^{-1}$, from the spectrum obtained at May 4, 2010, and 18.21$\pm$0.21 km\,s$^{-1}$, from the spectrum obtained at August 1, 2010. In fact, HE~1255-2324 presents a variation of its radial velocity. The radial velocity experiment \citep[RAVE;][]{kunder2017} gives for this object 20.747$\pm$2.184 km\,s$^{-1}$. Finally, for HE~2207-1746, there is no previous radial velocity determination; for this system, we obtained $-8.01\pm$0.21 km\,s$^{-1}$.

\subsection{Light elements}

\subsubsection{C, N, O, and carbon isotopic ratios}\label{sub:sub:cno}

\begin{figure}
    \centering
    \includegraphics{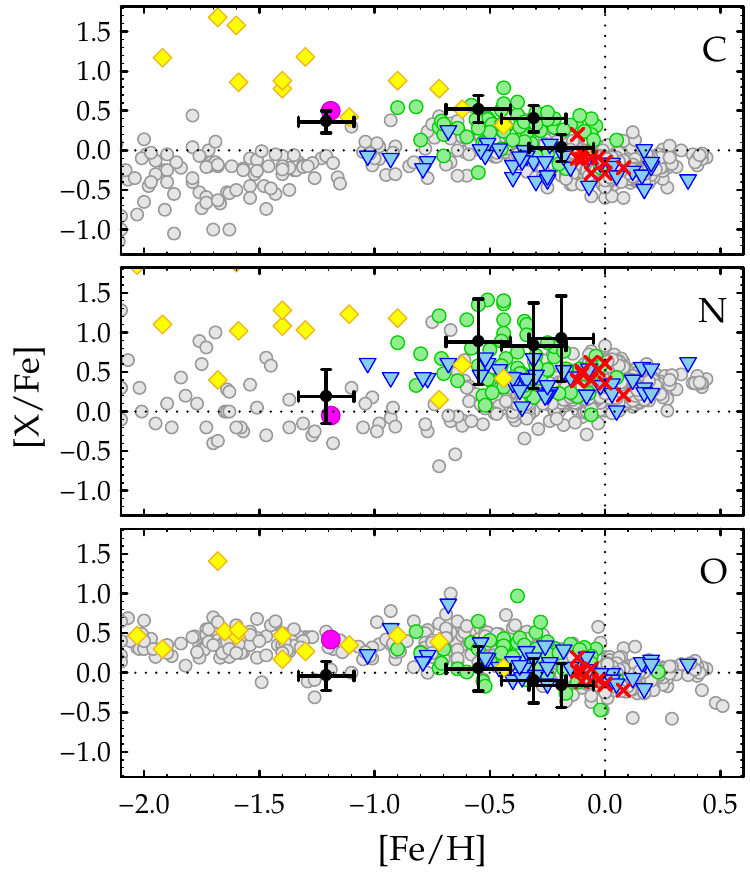}
    \caption{Observed [C, N, O/Fe] abundance ratios as a function of metallicity for the stars analyzed in this study (black dots with error bars) along with data for field giant stars \citep[grey dots, taken from][]{gratton1986, barbuy1988, barbuy1989, sneden1991, kraft1992, shetrone1996, gratton2000, carretta2000, simmerer2004, spite2005, mishenina2006, luck2007, smiljanic2009, alvesbrito2010, tautvaisiene2010, tautvaisiene2013, tautvaisiene2015, tautvaisiene2016, santrich2013, szigeti2018, takeda2019}, barium stars \citep[green dots, taken from][]{smith1984, barbuy1992, allen2006, pereira2009, karinkuzhi2018, shejeelammal2020}, CH stars \citep[yellow diamonds, taken from][]{vanture1992a, vanture1992b, pereira2009, pereira2012, goswami2016, purandardas2019}, S-type symbiotic stars \citep[blue triangles, taken from][]{galan2016, galan2017}, and M giant stars \citep[red crosses, taken from][]{smith1985}. The barium star HD~123396 is shown by the magenta dot (see discussion in the text).}
    \label{fig:cno_fe}
\end{figure}

Figure~\ref{fig:cno_fe} shows the [C/Fe], [N/Fe], and [O/Fe] ratios for the stars analyzed this work in comparison with data for field giant stars and other classes of chemically peculiar stars, as identified in the figure caption. Three out of the four program stars show carbon enrichment with [C/Fe] ratios above those observed for field stars, as typically found in barium stars and CH stars; they are BD$+03\degr$2688, HE~2207-1746, and HE~1255-2324, with [C/Fe] $=$ 0.36, 0.52, and 0.40, respectively. For HE~0457-1805 the [C/Fe] ratio is just slightly above the average trend of the field stars. The excess of carbon can be explained in terms of mass transfer from a previous carbon enriched AGB star, as a result of the helium burning and the occurrence of the third dredge-up.

From the middle panel of Figure~\ref{fig:cno_fe}, we see that the stars HE~0457-1805, HE~1255-2324, and HE~2207-1746 present a nitrogen abundance higher than the main trend of field stars, in accordance with high [N/Fe] ratios observed in some barium stars. The observed nitrogen overabundance may be due to the occurrence of the first dredge-up in the barium giant \citep[][]{barbuy1992, drake2008, allen2006}. Interesting, BD$+03\degr$2688 does not show a high [N/Fe] ratio, usually seen in chemically peculiar stars enriched by the mass transfer down to metallicity $\approx -1.0$. Nonetheless, its [N/Fe] ratio is similar to that of the other metal-poor barium star, HD~123396 \citep{allen2006}, thus indicating that in both stars mixing was not very efficient, similar to a phenomena already seen, but at lower metallicities, in possible single stars \citep[][]{spite2005, spite2006}. BD$+03\degr$2688 has also a [O/Fe] ratio lower than other field stars in the same range of metallicity, similar to the star BD+30$\degr$2611 ([O/Fe] $=+0.04$; \citealt{kraft1992}) of the same metallicity. Other chemically peculiar stars also have low [O/Fe] ratios considering their metallicities \citep{vanture1992b}.

\begin{figure}
    \centering
    \includegraphics{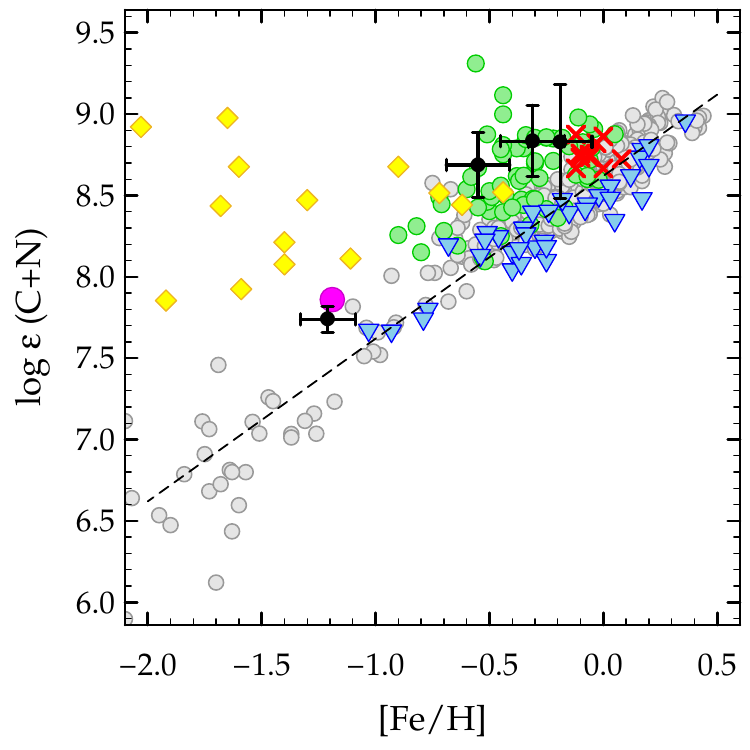}
    \caption{Observed C+N abundance in the notation of $\log \epsilon (\rm{C+N})$ as a function of metallicity. Symbols have the same meaning as in Figure~\ref{fig:cno_fe}. The black dashed line shows initial CN abundance for a given metallicity. The barium star HD~123396 is shown by the magenta dot (see discussion in the text).}
    \label{fig:c_plus_n}
\end{figure}

Dwarf stars do not exhibit nitrogen enrichment, and in the metallicity range between $-$2.0 and $+$0.3, the [N/Fe] ratio remains close to zero \citep{tomkin1984}. However, as the star becomes a red giant, the first dredge-up brings nuclear processed material from its interior layers, changing the stellar surface composition; as a result, there is a depletion of $^{12}\mathrm{C}$ and an enrichment of $^{14}\mathrm{N}$ \citep[][]{lambert1981}. Therefore, the sum $\log \epsilon $(C$+$N) should remain unchanged. The sample of giants stars analyzed by \cite{luck2007} shows that, in fact, the $\log \epsilon $(C$+$N) did not significantly changed as these stars became giants. For that sample, we found a mean of 8.66$\pm$0.17 for $\log \epsilon $(C$+$N), for a metallicity range between $-$0.6 and $+$0.3; this value is only $+$0.04 higher than the $\log \epsilon $(C$+$N)$_{\odot}$, for the solar carbon and nitrogen abundances given in Column 2 of Tables~\ref{tab:bd_and_he} and \ref{tab:he_and_he}. Figure~\ref{fig:c_plus_n} shows $\log \epsilon $(C$+$N) as a function of metallicity, where the black dashed line is the solar scaled composition, and includes data for $\log \epsilon $(C$+$N) from the giants analyzed by \citeauthor{luck2007}, symbiotic stars, M giants, barium stars, and CH stars. Figure~\ref{fig:c_plus_n} also shows how $\log \epsilon $(C$+$N) behaves for the stars studied in this work. HE~0457-1805, HE~1255-2324, and HE~2207-1746 present an excess of carbon and nitrogen, sharing the same position where we found barium stars in the diagram. Taken together, the excess of carbon is best explained in terms of mass transfer, while the excess of nitrogen is due the occurrence of the first dredge-up in the barium giant. The position of BD$+03\degr$2688 in this diagram also indicates that this star presents an excess of $\log \epsilon $(C$+$N), but this is due to excess of carbon only since it has a low [N/Fe] ratio like the barium star HD~123396 \citep[][]{allen2006}.

\begin{figure}
    \centering
    \includegraphics{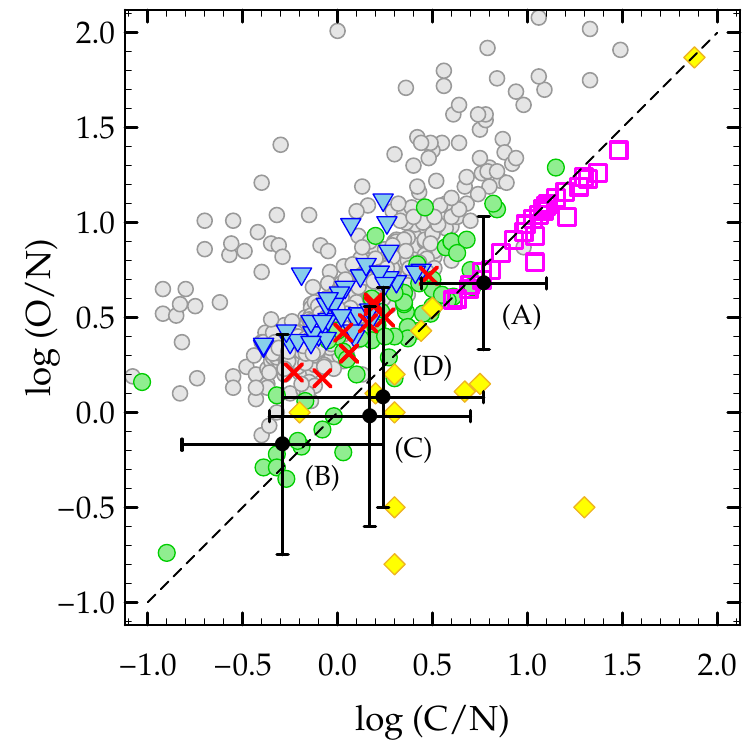}
    \caption{Abundance ratios by number O/N against C/N. Symbols have the same meaning as in Figure~\ref{fig:cno_fe}. We have added to this plot abundance data for the classical carbon stars \citep[magenta squares, taken from][]{lambert1986}. The program stars, BD$+03\degr$2688, HE~0457-1805, HE~1255-2324, and HE~2207-1746, are respectively labeled as (A), (B), (C) and (D).}
    \label{fig:on_vs_cn}
\end{figure}

Figure~\ref{fig:on_vs_cn} shows the stars analyzed in this work in the $\log {\rm O/N }$ ratio {\sl versus} the $\log {\rm C/N}$ ratio diagram for the same classes of stars shown in previous figures. Further, we added in Figure~\ref{fig:on_vs_cn} the classical Galactic carbon stars with the aim to show where the carbon enriched objects are located. Carbon rich objects lie below the black dashed line that represents C/O\,=\,1, from which we can better set the evolutionary status of the stars analyzed in this work. In this diagram, the star HE~0457-1805 (C/O\,=\,0.76), labeled as (B), shares the same region where other barium stars are found, while BD$+03\degr$2688 (C/O\,=\,1.23), HE~1255-2324 (C/O\,=\,1.55), and HE~2207-1746 (C/O\,=\,1.45), respectively labeled as (A), (C), and (D) in the diagram, lie on the side of carbon rich objects. Therefore, HE~2207-1746 (D) and HE~1255-2324 (C) are likely metal-rich CH stars, and spectroscopically can be confused as ''barium stars''. This explains why they may have the same position of the barium stars in previous Figure~\ref{fig:c_plus_n}.

\begin{figure}
    \centering
    \includegraphics{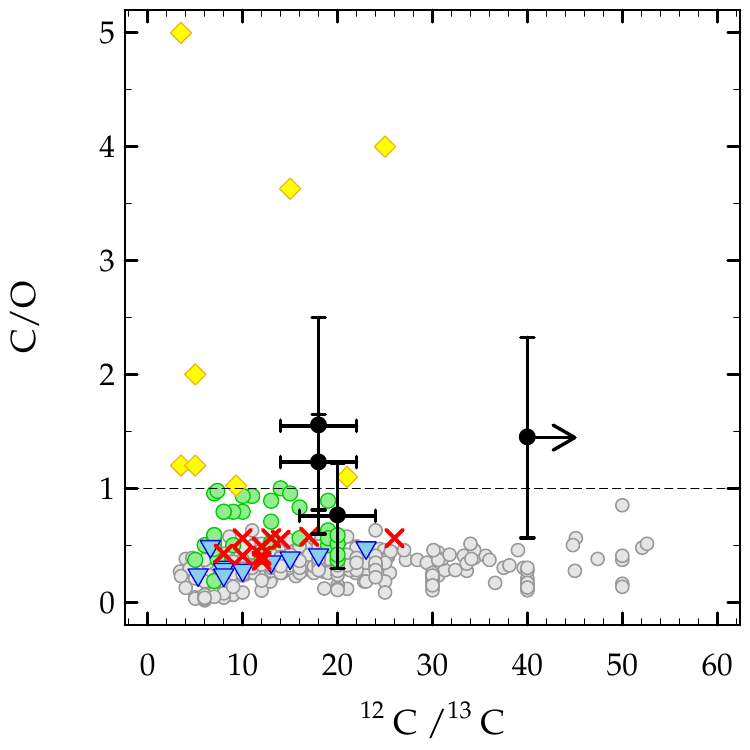}
    \caption{Carbon isotopic ratio, $^{12}\mathrm{C}/\,^{13}\mathrm{C}$, against C/N ratio for the program stars. Symbols have the same meaning as in Figure~\ref{fig:cno_fe}.}
    \label{fig:c12c13}
\end{figure}

In Figure~\ref{fig:c12c13}, we plot the C/O ratio {\sl versus} the $^{12}\mathrm{C}/\,^{13}\mathrm{C}$ isotopic ratio for the program stars, along with data for field giant stars and other chemically peculiar stars. The results determined for BD$+03\degr$2688, HE~0457-1805, and HE~1255-2324 are in accordance with other $^{12}\mathrm{C}/\,^{13}\mathrm{C}$ isotopic ratios already determined for barium stars \citep[][]{smith1984, barbuy1992, pereira2009, karinkuzhi2018}. Low $^{12}\mathrm{C}/\,^{13}\mathrm{C}$ isotopic ratios in combination with high [N/Fe] ratios follow the expectations given by the first dredge-up \citep[][]{lambert1981}. HE~2207-1746, on the other side, has a high nitrogen abundance ([N/Fe]=+0.88) and a $^{12}\mathrm{C}/\,^{13}\mathrm{C}$ isotopic ratio with value higher than 40, not commonly seen among the barium stars and CH stars, which may be due to the mass transfer from an AGB carbon star with high $^{12}\mathrm{C}/\,^{13}\mathrm{C}$.

According to the classical theory of stellar evolution \citep[][]{iben1964, iben1965, iben1967}, the $^{14}\mathrm{N}/\,^{12}\mathrm{C}$ abundance ratio and the $^{12}\mathrm{C}/\,^{13}\mathrm{C}$ isotopic ratio on the stellar surface change due to the dredging-up CN-processed material, caused by a deepening of the convective zone of the star approaching the red giant branch phase (first dredge-up). Comparison with observational data and further development of the theory of the first dredge-up are presented in \citet[][]{lambert1981, sneden1991iau, charbonnel1994, charbonnel1995, boothroyd1999, gratton2000}, among others. The origin of the nitrogen in extremely metal-poor giants was studied by \citet[][]{spite2005}. The authors found that the more evolved giants of their sample show evidence of CN cycling as well as strong Li dilution which confirms mixing by the first dredge-up at lower metallicities.

\subsubsection{Elements from sodium to nickel}

\begin{figure*}
    \centering
    \includegraphics{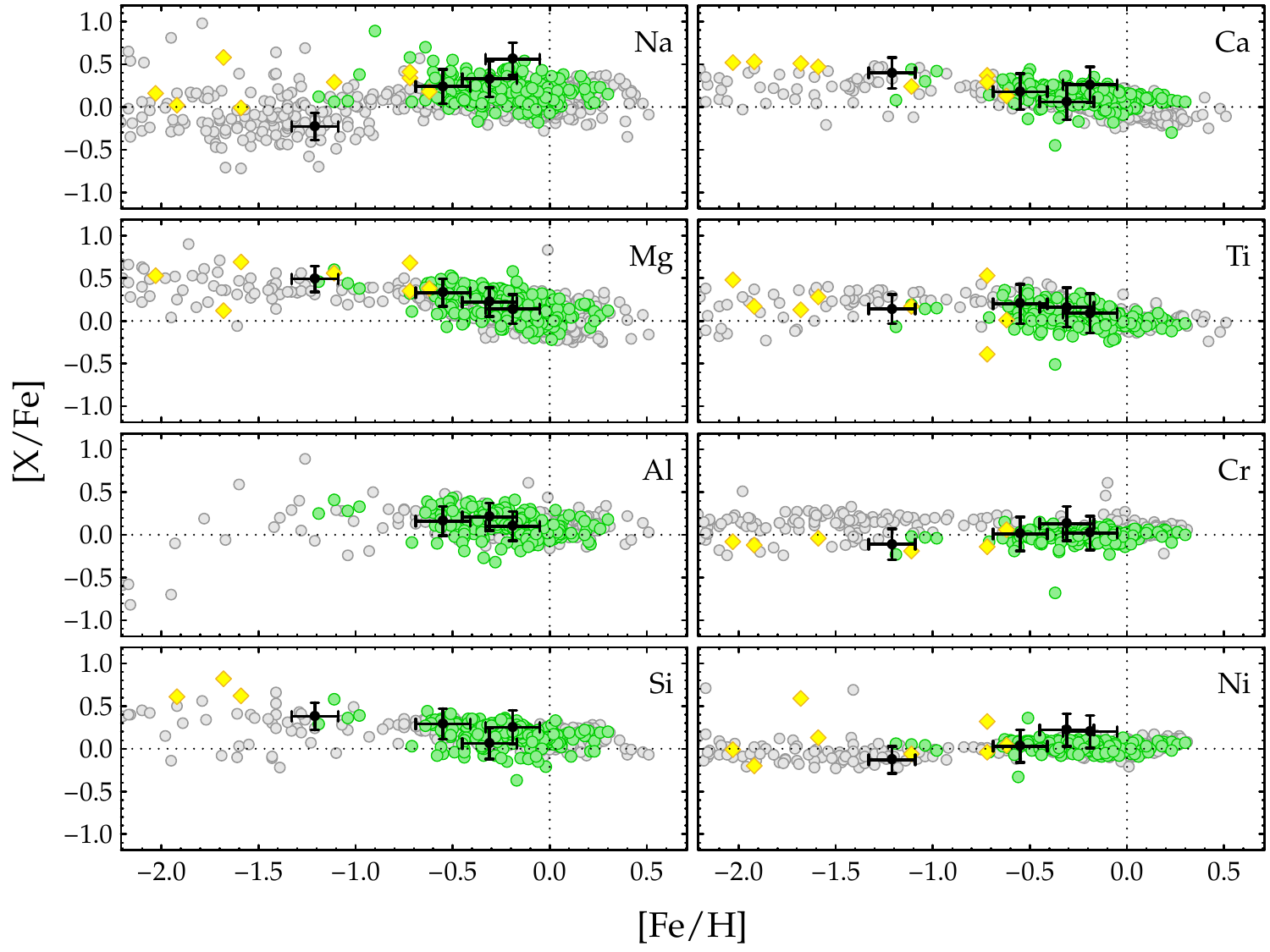}
    \caption{Observed [X/Fe] abundance ratios as a function of metallicity for the stars analyzed in this study (black dots with error bars) along with data for field giant stars \citep[grey dots, taken from][]{kraft1992, shetrone1996, gratton2000, carretta2000, johnson2002, mishenina2006, luck2007, zhang2009, smiljanic2009, alvesbrito2010, for2010, santrich2013, szigeti2018, ishigaki2018, takeda2019}, barium stars \citep[green dots, taken from][]{allen2006, pereira2011, decastro2016, karinkuzhi2018, shejeelammal2020}, and CH stars \citep[yellow diamonds, taken from][]{pereira2009, pereira2012, goswami2016, purandardas2019}.}
    \label{fig:light_elements}
\end{figure*}

In Figure~\ref{fig:light_elements}, we show the [X/Fe] abundance ratios for sodium, magnesium, aluminum, silicon, calcium, titanium, chromium, and nickel in comparison with data for field giants stars, barium stars, and CH stars. It can be seen that the target stars behave similar to the field stars, as well as the barium stars and CH stars. Sodium, aluminum, magnesium, silicon, calcium, and titanium are mainly produced in the Galaxy by hydrostatic burning in massive stars of initial masses roughly 10-30 $M_{\odot}$ \citep[][]{woosley1995}. Sodium, aluminum, and magnesium are produced by hydrostatic carbon burning, although magnesium can also be produced by neon burning. Silicon and calcium can be produced by hydrostatic oxygen burning. Type II supernova explosions may also originates large amounts of silicon and calcium. Titanium is produced by oxygen burning but can also be produced by Type Ia supernovae events \citep[][]{woosley1995}. Because of the atomic number of titanium, \citet{timmes1995} considered this element as an iron-peak element in their analysis of the chemical evolution of the Galaxy. Iron, chromium, and nickel are formed in Type I supernova events and also during the final stages of the massive stars \citep[][]{timmes1995}; therefore, these elements should follow the same trend as the iron abundance, as we can see in Figure~\ref{fig:light_elements}.

Compared to the field stars, HE~0457-1805 presents a high [Na/Fe] ratio of $+$0.56. Field stars with high sodium abundances, that is, those with [Na/Fe] ratios higher than $+$0.5 to $+$0.7 are rare \citep{pereira2019b}. Stars with [Na/Fe] ratios with values closer to $+$1.0 can be considered ''second generation of globular cluster star'' that has escaped from globular cluster and now is a field halo star \citep{pereira2019a}. Field giant stars that exhibit sodium enrichment up to $\approx +0.40$ have already been reported, but only in a few cases. In the samples of the studied stars by \citet{luck2007} and \citet{takeda2008}, only $3-4$\% of stars have [Na/Fe] ratios between $+$0.30 and $+$0.40. HD~188650, analyzed by \citeauthor{takeda2008}, with a metallicity of $-$0.67, has a high [Na/Fe] ratio of $+$0.62, which is only 0.3\% of the total sample. \citet{karinkuzhi2014} also reported a high [Na/Fe] ratio for this star ($+$0.76). These authors also rejected HD~188650 as a CH star due to its low heavy-element abundance. This star was classified as ''low velocity CH star'' by \citet{bidelman1957}, i.e. stars that have enhanced CH bands but with low radial velocities, contrary to the classical halo CH metal-poor stars which have high radial velocities. \citet{vanture1999}, in their analysis of the stars in Hertzsprung Gap, concluded that these low velocity CH stars with their mild metal deficient ''will enhance the appearance of the CH band on low-resolution spectra''. Since the stars that are in the Hertzsprung Gap have just left the main-sequence, probably their high [Na/Fe] ratios are owed to NeNa cycle \citep{denisenkov1987} in its hydrogen-burning core while the star was still on the main sequence. In barium stars, sodium enrichment at a similar quantity as seen is HE~0457-1805 has already been reported \citep{antipova2004, decastro2016, karinkuzhi2018}. Such enrichments can be currently explained through the reaction $^{22}$Ne($p$,$\gamma$)$^{23}$Na in the hydrogen burning shells \citep{mowlavi1999} in the previous AGB companion stars.

\subsection{Heavy elements}\label{sub:sub:heavy_elements}

The bulk of the chemical elements beyond the iron group is synthesized via neutron-capture nuclear reactions, as previously mentioned, since neutrons do not need to overcome the Coulomb barrier to penetrate into atomic nuclei. Depending on the timescales involved between neutron captures and $\beta$-decays of the radioactive nuclei, the mechanism can operates into two extreme regimes, the so called $s$-process and $r$-process, which require different astrophysical conditions. The physical conditions required by the $s$-process are found within the He-rich region between the H-burning shell and He-burning shell, which are alternately activated during the TP-AGB phase; the $s$-process is powered by neutron density of the order of $10^{7-12}$ cm$^{-3}$ \citep[e.g.][]{busso2001, abia2001, cristallo2011, vanraai2012, fishlock2014} and evolves close to the valley of $\beta$ stability. The $r$-process, on the other hand, is associated with explosive events, such as those found in supernovae and neutron star mergers \citep[see][for a recent review]{cowan2021}, where high neutron densities ($\gtrsim 10^{20}$ cm$^{-3}$) are needed. 

\begin{figure*}
    \centering
    \includegraphics{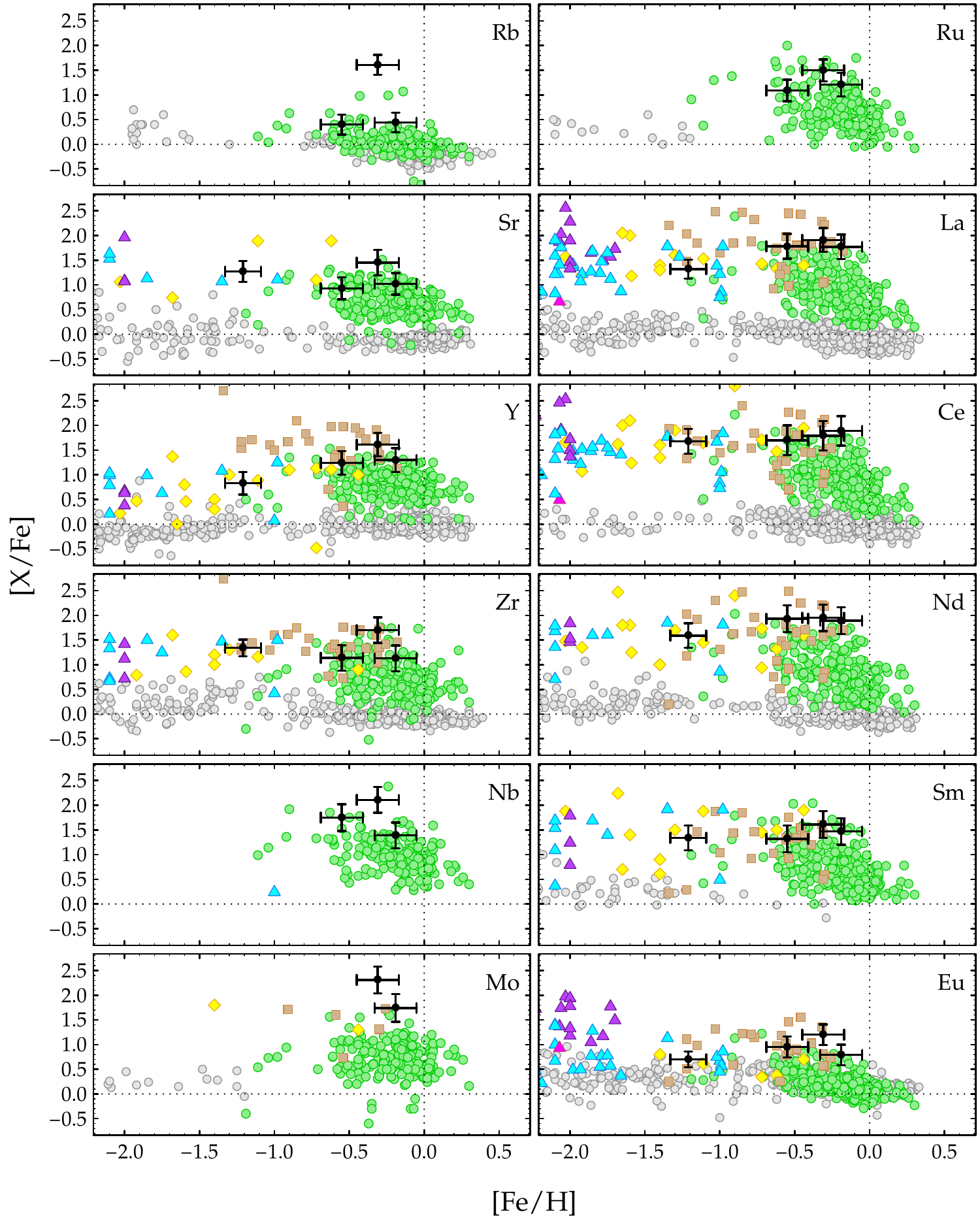}
    \caption{Observed [X/Fe] abundance ratios as a function of metallicity for the stars analyzed in this study (black dots with error bars) along with data for field giant stars \citep[grey dots, taken from][]{gratton1994, tomkin1999, burris2000, mishenina2001, mishenina2007, johnson2002, luck2007, zhang2009, for2010, santrich2013, hansen2014, ishigaki2018, forsberg2019, abia2021, takeda2021}, barium stars \citep[green dots, taken from][]{allen2006, allen2007, karinkuzhi2018, shejeelammal2020, roriz2021a, roriz2021b}, CH stars \citep[yellow diamonds, taken from][]{vanture1992c, pereira2009, pereira2012, goswami2016, purandardas2019}, and post-AGB stars \citep[brown squares, taken from][]{vanwinckel2000, reyniers2004, desmedt2012, desmedt2015, desmedt2016, vanaarle2013}. Triangles are data for CEMP-$s$ (cyan), CEMP-$r$ (magenta), and CEMP-$r/s$ (purple) stars, taken from \citet{masseron2010} and \citet{karinkuzhi2021}.}
    \label{fig:heavy_elements}
\end{figure*}

Both $s$-process and $r$-process contribute to the cosmic abundances of the elements from iron to bismuth. We show in Table~\ref{tab:s-fraction} the $s$-process contribution to the observed solar-system material for the chemical elements considered in the present study; these data were taken from different works in literature. Additionally, in Figure~\ref{fig:heavy_elements}, we compare the observed [X/Fe] abundance ratios derived for the targets of this work with data available in literature for field giant stars, barium stars, CH stars, CEMP stars, and post-AGB stars, where we can clearly see the $s$-rich nature of the program stars, i.e. large overabundances of $n$-capture elements compared to the normal giant stars of the Galaxy (grey dots in Figure~\ref{fig:heavy_elements}). Such characteristics will be discussed below. As we previously noted, these stars have not yet experienced the $s$-process nucleosynthesis in their interior, so that the presence of heavy elements in their atmospheres is probably a consequence of the mass transfer from a former AGB star.

\begin{table}
\centering
\caption{Contribution (\%) of the $s$-process to the solar-system material as provided by different works in literature for the neutron-capture elements considered in this study.} \label{tab:s-fraction}
    \begin{threeparttable}
        \begin{tabular}{lcccc}
        \toprule

        Species &   S96  & A99 & B00 & B14\\
	    \midrule
	    Rb  &     97 & 22      & 50  & 18 \\
        Sr  &     85 & 85      & 89  & 69 \\
        Y   &     72 & 92      & 72  & 72 \\
        Zr  &     83 & 83      & 81  & 66 \\
        Nb  &     68 & 85      & 68  & 56 \\
        Mo  &     68 & 50      & 68  & 39 \\
        Ru  &     39 & 32      & 39  & 29 \\
        La  &     75 & 62      & 75  & 76 \\
        Ce  &     77 & 77      & 81  & 84 \\
        Nd  &     47 & 56      & 47  & 58 \\
        Sm  &     24 & 29      & 34  & 31 \\
        Eu  &     3  & 6       & 3   & 6  \\
        Er  &     16 & 17      & 16  & 20 \\
        W   &     54 & 56      & 54  & 62 \\
        Tl  &     66 & 76      & 66  & 71 \\
        Pb  &     79 & 46      & 79  & 87 \\
    	\bottomrule
     
        \end{tabular}
        \textbf{References:} (S96) \citet{sneden1996}; (A99) \citet{arlandini1999}; (B00) \citet{burris2000}; (B14) \citet{bisterzo2014}.
    \end{threeparttable}
\end{table}

\begin{figure}
    \centering
    \includegraphics{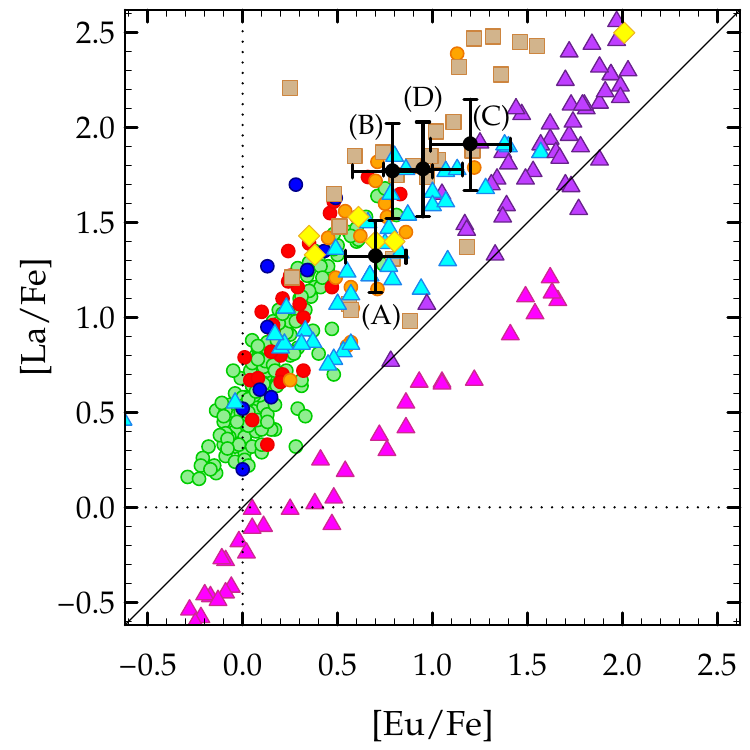}
    \caption{[La/Fe] {\sl versus} [Eu/Fe] plane showing the distribution of the program stars (black dots with error bars), BD$+03\degr$2688 (A), HE~0457-1805 (B), HE~1255-2324 (C), and HE~2207-1746 (D). Symbols have the same meaning as in Figure~\ref{fig:heavy_elements}. However, for the barium stars, we have adopt a color code to identify the objects analyzed by \citet[][red]{allen2006}, \citet[][orange]{karinkuzhi2018}, \citet[][blue]{shejeelammal2020}, and \citet[][green]{roriz2021b}.}
    \label{fig:la_vs_eu}
\end{figure}

In Figure~\ref{fig:la_vs_eu}, we show the [La/Fe] {\sl versus} [Eu/Fe] plane. The location of the stars in this diagram evidences their $s$-rich or $r$-rich nature, since La and Eu are representative elements of the $s$-process ($\sim$75\%) and $r$-process ($\gtrsim$ 95\%), respectively. We have also added to this plot abundance data available for barium stars, CH stars, CEMP stars, and post-AGB stars. As one can observe in Figure~\ref{fig:la_vs_eu}, the sub classes of the CEMP stars occupy different regions in this plot. While CEMP-$r$ stars (magenta triangles) are found on the $r$-rich side of this diagram, CEMP-$s$ stars (cyan triangles) are located on the $s$-rich side, lying close to barium stars, CH stars, and post-AGB stars. Note that CEMP-$r/s$ stars (purple triangles), in turn, represent a different population in this plane. The targets analyzed in this study, BD$+03\degr$2688 (A), HE~0457-1805 (B), HE~1255-2324 (C), and HE~2207-1746 (D), show high levels of $s$-enrichment, with [$s$/Fe] $=$ 1.36, 1.36, 1.77, and 1.38, respectively, where [$s$/Fe] is the average abundance of the $s$-elements, computed from the elemental abundances of Rb, Sr, Y, Zr, Nb, Mo, Ru, La, Ce, Nd, W, Tl, and Pb. As expected, these stars are found on the $s$-rich side of the [La/Fe] {\sl versus} [Eu/Fe] diagram. In particular, HE~1255-2324 (C) falls very close to the region where CEMP-$r/s$ stars are found. So far, similar characteristics were reported by \citet{karinkuzhi2018} for the barium star HD~100503 ([La/Fe] $=1.79$; [Eu/Fe] $=1.22$), with [$s$/Fe] $=1.46$ at [Fe/H] $=-0.72$.

Recently, \citet{karinkuzhi2021} have proposed a $d_{\rm{rms}}$ distance, in dex, of the observed abundance pattern from the standard solar $r$-process abundance profile, in order to constraint among the different classes of CEMP stars. The novelty of this metric lies in the fact that it is not based only on two elemental abundances (commonly La and Eu), but in a set of $N$ elemental abundances. As advantage, this new approach considers at the same time all the abundance data derived for the heavy elements, becoming more refined as more chemical elements are taken into account. By definition, the $d_{\rm{rms}}$ quantity is null for the $r$-element europium, that is assumed as a normalizing element. In summary, the more the $s$-process contribution dominates the stellar abundance pattern, the further away from the solar $r$-process profile the observed data set will be, reflecting in greater $d_{\rm{rms}}$ distance values. We refer the reader to Section 5 of \citeauthor{karinkuzhi2021} (their Equations 4 and 5) for further details. 

In \citet{roriz2021b}, we reported the average $d_{\rm{rms}}$ values for barium stars (1.15) and field stars (0.50). In the present study, we computed $d_{\rm{rms}}=$ 1.29, 1.25, 1.21, and 1.10 dex for BD$+03\degr$2688, HE~0457-1805, HE~1255-2324, and HE~2207-1746, respectively. These values, indeed, confirm the strong $s$-rich nature of the program stars. In Figure~\ref{fig:diagnostic}, similar to Figure 8 of \citet[][]{roriz2021b}, we show a plot of the [$s$/Fe] ratio against $d_{\rm{rms}}$ for the systems analyzed in this work, along with data for field stars, barium stars, CH stars, and post-AGB stars, where we can see that our stars exhibit [$s$/Fe] and $d_{\rm{rms}}$ on the top of the values observed in barium stars.

As far as the neutron source of the $s$-process is concerned, the $^{13}$C$(\alpha,n)^{16}$O and $^{22}$Ne$(\alpha,n)^{25}$Mg reactions provide the free neutrons needed to drive the $s$-process. For low-mass ($1 - 4$ M$_{\odot}$) AGB stars, the $^{13}$C$(\alpha,n)^{16}$O reaction is the main neutron source, effectively activated during the interpulse period, at $T \sim 10^{8}$ K, under radiative conditions, within a so called $^{13}$C pocket formed in the He-rich intershell region, whereas the $^{22}$Ne source is marginally activated for this mass range. For more massive ($4 - 8$ M$_{\odot}$) AGB stars, the $^{22}$Ne$(\alpha,n)^{25}$Mg reaction becomes the main neutron source, activated during the thermal pulses, at $T  \geq 3 \times 10^{8}$ K, under convective conditions, releasing higher neutron densities. The element rubidium is an useful chemical species to probe and constraint the main neutron source of the $s$-process. This is due to the fact that the amount of Rb produced depends on the unstable isotopes $^{85}$Kr ($T_{1/2} \sim 11$ years) and $^{86}$Rb ($T_{1/2} \sim 19$ days) which act as branching points along the $s$-process path \citep[see Figure 1 of][]{vanraai2012}. High neutron densities are able to open the branches, favoring the production of $^{87}$Rb, which has a magic number of neutrons and a low neutron-capture cross-section; therefore, once $^{87}$Rb is produced, it tends to accumulate. For low neutron densities, the branching points are closed and the $^{85}$Rb production is favored, which has a neutron-capture cross-section higher than $^{87}$Rb. The $^{85}$Rb/$^{87}$Rb isotopic ratio would give us direct insights about the physical conditions of the $s$-process, however, this quantity cannot be evaluated from spectroscopic data. However, the abundance ratios between Rb and its neighborhood elements, such as Sr and Zr, can be used with the same purpose; we will discuss this further in Section~\ref{sub:nucleosynthesis_models}, in the light of the $s$-process models. 

\begin{figure}
    \centering
    \includegraphics{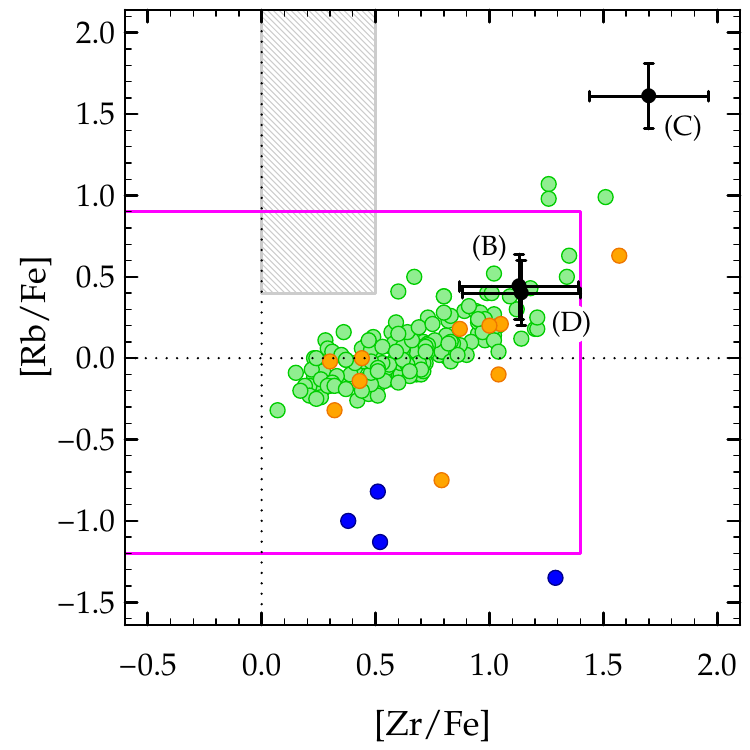}
    \caption{[Rb/Fe] {\sl versus} [Zr/Fe] plane showing the distribution of the program stars (black dots with error bars), HE~0457-1805 (B), HE~1255-2324 (C), and HE~2207-1746 (D), along with data available for barium stars. Symbols have the same meaning as in Figure~\ref{fig:la_vs_eu}. The magenta box delimits the ranges of the [Rb/Fe] and [Zr/Fe] ratios observed for low-mass AGB stars of the Galaxy and the Magellanic Clouds, whereas the grey shaded area shows the ranges observed for intermediate-mass AGB stars \citep{vanraai2012}.}
    \label{fig:rb_vs_zr}
\end{figure}

In Figure~\ref{fig:rb_vs_zr}, where we consider the Rb and Zr abundances, the data derived in this study are compared with data recently reported in literature for barium stars. Additionally, Figure~\ref{fig:rb_vs_zr} shows the observed ranges of the [Rb/Fe] and [Zr/Fe] ratios for AGB stars of the Galaxy and Magellanic Clouds. The magenta box and grey shaded region drawn in this plot delimit the observed values for AGB stars of low- and intermediate-mass, respectively \citep{garcia2006,garcia2009}. In the light of the $s$-process nucleosynthesis models, it is known that low-mass ($\sim 2-3$\,M$_{\odot}$) polluting AGB stars are able to reproduce the abundance profiles observed in barium stars \citep[e.g.][]{karinkuzhi2018, shejeelammal2020, roriz2021a, stancliffe2021, cseh2022}. A discussion of the physics of mass transfer is beyond the scope of this study, nonetheless, the interested reader is referred to the works of \citet{liu2017} and \citet{chen2020}, for instance, who performed 3D hydrodynamical simulations to predict the mass-accretion efficiency on interacting binary systems for different values of mass-ratio. As seen in Figure~\ref{fig:rb_vs_zr}, the [Rb/Fe] and [Zr/Fe] ratios derived for the targets HE~0457-1805 (B) and HE~2207-1746 (D) are consistent with the values observed in low-mass AGB stars. For the star HE~1255-2324 (C), the high [Rb/Fe] ratio is accompanied by similar level of [Zr/Fe], indicating large $s$-process efficiency in the processed material.

\begin{figure}
    \centering
    \includegraphics{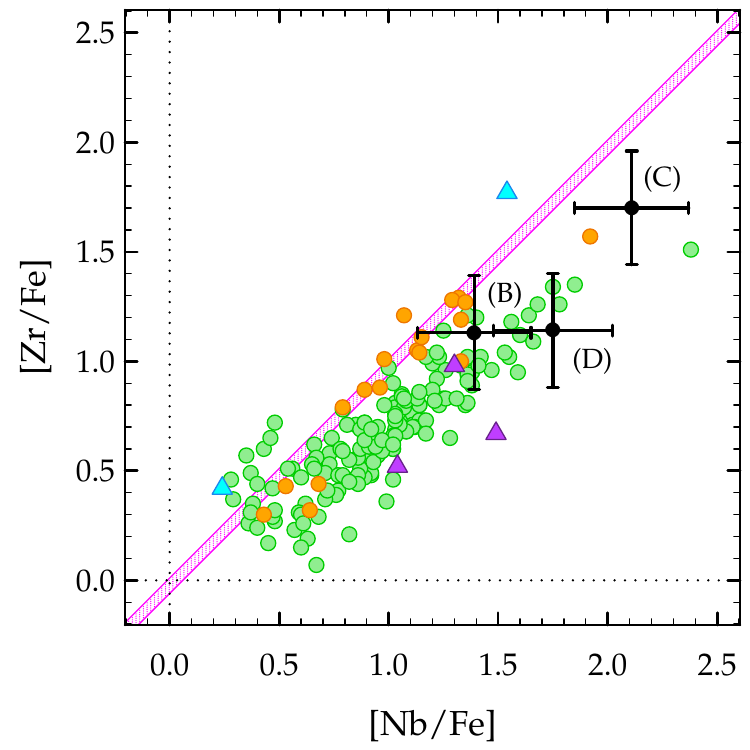}
    \caption{[Zr/Fe] {\sl versus} [Nb/Fe] plane showing the distribution of the program stars (black dots with error bars), HE~0457-1805 (B), HE~1255-2324 (C), and HE~2207-1746 (D), along with data available for barium stars and CEMP stars. Symbols have the same meaning as in Figure~\ref{fig:la_vs_eu}. The magenta region in this plot represents the analytical prediction reported by \citet{neyskens2015} for extrinsic stars.}
    \label{fig:nb_vs_zr}
\end{figure}

In Figure~\ref{fig:nb_vs_zr}, we have examined the [Zr/Fe] and [Nb/Fe] ratios. As demonstrated by \citet{neyskens2015}, under certain conditions, the [Zr/Fe] and [Nb/Fe] pair of abundance ratios observed in extrinsic stars can be used as a temperature diagnostic of the $s$-process that operated inside the former donor AGB star, thus constraining its main neutron source. This analytical approach is based on the fact that Nb has only one stable isotope, $^{93}$Nb, created exclusively from the $\beta$-decay of the $^{93}$Zr ($T_{1/2}=1.53 \times 10^{6}$ years). From the end of the $s$-process to the end of mass transfer, all the $^{93}$Zr received by the evolved secondary star had time to decay into $^{93}$Nb. Therefore, the Zr/Nb ratio, currently observed in the extrinsic star, should be equal to the Zr/$^{93}$Zr ratio in the atmosphere of the former AGB star. Assuming local equilibrium, in which the product $\langle \sigma_{i} \rangle N$, where $\langle \sigma_{i} \rangle$ is the Maxwellian-averaged neutron-capture cross section, remains constant along an isotopic chain, the Zr/$^{93}$Zr can be written in terms of the $\langle \sigma_{i} \rangle$ for the Zr isotopes. The neutron-capture cross sections in turn are temperature dependent; therefore, the current Zr/Nb ratio is a thermometer of the $s$-process responsible for the observed pattern. 

For the temperature range of the $s$-process operation, i.e. $(1-3) \times 10^{8}$ K, corresponding to $k_{B}T=10-30$ keV, the analytical prediction reported by \citet{neyskens2015} defines a region in the [Zr/Fe] {\sl versus} [Nb/Fe] plane, as shown in Figure~\ref{fig:nb_vs_zr}, where it is expected to find extrinsic stars. Latter, from more refined nucleosynthesis predictions computed by the STAREVOL models, for [Fe/H] $=-0.5$, \citet{karinkuzhi2018} found a close agreement with the analytical model of \citeauthor{neyskens2015} for low-mass ($1-3$\,M$_{\odot}$) AGB stars. On the other hand, for more massive ($> 4$ M$_{\odot}$) AGB stars, the models predictions computed by these authors deviate from the analytical approach.

\citet{karinkuzhi2018} and \citet{roriz2021b} applied the method of \citet{neyskens2015} for their respective sample of barium stars. While the targets considered of \citeauthor{karinkuzhi2018} (orange dots) fall partially in the predicted magenta region in Figure~\ref{fig:nb_vs_zr}, the stars of \citeauthor{roriz2021b} (green dots) are found systematically below the predicted region, on the Nb-rich side of the plane. In this study, we examined the positions of the program stars in the [Zr/Fe] and [Nb/Fe] plane; we note that the stars HE~0457-1805 (B), HE~1255-2324 (C), and HE~2207-1746 (D), for which we were able to derive the [Zr/Fe] and [Nb/Fe] ratios, behave a similar trend as found by \citeauthor{roriz2021b}. In particular, HE~1255-2324 (C) stands out from other objects, with [Zr/Fe] $=1.70$ and [Nb/Fe] $=2.08$. The observed excess in [Nb/Fe] relative to the production model of \citeauthor{neyskens2015}, however, seems to indicate the presence of an extra contribution to $^{93}$Nb. This may derived from, e.g., $^{93}$Y, if all the branching points at $^{90,91,92}$Y may be open in conditions of relatively high neutron densities. For example, to open the branching point at $^{93}$Y, with a half-life of 3.54 hours, sustained neutron densities above 10$^{12}$ cm$^{-3}$ would be required. New detailed models are needed to investigate the production of Nb in AGB stars.

\begin{figure}
    \centering
    \includegraphics{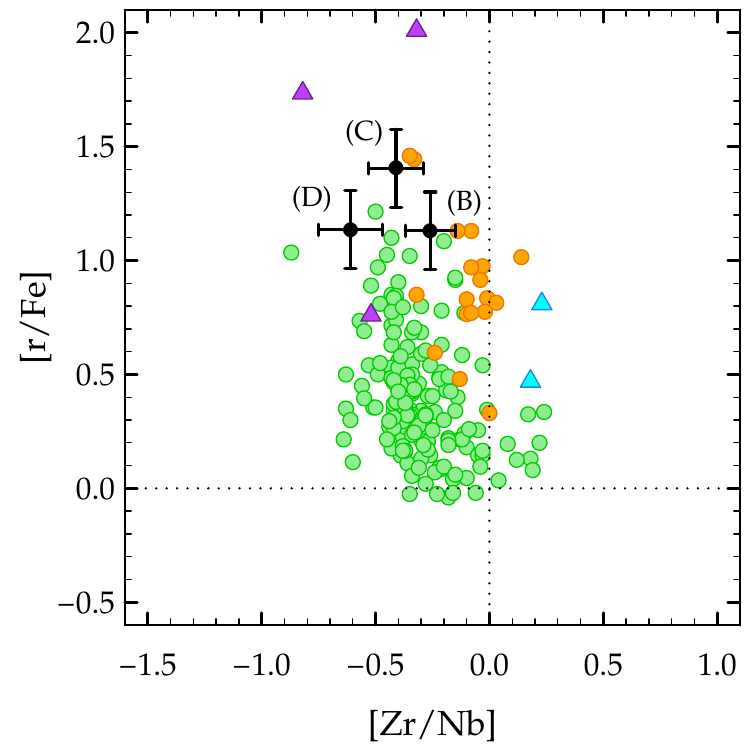}
    \caption{[$r$/Fe] {\sl versus} [Zr/Nb] plane showing the distribution of the program stars (black dots with error bars), HE~0457-1805 (B), HE~1255-2324 (C), and HE~2207-1746 (D), where [$r$/Fe] $=(\textrm{[Sm/Fe]}+\textrm{[Eu/Fe]})/2$ is the mean of the $r$-process elements. For comparison, we added to this plot data available for barium stars and CEMP stars. Symbols have the same meaning as in Figures~\ref{fig:heavy_elements} and \ref{fig:la_vs_eu}.}
    \label{fig:r-zrnb}
\end{figure}

In addition to the $s$-enrichment observed in our targets, as we have discussed, these stars also show atmospheres with high levels of the $r$-elements, displaying [$r$/Fe] $=$ 1.02, 1.13, 1.41, and 1.14, respectively for BD$+03\degr$2688, HE~0457-1805, HE~1255-2324, and HE~2207-1746, where [$r$/Fe] is the averaged abundance of the $r$-process elements, computed from the Sm and Eu abundances. However, if we also consider the $r$-element erbium in the [$r$/Fe] ratios, we then found higher $r$-process means, with [$r$/Fe] $=$ 1.41, 1.85, and 1.35, respectively for HE~0457-1805, HE~1255-2324, and HE~2207-1746. For BD$+03\degr$2688, we were not able to derive its Er abundance. In Figure~\ref{fig:r-zrnb}, we plot the [$r$/Fe] ratios as a function of [Zr/Nb], and compare them with data for barium stars. 
As evidenced in Figure~\ref{fig:r-zrnb}, the program stars show high levels of the [$r$/Fe] ratios, reaching values on the top of those commonly observed in barium stars. For HE~1255-2324, we stress that this system presents a [$r$/Fe] ratio similar to that found in the barium star HD~100503 ([$r$/Fe] $=1.45$), previously reported by \citet[][]{karinkuzhi2018}, who labeled tis object as a possible analogue of CEMP-$r/s$ star. As we have mentioned, these stars lie close to the CEMP-$r/s$ stars in the [La/Fe] {\sl versus} [Eu/Fe] diagram (Figure~\ref{fig:la_vs_eu}).

\subsubsection{Tungsten and thallium}

\begin{figure*}
    \centering
    \includegraphics{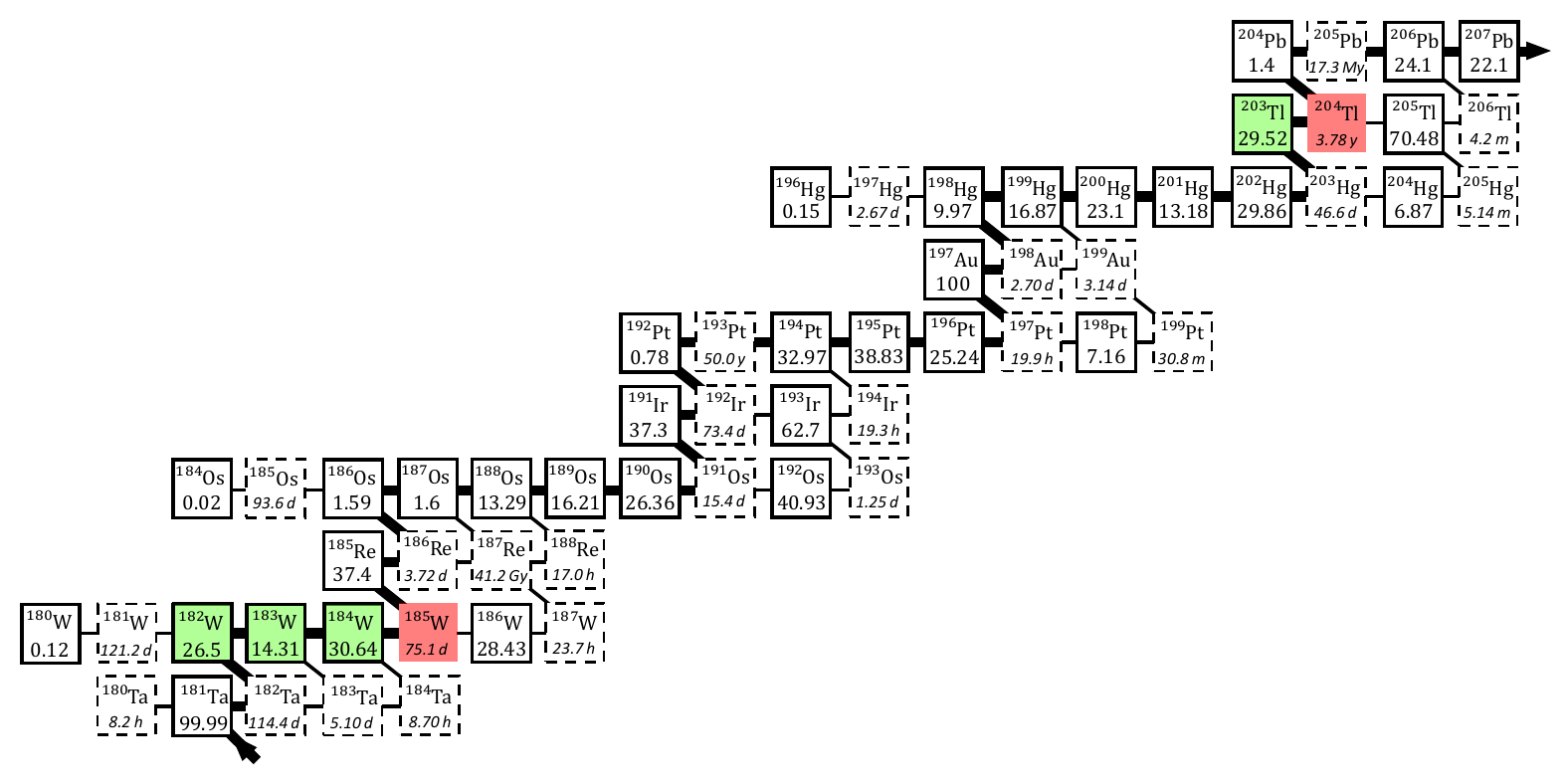}
    \caption{A portion of the nuclide chart showing the main (bold lines) and secondaries (thin lines) $s$-process paths from Ta to Pb. For unstable isotopes (dashed line boxes), we show their half-lives in italic, whereas for the stable ones (bold line boxes), we show their percent abundances. Data for this figure were taken from \citet[][]{dillman2006}. The branching points at $^{185}$W and $^{204}$Tl are identified in red boxes, while the green boxes are the stable isotopes of the W and Tl nuclide along the main $s$-process path.}
    \label{fig:nuclide_chart}
\end{figure*}

Tungsten and thallium have a relatively high $s$-process contribution in the Solar System (see Table~\ref{tab:s-fraction}), and because they are located between the second and third $s$-process peaks, their abundances (relative to the peaks) during the $s$-process flux are mostly determined by their neutron-capture cross sections. Once the flux reaches the second peak, the absolute amount of the elements beyond it that can be produced is a function of the neutron exposure, i.e., how many free neutrons are available. However, the distribution of such elements from Ba to Pb is controlled by the neutron capture cross sections. In other words, if the final abundance of Pb is high or low, then the abundances of these elements are also high or low, as their relative fraction cannot vary strongly. This property was used by \citet{lugaro2015} to analyze the abundance pattern observed in post-AGB stars (including elements such as W and Tl, and others in the same mass regions) and demonstrated that could not be a product of the $s$-process. Subsequently, \citet{hampel2019} analyzed these stars in the light of the intermediate $i$-process and obtained abundance patterns that could match them. This is because during the $i$-process the neutron-capture flux moves away from the valley of $\beta$ stability, and it is possible in this case to enhance the abundances of e.g. W and Tl, without increasing that of Pb. In Figure~\ref{fig:nuclide_chart}, we show a portion of the nuclide chart close to W and Tl, where we illustrate the $s$-process path throughout these species.

Because there are few data in literature for tungsten and lead, and no data for thallium, we do not plot them in Figure~\ref{fig:heavy_elements}. However, we can discuss the [W/hs], [Tl/hs] and [Pb/W], [Pb/Tl] ratios of our stars as compared to nucleosynthesis models in the next section.

\subsection{Comparison to nucleosynthesis models}\label{sub:nucleosynthesis_models}

\begin{figure}
    \centering
    \includegraphics{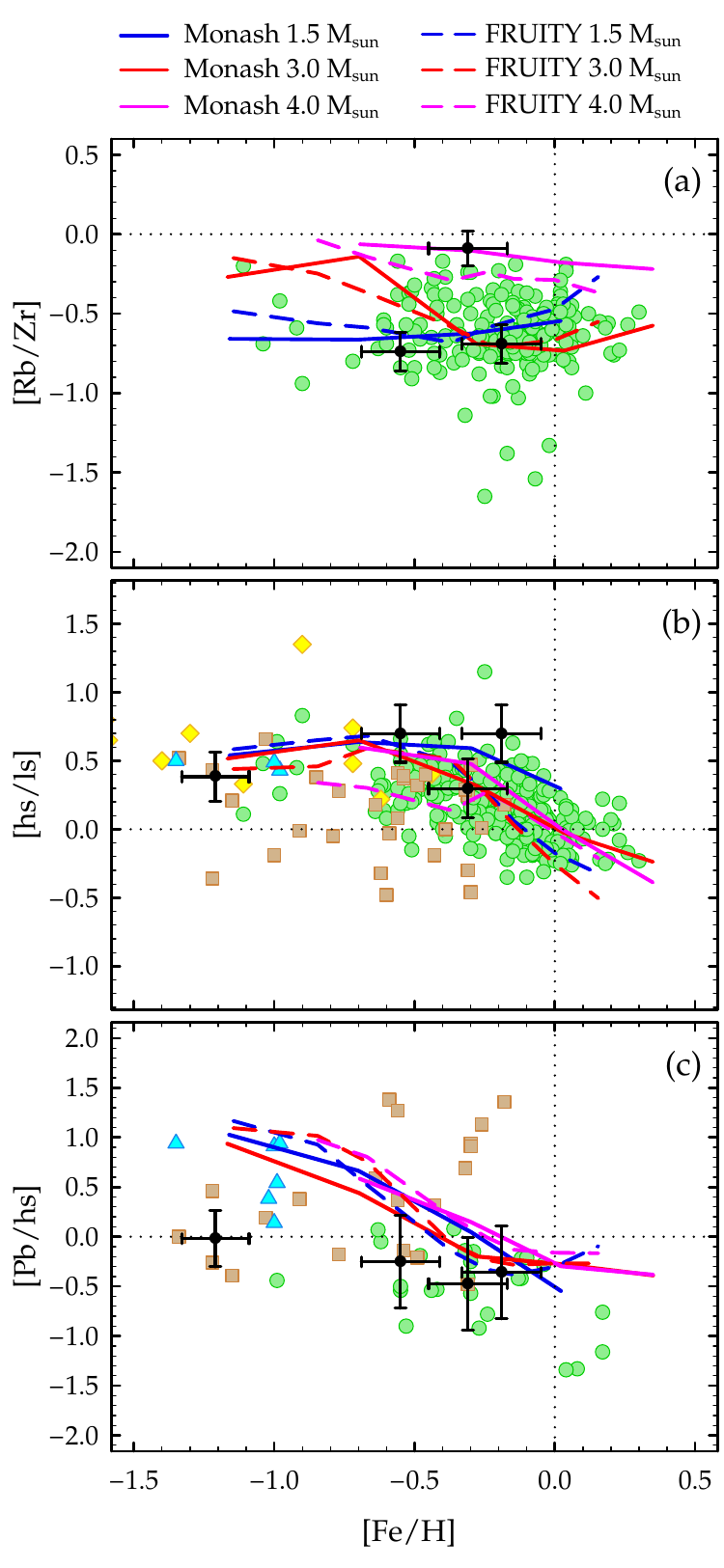}
    \caption{Comparison between the observed [Rb/Zr] (panel a), [hs/ls] (panel b), and [Pb/hs] (panel c) ratios with theoretical predictions from the Monash and {\sc fruity} $s$-process models for low-mass AGB stars. Symbols have the same meaning as in Figure~\ref{fig:heavy_elements}.}
    \label{fig:s-index}
\end{figure}

In Figure~\ref{fig:s-index}, panels (a), (b), and (c), we plot different abundance ratios, and compare them with predictions from two different sets of the $s$-process models, in order to probe the physical conditions that would originate the $s$-pattern observed in the program stars. We also added in this figure data for barium stars, CH stars, and post-AGB stars, when available. The Monash models are provided by \citet{fishlock2014}, \citet{karakas2016}, and \citet{karakas2018}, with the standard choice of the partial mixing zone leading to the formation of the $^{13}$C pocket. The {\sc fruity} (FUll-Network Repository of Updated Isotopic Tables \& Yields) nucleosynthesis models \citep{cristallo2011} are provided online by the {\sc fruity} database\footnote{Available at: \url{http://fruity.oa-teramo.inaf.it/}.}, not including rotation, and with the standard choice of the $^{13}$C pocket. However, because the models do not account the mass transfer and the further dilution of the AGB material deposited in the convective atmosphere of the observed star, we can not directly compare the models predictions with the data for [X/Fe]. Instead, we should consider only ratios that cancel the dilution effects.

In panel (a) of Figure~\ref{fig:s-index}, we plot the [Rb/Zr] ratios {\sl versus} [Fe/H]. As we have argued in the previous section, rubidium is a key element for neutron density diagnostic of the $s$-process, thanks to the branch points at $^{85}$Kr and $^{86}$Rb along the $s$-process path. In fact, the abundance ratios between Rb and its neighboring elements Sr and Zr, [Rb/Sr, Zr], can be also used as a probe to constraint the main neutron source of the $s$-process, and consequently the mass of the polluting AGB star. In the light of the $s$-process models, it is expected [Rb/Sr, Zr] $< 0$ for low-mass AGB stars, in which $^{13}$C is the main neutron source of the $s$-process, and [Rb/Sr, Zr] $> 0$ for more massive AGB stars, in which $^{22}$Ne becomes the main neutron source. This fact has been used for many authors to constrain the AGB mass \citep[e.g.][]{lambert1995, abia2001, garcia2006, garcia2009, vanraai2012, shejeelammal2020, roriz2021a}. For the stars HE~0457-1805, HE~1255-2324, and HE~2207-1746, we found [Rb/Zr] $<0$. For the star BD$+03\degr$2688, however, we were not able to compute its rubidium abundance, since the Rb line was too weak to be detected in its spectrum.

In panels (b) and (c) of Figure~\ref{fig:s-index}, we investigate the behavior of the $s$-process peaks, which occur close to the elements with magic number of neutrons; these elements tend to accumulate due to their low neutron-capture cross-section. The light $s$-elements (Sr, Y, and Zr) form the first peak, the heavy $s$-elements (La, Ce, and Nd) form the second peak, and the third peak occurs close to Pb. The index [hs/ls]$=$[hs/Fe]$-$[ls/Fe], where [hs/Fe] and [ls/Fe] are, respectively, the means of the heavy and light $s$-elements, is commonly used as a probe of the $s$-process efficiency \citep[][]{busso2001}. From panel (b), we can see that the observed data are in good agreement with the models predictions. The stars HE~0457-1805 and HE~2207-1746 show [hs/ls] ratios in the top range of the values commonly observed in barium stars in the same range of metallicity, indicating higher neutron exposures where the $s$-rich material of these stars was previously processed. In panel (c), we consider the [Pb/hs] ratios. The stars HE~0457-1805, HE~1255-2324, and HE~2207-1746 show typical values found in barium stars, which are slightly lower than the models. For metallicities near to BD$+03\degr$2688, the models predict [Pb/hs] $\sim 1$, 
because more neutrons per iron seed are available for the $s$-process. This is much larger than the observed in this object (i.e. [Pb/hs] $\sim 0$). This is similar to the case of post-AGB stars of similar metallicity showing Pb abundances much lower than expected by the models \citep{lugaro2015} and probably due to the operation of the $i$-process \citep{hampel2019}. Taken together, since we observe [Pb/hs] $\lesssim 0$ for the target stars, we can conclude that the $s$-process was not efficient in producing Pb. This is predicted by the AGB models for metallicities around solar, which also show [Pb/hs] $\lesssim 0$.

\begin{figure}
    \centering
    \includegraphics{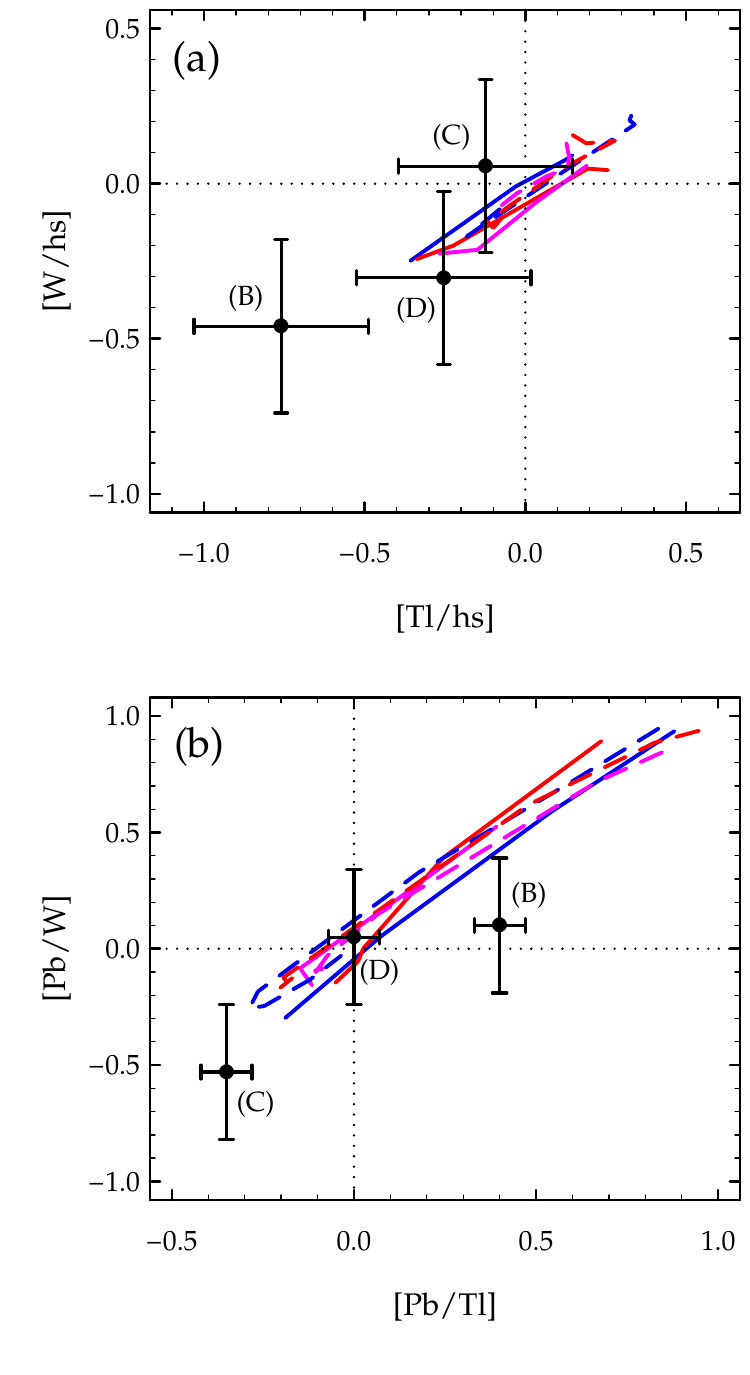}
    \caption{Different abundance ratios among the hs index and the heavy elements W, Tl, and Pb derived for the program stars (black dots with error bars) HE~0457-1805 (B), HE~1255-2324 (C), and HE~2207-1746 (D). The observed data set are compared with the same set of the $s$-process nucleosynthesis models showed in Figure~\ref{fig:s-index}.}
    \label{fig:w_tl_pb}
\end{figure}

We can infer more information from the W and Tl abundances and their relationship with the hs elements and with Pb (Figure~\ref{fig:w_tl_pb}). Relative to each other and to the second peak elements, the models show small variations, because in the $s$-process W and Tl follows the hs elements according to their neutron-capture cross sections, and therefore their ratios are typically solar. On the other hand, the Pb abundance increases, relative to W and Tl, with decreasing metallicity. The three stars for which W and Tl could be observed do not significantly deviate from the model predictions, except for HE~0457-1805 (B) for which Tl may have been slightly under estimated by the observations. It will be interesting to derive W and Tl for BD$+03\degr$2688 to investigate if the lower [Pb/hs] seen in this star to be lower than the model prediction is accompanied by an under-abundance of these two elements as well.

\section{Conclusions}\label{sec:conclusions}

Based on high-resolution optical spectroscopic data, we reported the first detailed chemical abundance analysis for a sample of four new chemically peculiar stars, BD$+03\degr$2688, HE~0457-1805, HE~1255-2324 and HE~2207-1746, for which we have derived atmospheric parameters and abundances for several chemical elements. Although these targets show in their atmospheres high levels of the $s$-process elements, with [$s$/Fe] $>1.0$, they are not luminous enough to be intrinsic AGB stars and experience internal nucleosyntheis of the $s$-process; in fact, these systems are first-ascent giant branch stars. 

The abundance ratios obtained in this work were compared with data for barium stars, CH stars, post-AGB stars and field giant stars, evidencing the $s$-rich nature of the program stars. HE~0457-1805 and HE~1255-2324 are binary stars; therefore, we can attribute their chemical peculiarities to mass transfer mechanism. We have identified HE~0457-1805 as a new barium star, since its C/O ratio is lesser the unity (C/O = 0.76), while HE~1255-2324 is a new CH star (C/O = 1.55). BD$+03\degr$2688 and HE~2207-1746 are probably CH stars, with C/O = 1.23 and 1.45, respectively, although this conclusion awaits an extensive radial-velocity monitoring.

Further to the $s$-enhancement ([$s$/Fe] $=1.77$) observed in HE~1255-2324, this star also shows a large amount of the $r$-elements Sm and Eu, with [$r$/Fe] $=1.41$, at [Fe/H] $=-0.31$. So far, the first object reported in literature sharing similar characteristics was the barium star HD~100503 ([Fe/H] $=-0.72$), analyzed by \citet{karinkuzhi2018}, who argued that HD~100503 would be an analogue of CEMP-$r/s$ stars. Therefore, we conclude that the new CH star HE~1255-2324 is a higher metallicity analogue of CEMP-$r/s$ stars.   

We compare the observed data for the $s$-elements with predictions from the Monash and {\sc fruity} $s$-process models. We derived for the sample stars [Rb/Zr] $<0$; from the nucleosynthesis models, this ratio is expected to be negative for low-mass ($1-4$\,M$_{\odot}$) AGB stars. Furthermore, the low-mass AGB nucleosynthesis models are able to reproduce in good agreement the observed [hs/ls] ratios. On the other hand, models predict somewhat higher [Pb/hs] ratios than the observed for the targets of this study and other barium stars collected from literature; at the lowest metallicity of BD$+03\degr$2688 ([Fe/H] $=-1.21$), this mismatch is the most evident. Tungsten and thallium abundances were derived for the first time in the target systems of this study, and the new abundance ratios are in general agreement with the $s$-process nucleosynthesis models.

\section*{Acknowledgements}

M.P.R. acknowledges Conselho Nacional de Desenvolvimento Cient{\'i}fico e Tecnol{\'o}gico (CNPq) for the postdoctoral fellowship No 313426/2022-8. M.L. acknowledges the support of the Hungarian National Research, Development and Innovation Office (NKFI), grant KH\_18 130405. N.A.D. acknowledges Russian Foundation for Basic Research (RFBR) according to the research projects 18-02-00554 and 18-02-06004. C. S. thanks the U.S. National Science Foundation for support under grant AST 1616040. The authors thank the anonymous referee for the comments that helped to improve the manuscript. This work is based on the observations made with the 2.2 m telescope at the European Southern Observatory (La Silla, Chile) under the agreement with Observat\'orio Nacional and under agreement between Observat\'orio Nacional and Max-Planck Institute f\"ur Astronomie. This research has made use of NASA’s Astrophysics Data System Bibliographic Services, and the SIMBAD \citep[][]{wenger2000}, {\sc vald} \citep[][]{ryabchikova2015}, and {\sc nist} \citep[][]{kramida2020} databases. This research has made also use of the NASA/IPAC Infrared Science Archive \citep[][]{irsa}, which is funded by the National Aeronautics and Space Administration and operated by the California Institute of Technology.

\section*{Data Availability}
The data underlying this article will be shared on reasonable request to the corresponding author.




\appendix
\onecolumn
\section{Observed atomic lines in the spectra of the program stars.}\label{app:obs_lines}

\begin{longtable}{lccccccc}
\caption{Equivalent width measurements of the Fe\,{\sc i} and Fe\,{\sc ii} lines considered in this study.}\label{tab:iron_lines}\\
\toprule
\multicolumn{1}{l}{} &
\multicolumn{1}{l}{} & 
\multicolumn{1}{l}{} & 
\multicolumn{1}{l}{} & 
\multicolumn{4}{c}{Equivalent widths (m\AA)}\\
\cline{5-8}
Element & Wavelength (\AA) & $\chi$\,(eV) & $\log$ \textit{gf} & BD$+03\degr$2688 & HE~0457-1805 & HE~1255-2324 & HE~2207-1746\\
\endfirsthead

\multicolumn{8}{c}%
{{\tablename\ \thetable{} -- Continued from previous page}} \\
\toprule
\multicolumn{1}{l}{} &
\multicolumn{1}{l}{} &
\multicolumn{1}{l}{} &
\multicolumn{1}{l}{} &
\multicolumn{4}{c}{Equivalent widths (m\AA)} \\
 \cline{5-8}
Element & Wavelength (\AA) & $\chi$\,(eV) & $\log$ \textit{gf} & BD$+03\degr$2688 & HE~0457-1805 & HE~1255-2324 & HE~2207-1746\\
\midrule
\endhead

\midrule
\multicolumn{8}{r}{{\textit{Continued on next page}}} \\
\endfoot

\bottomrule
\endlastfoot

\midrule

Fe\,{\sc i}   &  5\,198.71   &   2.22  &   $-$2.14   & 110 & --- & 131 & 128\\
              &  5\,242.49  &  3.63  &  $-$0.97   &  82 & 114 & --- & 100 \\
              &  5\,250.21  &  0.12  &  $-$4.92   & 118 & --- & --- & --- \\
              &  5\,253.03  &  2.28  &  $-$3.79   &  26 & --- & --- & --- \\
              &  5\,281.79  &  3.04  &  $-$0.83   & 123 & --- & --- & --- \\
              &  5\,288.52  &  3.69  &  $-$1.51   &  44 &  90 &  74 &  79 \\ 
              &  5\,307.36  &  1.61  &  $-$2.97   & 103 & --- & 120 & 121 \\
              &  5\,322.04  &  2.28  &  $-$2.84   &  74 & --- &  93 &  89 \\
              &  5\,364.87  &  4.45  &  $+$0.23   &  88 & --- & --- & --- \\
              &  5\,367.47  &  4.42  &  $+$0.43   &  96 & --- & --- & 124 \\
              &  5\,369.96  &  4.37  &  $+$0.54   & 104 & --- & --- & --- \\
              &  5\,373.71  &  4.47  &  $-$0.71   &  39 & --- &  72 & --- \\
              &  5\,389.48  &  4.42  &  $-$0.25   &  75 & 126 & 118 &  98 \\
              &  5\,393.17  &  3.24  &  $-$0.72   & 119 & --- & --- & --- \\
              &  5\,400.50  &  4.37  &  $-$0.10   &  89 & --- & --- & --- \\
              &  5\,410.91  &  4.47  &  $+$0.40   &  91 & 138 & --- & --- \\
              &  5\,417.03  &  4.42  &  $-$1.53   & --- & --- & --- & --- \\
              &  5\,445.04  &  4.39  &  $+$0.04   &  90 & --- & --- & --- \\
              &  5\,522.45  &  4.21  &  $-$1.40   & --- & --- & --- & --- \\
              &  5\,560.21  &  4.43  &  $-$1.04   &  34 &  67 & --- & --- \\
              &  5\,567.39  &  2.61  &  $-$2.56   &  87 & --- & --- & --- \\
              &  5\,624.02  &  4.39  &  $-$1.33   & --- &  60 & --- & --- \\
              &  5\,635.82  &  4.26  &  $-$1.74   & --- & --- &  53 & --- \\
              &  5\,686.53  &  4.55  &  $-$0.45   &  69 & --- & --- &  94 \\
              &  5\,691.50  &  4.30  &  $-$1.37   &  36 & --- & --- &  63 \\
              &  5\,705.47   & 4.30  &  $-$1.36   &  25 &  56 &  50 &  42 \\
              &  5\,717.83   & 4.28  &  $-$0.97   &  47 & --- & --- &  75 \\
              &  5\,731.76   & 4.26  &  $-$1.15   &  54 & --- & --- & --- \\
              &  5\,762.99   & 4.21  &  $-$0.41   &  81 & --- & --- & 104 \\
              &  5\,791.02   & 3.21  &  $-$2.27   &  63 & --- & --- & --- \\
              &  5\,806.73   & 4.61  &  $-$0.90   &  29 &  74 &  67 &  57 \\
              &  5\,814.81   & 4.28  &  $-$1.82   &  16 & --- & --- &  44 \\
              &  5\,852.22   & 4.55  &  $-$1.18   & --- &  73 &  53 &  46 \\
              &  5\,883.82   & 3.96  &  $-$1.21   &  51 & --- &  89 & --- \\
              &  5\,916.25   & 2.45  &  $-$2.99   &  61 & --- &  84 &  94 \\
              &  6\,024.06   & 4.55  &  $-$0.06   &  80 & --- & --- & --- \\
              &  6\,027.05   & 4.08  &  $-$1.09   &  48 &  99 &  81 &  82 \\
              &  6\,056.01   & 4.73  &  $-$0.40   &  43 &  75 &  78 &  67 \\
              &  6\,065.48   & 2.61  &  $-$1.53   & 129 & --- & --- & --- \\
              &  6\,079.01   & 4.65  &  $-$0.97   &  24 &  62 &  57 &  44 \\
              &  6\,082.71   & 2.22  &  $-$3.58   &  41 & --- & --- &  61 \\
              &  6\,093.64   & 4.61  &  $-$1.35   & --- & --- &  37 & --- \\
              &  6\,096.66   & 3.98  &  $-$1.78   &  21 &  62 & --- &  41 \\
              &  6\,136.61   & 2.45  &  $-$1.40   & 138 & --- & --- & --- \\
              &  6\,137.69   & 2.59  &  $-$1.40   & 135 & --- & --- & --- \\
              &  6\,151.62   & 2.18  &  $-$3.29   &  59 & 104 &  84 &  81 \\
              &  6\,165.36   & 4.14  &  $-$1.47   &  30 &  80 &  63 &  56 \\
              &  6\,173.34   & 2.22  &  $-$2.88   &  86 & --- & --- & --- \\
              &  6\,187.99   & 3.94  &  $-$1.57   &  35 &  90 & --- & --- \\
              &  6\,191.56   & 2.43  &  $-$1.42   & 143 & --- & --- & --- \\
              &  6\,200.31   & 2.60  &  $-$2.44   &  75 & --- & --- & 105 \\
              &  6\,213.43   & 2.22  &  $-$2.48   &  99 & --- & --- & 131 \\
              &  6\,252.56   & 2.40  &  $-$1.72   & 131 & --- & --- & --- \\
              &  6\,254.26   & 2.28  &  $-$2.44   & 116 & --- & --- & --- \\
              &  6\,265.13   & 2.18  &  $-$2.55   & 109 & --- & --- & --- \\
              &  6\,311.50   & 2.83  &  $-$3.23   & --- & --- & --- &  38 \\
              &  6\,322.69   & 2.59  &  $-$2.43   &  80 & --- & 121 & 109 \\
              &  6\,380.74   & 4.19  &  $-$1.32   &  47 & --- & --- & --- \\
              &  6\,392.54   & 2.28  &  $-$4.03   & --- &  70 &  50 &  49 \\
              &  6\,393.60   & 2.43  &  $-$1.43   & 142 & --- & --- & --- \\
              &  6\,411.65   & 3.65  &  $-$0.66   & 104 & --- & --- & 129 \\
              &  6\,419.95   & 4.73  &  $-$0.09   &  61 & --- & 115 & --- \\
              &  6\,421.35   & 2.28  &  $-$2.01   & 128 & --- & --- & --- \\
              &  6\,430.85   & 2.18  &  $-$2.01   & 137 & --- & --- & --- \\
              &  6\,518.37   & 2.83  &  $-$2.30   & --- & 106 &  93 & --- \\
              &  6\,551.68   & 0.99  &  $-$5.79   & --- & --- & --- &  40 \\
              &  6\,574.23   & 0.99  &  $-$5.02   & --- & --- &  96 &  92 \\
              &  6\,592.91   & 2.72  &  $-$1.47   & 117 & --- & 137 & --- \\
              &  6\,593.87   & 2.44  &  $-$2.42   &  95 & --- & 125 & --- \\
              &  6\,597.56   & 4.79  &  $-$0.92   &  23 & --- &  46 &  48 \\
              &  6\,608.03   & 2.28  &  $-$4.03   & --- &  62 &  43 &  32 \\
              &  6\,609.11   & 2.56  &  $-$2.69   &  73 & --- & 103 &  93 \\
              &  6\,704.48   & 4.22  &  $-$2.66   & --- &  25 & --- & --- \\
              &  6\,739.52   & 1.56  &  $-$4.95   & --- &  58 &  37 &  30 \\
              &  6\,793.26   & 4.07  &  $-$2.47   & --- & --- & --- & --- \\
              &  6\,750.15   & 2.42  &  $-$2.62   &  91 & --- & --- & --- \\
              &  6\,752.71   & 4.64  &  $-$1.20   &  28 & --- & --- & --- \\
              &  6\,804.00   & 4.65  &  $-$1.67   & --- & --- & --- &  26 \\
              &  6\,810.26   & 4.61  &  $-$0.99   &  25 &  76 &  65 &  63 \\
              &  6\,820.37   & 4.64  &  $-$1.17   &  27 & --- & --- & --- \\
              &  6\,858.15   & 4.61  &  $-$0.93   &  35 & --- &  80 & --- \\
\midrule
Fe\,{\sc ii}  &  4\,993.35   & 2.81  &  $-$3.67   & 46  & 65  & --- & --- \\
              &  5\,234.62   & 3.22  &  $-$2.24   & 87  &  94 &  85 &  82 \\
              &  5\,284.10   & 2.89  &  $-$3.01   & 66  &  87 &  72 & --- \\
              &  5\,325.56   & 3.22  &  $-$3.17   & 38  & --- & --- & --- \\
              &  5\,425.25   & 3.20  &  $-$3.21   & 48  & --- & --- & --- \\
              &  5\,534.83   & 3.25  &  $-$2.77   & 54  & --- & --- & --- \\
              &  5\,991.37   & 3.15  &  $-$3.56   & 28  & --- & --- & --- \\
              &  6\,084.10   & 3.20  &  $-$3.80   & 18  & --- & --- & --- \\
              &  6\,149.25   & 3.89  &  $-$2.72   & 30  & --- & --- &  41 \\  
              &  6\,247.55   & 3.89  &  $-$2.34   & 42  &  51 &  51 &  45 \\
              &  6\,416.92   & 3.89  &  $-$2.68   & 30  &  46 &  50 &  38 \\
              &  6\,432.68   & 2.89  &  $-$3.58   & 37  &  59 &  53 &  46 \\
\end{longtable}

\begin{longtable}{lcccccccc}
\caption{Equivalent width measurements of other atomic lines considered in this study.}\label{tab:other_lines}\\
\toprule
\multicolumn{5}{l}{} & 
\multicolumn{4}{c}{Equivalent widths (m\AA)}\\
\cline{6-9}
Element  &  Wavelength (\AA) & $\chi$\,(eV) & $\log$ \textit{gf} &  Ref & BD$+03\degr$2688 & HE~0457-1805 & HE~1255-2324 & HE~2207-1746\\
\endfirsthead

\multicolumn{9}{c}%
{{\tablename\ \thetable{} -- Continued from previous page}} \\
\toprule
\multicolumn{5}{l}{} & 
\multicolumn{4}{c}{Equivalent widths (m\AA)} \\
 \cline{6-9}
Element  &  Wavelength (\AA) & $\chi$\,(eV) & $\log$ \textit{gf} &  Ref & BD$+03\degr$2688 & HE~0457-1805 & HE~1255-2324 & HE~2207-1746\\
\midrule
\endhead

\midrule
\multicolumn{9}{r}{{\textit{Continued on next page}}} \\
\endfoot

\bottomrule
\endlastfoot

\midrule
Na\,{\sc i} & 5\,682.65  &  2.10 & $-$0.70 & PS  & 37 & --- & 126 & 105 \\
            & 5\,688.22  &  2.10 & $-$0.40 & PS  & 61 & 169 & 145 & 133 \\
            & 6\,154.22  &  2.10 & $-$1.57 & R03 & 12 & 103 &  42 &  47 \\
            & 6\,160.75  &  2.10 & $-$1.27 & R03 & 15 & 110 & 106 & --- \\
\midrule
Mg\,{\sc i} & 4\,730.04  &  4.34 & $-$2.39 & R03     & --- & --- & --- & --- \\
            & 5\,711.10  &  4.34 & $-$1.68 & R99     & --- & 132 & 123 & 121 \\
            & 6\,319.24  &  5.11 & $-$2.16 & Ca2007  & --- & --- &  43 &  42 \\
            & 6\,319.49  &  5.11 & $-$2.67 & Ca2007  &  11 & --- &  26 & --- \\
            & 8\,712.69  &  5.93 & $-$1.26 & Ca2007  & --- & --- & --- & --- \\
            & 8\,717.83  &  5.91 & $-$0.97 & WSM     & --- & --- & --- &  72 \\ 
            & 8\,736.04  &  5.94 & $-$0.34 & WSM     &  65 & 120 & 129 & 117 \\
\midrule
Al\,{\sc i} & 6\,696.03  &  3.14 & $-$1.481 & MR94 & --- &  67 &  62 &  54 \\
            & 7\,835.32  &  4.04 & $-$0.580 & R03  &  ---&  76 &  57 &  42 \\
            & 7\,836.13  &  4.02 & $-$0.400 & R03  & --- & --- & --- &  52 \\
            & 8\,772.88  &  4.02 & $-$0.250 & R03  & --- &  82 &  90 & --- \\
            & 8\,773.91  &  4.02 & $-$0.070 & R03  & --- & 110 & 102 & --- \\
\midrule
Si\,{\sc i} & 5\,793.08  &  4.93 & $-$2.06 & R03 &  29 & --- & --- & --- \\
            & 6\,131.58  &  5.62 & $-$1.69 & E93 & --- &  47 & --- & --- \\
            & 6\,145.08  &  5.62 & $-$1.43 & E93 &  20 &  41 &  39 &  32 \\
            & 6\,155.14  &  5.62 & $-$0.77 & E93 &  43 &  97 & --- &  83 \\
            & 8\,728.01  &  6.18 & $-$0.36 & E93 & --- &  77 &  63 &  73 \\
            & 8\,742.45  &  5.87 & $-$0.51 & E93 &  67 & --- &  77 &  76 \\
\midrule
Ca\,{\sc i} & 5\,601.29  &  2.52 & $-$0.52 & C2003 & --- & 147 & --- & --- \\
            & 5\,857.46  &  2.93 & $+$0.11 & C2003 & 112 & 155 & --- & 141 \\
            & 5\,867.57  &  2.93 & $-$1.61 & C2003 &  15 & --- &  40 &  31 \\
            & 6\,102.73  &  1.88 & $-$0.79 & D2002 & 122 & --- & --- & --- \\
            & 6\,166.44  &  2.52 & $-$1.14 & R03   &  70 & --- &  93 &  86 \\
            & 6\,169.04  &  2.52 & $-$0.80 & R03   &  88 & 127 & 107 & 116 \\
            & 6\,169.56  &  2.53 & $-$0.48 & DS91  & 105 & 148 & 129 & 126 \\
            & 6\,449.82  &  2.52 & $-$0.50 & C2003 & --- & 160 & --- & --- \\
            & 6\,455.60  &  2.51 & $-$1.29 & R03   &  61 & --- & --- & --- \\
            & 6\,471.66  &  2.51 & $-$0.69 & S86   & 101 & --- & --- & --- \\
            & 6\,493.79  &  2.52 & $-$0.11 & DS91  & 127 & --- & 156 & 144 \\
            & 6\,499.65  &  2.52 & $-$0.81 & C2003 &  88 & 130 & --- & --- \\
            & 6\,717.69  &  2.71 & $-$0.52 & C2003 & 101 & --- & --- & --- \\
\midrule
Ti\,{\sc i} & 4\,512.74  &  0.84 &  $-$0.48  & {\sc nist}  & --- & --- & 114 & --- \\
            & 4\,534.78  &  0.84 &  $+$0.28  & {\sc nist}  & --- & --- & --- & 139 \\
            & 4\,617.28  &  1.75 &  $+$0.39  & {\sc nist}  & --- & 119 & 112 & --- \\
            & 4\,562.64  &  0.02 &  $-$2.66  & {\sc nist}  & --- & --- & --- &  69 \\
            & 4\,758.12  &  2.25 &  $+$0.43  & {\sc nist}  &  56 & --- & --- & --- \\
            & 4\,778.26  &  2.24 &  $-$0.33  & {\sc nist}  & --- & --- &  45 &  44 \\
            & 4\,997.10  &  0.00 &  $-$2.12  & {\sc nist}  &  76 & --- & --- & 102 \\
            & 5\,009.66  &  0.02 &  $-$2.26  & {\sc nist}  &  63 & --- & --- & --- \\
            & 5\,016.17  &  0.85 &  $-$0.57  & {\sc nist}  &  95 & --- & --- & 112 \\
            & 5\,022.87  &  0.83 &  $-$0.43  & {\sc nist}  &  99 & 137 & --- & --- \\
            & 5\,039.96  &  0.02 &  $-$1.13  & {\sc nist}  & --- & 152 & 137 & 136 \\
            & 5\,062.10  &  2.16 &  $-$0.46  & {\sc nist}  &  21 &  67 &  47 & --- \\
            & 5\,087.06  &  1.43 &  $-$0.84  & E93         &  46 & --- & --- & --- \\
            & 5\,113.45  &  1.44 &  $-$0.88  & E93         &  30 &  82 & --- & --- \\
            & 5\,147.48  &  0.00 &  $-$2.01  & {\sc nist}  &  88 & 121 & --- & --- \\
            & 5\,152.19  &  0.02 &  $-$2.02  & {\sc nist}  &  72 & --- &  93 &  99 \\
            & 5\,210.39  &  0.05 &  $-$0.88  & {\sc nist}  & --- & --- & --- & 151 \\
            & 5\,219.71  &  0.02 &  $-$2.29  & {\sc nist}  &  74 & --- & --- & --- \\
            & 5\,223.63  &  2.09 &  $-$0.56  & {\sc nist}  & --- & --- & --- &  46 \\
            & 5\,295.78  &  1.05 &  $-$1.63  & {\sc nist}  &  24 &  72 &  52 &  51 \\
            & 5\,490.16  &  1.46 &  $-$0.94  & {\sc nist}  &  36 &  84 &  65 &  63 \\
            & 5\,503.90  &  2.58 &  $-$0.19  & {\sc nist}  & --- &  46 & --- & --- \\
            & 5\,662.16  &  2.32 &  $-$0.11  & {\sc nist}  &  27 &  69 &  63 &  51 \\
            & 5\,866.46  &  1.07 &  $-$0.87  & E93         &  74 & --- & --- & --- \\
            & 5\,922.12  &  1.05 &  $-$1.47  & {\sc nist}  &  31 & --- & --- &  61 \\
            & 5\,978.55  &  1.87 &  $-$0.50  & {\sc nist}  &  41 & --- & --- & --- \\
            & 6\,091.18  &  2.27 &  $-$0.37  & R03         & --- & --- &  50 &  54 \\
            & 6\,258.11  &  1.44 &  $-$0.36  & {\sc nist}  &  67 & 117 & --- & --- \\
            & 6\,261.10  &  1.43 &  $-$0.48  & {\sc nist}  &  67 & --- & --- & 108 \\
            & 6\,554.24  &  1.44 &  $-$1.22  & {\sc nist}  &  32 & --- & --- & --- \\
\midrule
Cr\,{\sc i} & 4\,964.93  & 0.94 & $-$2.53  & {\sc nist}  &  48 & --- & --- &  74 \\
            & 5\,193.50  & 3.42 & $-$0.90  & {\sc nist}  & --- & --- &  31 & --- \\
            & 5\,214.13  & 3.37 & $-$0.74  & {\sc nist}  & --- & --- &  32 & --- \\
            & 5\,214.61  & 3.32 & $-$0.66  & {\sc nist}  & --- &  46 & --- &  35 \\
            & 5\,238.96  & 2.71 & $-$1.30  & {\sc nist}  & --- & --- &  38 & --- \\
            & 5\,247.57  & 0.96 & $-$1.63  & {\sc nist}  &  90 & --- & --- & 112 \\
            & 5\,296.70  & 0.98 & $-$1.39  & GS88        & 103 & --- & --- & --- \\
            & 5\,298.28  & 0.98 & $-$1.17  & {\sc nist}  & 114 & 164 & 149 & --- \\
            & 5\,300.75  & 0.98 & $-$2.13  & R03         &  63 & 121 & --- &  89 \\
            & 5\,318.77  & 3.44 & $-$0.69  & {\sc nist}  & --- & --- &  41 & --- \\
            & 5\,304.18  & 3.46 & $-$0.71  & {\sc nist}  & --- & --- & --- &  15 \\
            & 5\,318.77  & 3.44 & $-$0.69  & {\sc nist}  & --- &  38 & --- & --- \\
            & 5\,348.33  & 1.00 & $-$1.29  & {\sc nist}  & 106 & 159 & 134 & --- \\
            & 5\,628.65  & 3.42 & $-$0.77  & {\sc nist}  & --- &  35 &  25 & --- \\
            & 5\,783.07  & 3.32 & $-$0.29  & {\sc nist}  &  14 & --- & --- &  39 \\
            & 5\,784.97  & 3.32 & $-$0.38  & {\sc nist}  &  16 &  72 &  63 &  49 \\
            & 5\,785.73  & 3.32 & $-$0.51  & {\sc nist}  & --- & --- & --- &  48 \\
            & 5\,787.93  & 3.32 & $-$0.08  & R03         &  37 & --- & --- &  75 \\
            & 6\,330.10  & 0.94 & $-$2.92  & R03         &  49 & --- & --- & --- \\
\midrule
Ni\,{\sc i} & 4\,513.00  & 3.71 & $-$1.52  & {\sc nist} & --- & --- &  36 & --- \\
            & 4\,740.17  & 3.48 & $-$1.78  & {\sc nist} & --- &  47 & --- & --- \\
            & 4\,904.42  & 3.54 & $-$0.19  & {\sc nist} &  71 & --- & 108 & --- \\
            & 4\,913.98  & 3.74 & $-$0.60  & {\sc nist} &  33 & --- & --- & --- \\
            & 4\,935.83  & 3.94 & $-$0.34  & {\sc nist} &  46 & --- & --- &  69 \\
            & 4\,967.52  & 3.80 & $-$1.60  & {\sc nist} & --- &  37 & --- & --- \\
            & 4\,995.66  & 3.63 & $-$1.61  & {\sc nist} & --- &  53 & --- &  25 \\
            & 5\,003.75  & 1.68 & $-$3.13  & {\sc nist} & --- &  82 & --- & --- \\
            & 5\,010.94  & 3.63 & $-$0.90  & {\sc nist} & --- & --- &  71 & --- \\
            & 5\,048.85  & 3.85 & $-$0.37  & {\sc nist} &  45 & --- & --- & --- \\
            & 5\,084.11  & 3.68 & $-$0.18  & E93        &  64 & --- & --- & --- \\
            & 5\,115.40  & 3.83 & $-$0.28  & R03        &  52 & --- &  92 & --- \\
            & 5\,197.17  & 3.90 & $-$1.14  & {\sc nist} & --- &  62 & --- & --- \\
            & 5\,589.37  & 3.90 & $-$1.15  & {\sc nist} & --- & --- & --- &  33 \\
            & 5\,643.09  & 4.17 & $-$1.25  & {\sc nist} & --- & --- &  35 & --- \\
            & 5\,709.56  & 1.68 & $-$2.14  & {\sc nist} &  89 & 130 & --- & --- \\
            & 5\,748.36  & 1.68 & $-$3.25  & {\sc nist} &  37 &  96 &  79 & --- \\
            & 5\,760.84  & 4.11 & $-$0.81  & {\sc nist} &  23 &  71 & --- & --- \\
            & 5\,805.23  & 4.17 & $-$0.60  & {\sc nist} &  13 & --- & --- &  45 \\
            & 5\,847.01  & 1.68 & $-$3.44  & {\sc nist} & --- &  86 &  69 &  54 \\
            & 5\,996.74  & 4.24 & $-$1.06  & {\sc nist} & --- &  36 & --- &  23 \\
            & 6\,086.29  & 4.27 & $-$0.47  & {\sc nist} &  21 & --- & --- & --- \\
            & 6\,108.12  & 1.68 & $-$2.49  & {\sc nist} &  73 & --- & --- & 104 \\
            & 6\,111.08  & 4.09 & $-$0.83  & {\sc nist} & --- &  70 & --- & --- \\
            & 6\,128.98  & 1.68 & $-$3.39  & {\sc nist} &  27 & --- &  67 & --- \\
            & 6\,130.14  & 4.27 & $-$0.98  & {\sc nist} & --- & --- & --- &  22 \\
            & 6\,176.82  & 4.09 & $-$0.26  & R03        &  41 & --- & --- & --- \\
            & 6\,177.25  & 1.83 & $-$3.60  & {\sc nist} &  21 &  71 & --- & --- \\
            & 6\,186.72  & 4.11 & $-$0.90  & {\sc nist} & --- & --- &  50 & --- \\
            & 6\,230.10  & 4.11 & $-$1.20  & {\sc nist} & --- &  40 & --- & --- \\
            & 6\,322.17  & 4.15 & $-$1.21  & {\sc nist} & --- & --- &  29 &  20 \\
            & 6\,327.60  & 1.68 & $-$3.09  & {\sc nist} &  57 & --- & --- & --- \\
            & 6\,378.26  & 4.15 & $-$0.82  & {\sc nist} & --- &  66 & --- & --- \\
            & 6\,384.67  & 4.15 & $-$1.00  & {\sc nist} & --- &  45 & --- & --- \\
            & 6\,482.80  & 1.94 & $-$2.85  & {\sc nist} &  46 & --- & --- & --- \\
            & 6\,532.88  & 1.94 & $-$3.42  & {\sc nist} &  17 &  59 & --- & --- \\
            & 6\,586.33  & 1.95 & $-$2.79  & {\sc nist} &  43 & 104 &  88 &  76 \\
            & 6\,598.61  & 4.24 & $-$0.93  & {\sc nist} & --- & --- & --- &  26 \\
            & 6\,635.14  & 4.42 & $-$0.75  & {\sc nist} & --- &  51 & --- & --- \\
            & 6\,767.77  & 1.83 & $-$2.11  & {\sc nist} &  91 & --- & 130 & 115 \\
            & 6\,772.32  & 3.66 & $-$0.97  & R03        &  29 &  79 & --- &  58 \\
            & 6\,842.04  & 3.66 & $-$1.44  & {\sc nist} &  17 &  68 &  50 & --- \\
\midrule
Sr\,{\sc i} & 4\,607.33 & 0.00 & $+$0.28 & SN96        &  70 & 153 & 148 & 113 \\
            & 4\,872.49 & 1.80 & $-$0.07 & {\sc nist}  &  18 & --- &  58 & --- \\
            & 7\,070.07 & 1.85 & $-$0.02 & K2018       &  22 &  48 &  62 &  32 \\
\midrule
Y\,{\sc ii} & 4\,883.68 & 1.08 & $+$0.07 & SN96 & 132 & --- & --- & --- \\
            & 5\,087.43 & 1.08 & $-$0.17 & SN96 & 115 & --- & 167 & --- \\
            & 5\,200.41 & 0.99 & $-$0.57 & SN96 & 100 & 141 & 133 & 122 \\
            & 5\,205.72 & 1.03 & $-$0.34 & SN96 & 105 & 145 & --- & 123 \\
            & 5\,289.81 & 1.03 & $-$1.85 & R04  &  38 & --- &  87 &  68 \\
            & 5\,402.78 & 1.84 & $-$0.44 & R03  &  54 &  91 &  99 &  79 \\
\midrule
Zr\,{\sc i} & 4\,772.30 & 0.62 & $-$0.06 & A04  & --- & --- &  89  & --- \\
            & 4\,784.94 & 0.69 & $-$0.60 & A04  & --- &  64 &  67  &  46 \\     
            & 4\,805.87 & 0.69 & $-$0.58 & A04  & --- &  76 &  73  &  51 \\
            & 4\,809.47 & 1.58 &    0.35 & A04  & --- &  54 &  66  & --- \\
            & 4\,815.05 & 0.65 & $-$0.38 & A04  & --- & --- &  --- &  46 \\
            & 4\,815.63 & 0.60 & $-$0.27 & A04  & --- & --- &  --- & --- \\
            & 4\,828.05 & 0.62 & $-$0.75 & A04  & --- &  57 &  69  &  35 \\
            & 5\,046.55 & 1.53 & $-$0.18 & A04  & --- &  36 &  50  & --- \\
            & 5\,385.13 & 0.52 & $-$0.64 & A04  & --- &  76 &  --- &  51 \\
            & 5\,437.77 & 0.15 & $-$2.12 & A04  & --- &  18 &  --- & --- \\
            & 5\,620.13 & 0.52 & $-$1.09 & A04  & --- &  45 &  --- &  33 \\
            & 5\,680.91 & 0.54 & $-$0.86 & A04  & --- & --- &  --- & --- \\
            & 5\,879.79 & 0.15 & $-$1.03 & A04  & --- &  83 &  89  &  53 \\     
            & 5\,885.62 & 0.07 & $-$1.73 & A04  & --- &  44 &  --- &  20 \\     
            & 5\,955.34 & 0.00 & $-$1.70 & A04  & --- &  53 &  --- &  31 \\     
            & 6\,032.60 & 1.48 & $-$0.35 & A04  & --- &  37 &  36  &  19 \\
            & 6\,127.46 & 0.15 & $-$1.06 & A04  & --- &  99 &  83  &  64 \\
            & 6\,134.57 & 0.00 & $-$1.28 & A04  & --- & 103 &  92  &  67 \\
            & 6\,140.46 & 0.52 & $-$1.41 & S96  &  16 & --- & ---  & --- \\
            & 6\,143.18 & 0.07 & $-$1.10 & S96  &  19 & --- & ---  & --- \\
\midrule
Mo\,{\sc i} & 5\,238.20 & 3.21 & $+$0.14 & {\sc vald} & --- &  17 & --- & --- \\
            & 5\,360.51 & 3.24 & $+$0.00 & {\sc vald} & --- &  15 & --- & --- \\
            & 5\,506.49 & 1.33 & $+$0.06 & V2015      & --- & 101 & --- & --- \\
            & 5\,533.03 & 1.33 & $-$0.07 & V2015      & --- & --- & 120 & --- \\
            & 5\,858.27 & 1.47 & $-$0.99 & {\sc vald} & --- & --- &  72 & --- \\
            & 6\,030.63 & 1.53 & $-$0.45 & V2015      & --- &  90 &  94 & --- \\
\midrule
Ru\,{\sc i} & 4\,869.15 & 0.93 & $-$0.83 & {\sc vald} & --- &  63 &  49 &  36 \\
            & 5\,309.27 & 0.93 & $-$1.39 & {\sc vald} & --- &  26 &  26 &  13 \\
            & 5\,636.24 & 1.06 & $-$1.07 & {\sc vald} & --- &  46 &  34 &  15 \\
\midrule
La\,{\sc ii} & 5\,303.53 & 0.32 & $-$1.35 & L01  &  77 & 114 & 107 &  99 \\ 
             & 5\,805.77 & 0.13 & $-$1.56 & L01  &  81 & 111 &  92 &  99 \\
             & 6\,262.29 & 0.40 & $-$1.22 & L01  & 185 & --- & --- & --- \\
             & 6\,320.43 & 0.17 & $-$1.52 & SN96 & 110 & 183 & 162 & 145 \\
             & 6\,774.33 & 0.13 & $-$1.71 & VWR  & 117 & 193 & 185 & 166 \\
\midrule
Ce\,{\sc ii} & 4\,418.79 & 0.86 & $+$0.27 & L09 & 106 & 118 & --- & --- \\
             & 4\,486.91 & 0.29 & $-$0.18 & L09 & 108 & --- & 100 &  96 \\
             & 4\,539.74 & 0.33 & $-$0.08 & L09 & --- & 131 & 110 & 107 \\
             & 4\,562.37 & 0.48 & $+$0.21 & L09 & 122 & 139 & 120 & 122 \\
             & 5\,187.45 & 1.21 & $+$0.17 & L09 &  81 & 105 & 100 &  86 \\
             & 5\,274.24 & 1.04 & $+$0.13 & L09 &  86 & 106 &  93 &  90 \\
             & 5\,330.58 & 0.87 & $-$0.40 & L09 &  73 &  93 &  88 &  72 \\
             & 5\,393.39 & 1.10 & $-$0.06 & L09 & --- & 107 & --- & --- \\
             & 5\,975.82 & 1.33 & $-$0.45 & L09 &  41 &  65 &  62 &  58 \\ 
             & 6\,043.37 & 1.21 & $-$0.48 & L09 &  50 & --- & --- &  71 \\
\midrule
Nd\,{\sc ii} & 4\,709.72  & 0.18 &  $-$0.97 & DH & --- & --- & 105 & --- \\
             & 4\,763.62  & 0.38 &  $-$1.27 & DH &  59 &  84 &  86 &  77 \\
             & 4\,786.11  & 0.18 &  $-$1.42 & DH & --- & --- &  76 & --- \\
             & 4\,797.15  & 0.56 &  $-$0.69 & DH & --- & 100 &  92 & --- \\
             & 4\,825.48  & 0.18 &  $-$0.42 & DH & --- & 139 & 121 & 119 \\
             & 4\,859.03  & 0.32 &  $-$0.44 & DH & 108 & --- & --- & 124 \\
             & 4\,902.04  & 0.06 &  $-$1.34 & DH & --- & 107 &  93 &  97 \\
             & 4\,914.38  & 0.38 &  $-$0.70 & DH &  93 & --- & 114 & 110 \\
             & 5\,063.72  & 0.98 &  $-$0.62 & DH &  76 &  80 &  91 &  77 \\
             & 5\,092.80  & 0.38 &  $-$0.61 & DH &  97 & 116 & 101 &  99 \\
             & 5\,192.61  & 1.14 &  $-$0.27 & DH &  90 & --- & 101 & 100 \\
             & 5\,212.36  & 0.20 &  $-$0.96 & DH &  92 & --- & 104 & 112 \\
             & 5\,234.19  & 0.55 &  $-$0.51 & DH & --- & 115 & 112 &  99 \\
             & 5\,249.58  & 0.98 &     0.20 & DH &  98 & 122 & --- & 104 \\
             & 5\,255.51  & 0.20 &  $-$0.67 & DH & 104 & --- & 114 & --- \\
             & 5\,293.16  & 0.82 &     0.10 & DH & --- & 134 & 118 & 109 \\
             & 5\,306.46  & 0.86 & $-$0.97 & DH  &  51 &  74 &  71 &  75 \\
             & 5\,311.46  & 0.98 & $-$0.42 & DH  &  76 & --- & --- &  96 \\
             & 5\,319.81  & 0.55 & $-$0.14 & DH  & 106 & 133 & 119 & 114 \\
             & 5\,356.97  & 1.26 & $-$0.28 & DH  &  70 &  92 &  85 & --- \\
             & 5\,371.93  & 1.41 & $-$0.00 & DH  &  79 & --- & --- & --- \\
             & 5\,485.70  & 1.26 & $-$0.12 & DH  & --- & --- &  80 & --- \\
             & 5\,698.92  & 1.54 & $-$0.67 & DH  & --- & --- & --- & --- \\
             & 5\,740.88  & 1.16 & $-$0.53 & DH  &  56 &  82 &  78 &  78 \\
             & 5\,811.57  & 0.86 & $-$0.86 & DH  &  58 &  85 &  81 &  73 \\
\midrule
Sm\,{\sc ii} & 4\,467.34  & 0.66 & $+$0.15 & L06 &  77 &  99 &  89 &  81 \\
             & 4\,499.48  & 0.25 & $-$0.87 & L06 &  70 &  77 & --- &  61 \\
             & 4\,523.91  & 0.43 & $-$0.39 & L06 & --- &  89 & --- &  89 \\
             & 4\,566.20  & 0.33 & $-$0.59 & L06 &  68 &  89 &  82 & --- \\
             & 4\,577.69  & 0.25 & $-$0.65 & L06 &  73 & --- & --- & --- \\
             & 4\,676.90  & 0.04 & $-$0.87 & L06 &  88 &  88 &  89 & --- \\
             & 4\,704.40  & 0.00 & $-$0.86 & L06 &  92 & --- & --- & --- \\
             & 4\,791.60  & 0.10 & $-$1.44 & L06 &  47 & --- &  66 &  57 \\
\midrule
Er\,{\sc ii} & 4\,675.62  & 1.32 & $-$0.05 & DeS       & --- &  81 & --- & --- \\
             & 4\,759.65  & 0.00 & $-$1.90 & {\sc vald} & --- &  68 & --- &  58 \\
             & 4\,831.15  & 1.62 & $-$0.62 & DeS        & --- & --- &  80 & --- \\
             & 4\,872.09  & 1.61 & $-$0.39 & {\sc vald} & --- & --- &  82 & --- \\
             & 4\,879.89  & 1.62 & $-$1.15 & DeS        & --- & --- &  48 & --- \\
             & 6\,006.78  & 0.00 & $-$2.82 & {\sc vald} & --- &  60 &  58 & --- \\
\end{longtable}
\textbf{References:} (A04) \citet{antipova2004}; (C2003) \citet{chen2003}; (Ca2007) \citet{carretta2007}; (D2002) \citet{depagne2002}; (DeS) \citet{desmedt2016}; (DH) \citet{denhartog2003}; (DS91) \citet{drake1991}; (E93) \citet{edvardsson1993}; (K2018) \citet[][]{karinkuzhi2018}; (GS88) \citet{gratton1988}; (L01) \citet{lawler2001a}; (L06) \citet{lawler2006}; (L09) \citet{lawler2009}; (MR94) \citet{mcwilliam1994}; ({\sc nist}) \citet{kramida2020}; (PS) \citet{preston2001}; (R03) \citet{reddy2003}; (R04) \citet{reyniers2004}; (R99) \citet{reddy1999}; (S86) \citet{smith1986}; (S96) \citet{smith1996}; (SN96) \citet{sneden1996}; ({\sc vald}) \citet{ryabchikova2015}; (VWR) \citet{vanwinckel2000}; (V2015) \citet{veklich2015}; (WSM) \citet{wiese1969}.

\twocolumn
\section{Supplementary figures}\label{app:extra_figures}

\begin{figure}
    \centering
    \includegraphics[width=\columnwidth]{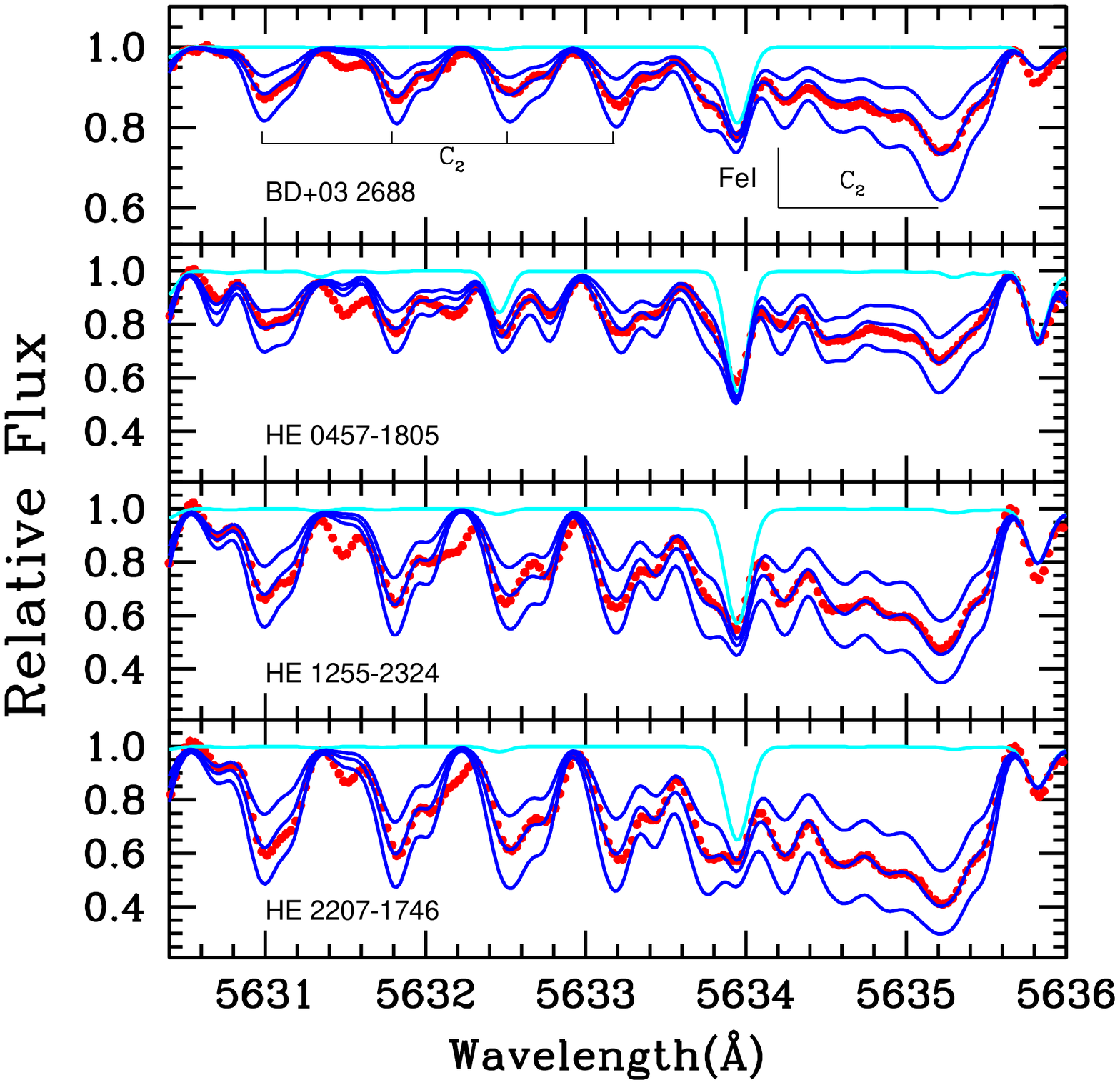}
    \caption{Observed (red dots) and synthetic (curves) spectra close to the region of the C$_{2}$ molecular band at 5\,635~\AA\ for the stars BD$+03\degr$2688, HE~0457-1805, HE~1255-2324, and HE~2207-1746. The upper cyan curves shown in the panels are synthetic spectra calculated without contribution from the C$_{2}$ lines. The middle blue curves are synthetic spectra calculated for the adopted solutions, $\log \epsilon $(C)\,=\,7.67, 8.36, 8.61, and 8.49 dex, that provide the best fits for BD$+03\degr$2688, HE~0457-1805, HE~1255-2324, and HE~2207-1746, respectively; the upper and lower blue curves show the spectral synthesis for $\Delta \log \epsilon (\textrm{C})= \pm 0.1$ dex around the adopted solutions.}
    \label{fig:syn_c2}
\end{figure}

\begin{figure}
    \centering
    \includegraphics[width=\columnwidth]{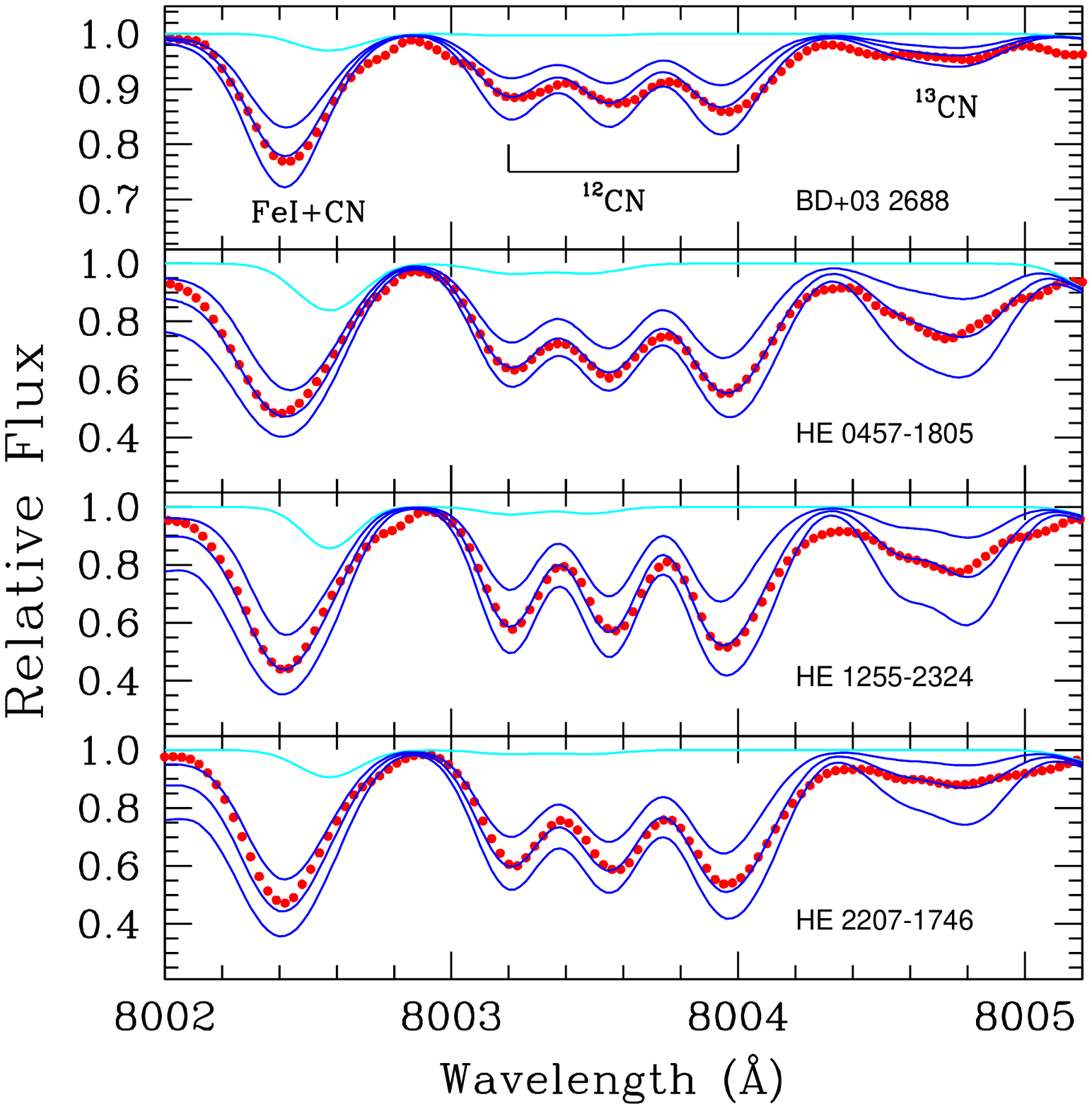}
    \caption{Observed (red dots) and synthetic (curves) spectra close to the region of the CN molecular band between 8\,002~\AA, and 8\,005~\AA\ for the stars BD$+03\degr$2688, HE~0457-1805, HE~1255-2324, and HE~2207-1746. The upper cyan curves shown in the panels are synthetic spectra calculated without contribution from the CN lines. The middle blue curves are synthetic spectra calculated for the adopted solutions, $\log \epsilon$(N)\,=\,6.90, 8.65, 8.44, and 8.27 dex, that provide the best fits for BD$+03\degr$2688, HE~0457-1805, HE~1255-2324, and HE~2207-1746, respectively; the upper and lower blue curves show the spectral synthesis for $\Delta \log \epsilon (\textrm{N})= \pm 0.3$ dex around the adopted solutions.}
    \label{fig:syn_cn}
\end{figure}

\begin{figure}
    \centering
    \includegraphics[width=\columnwidth]{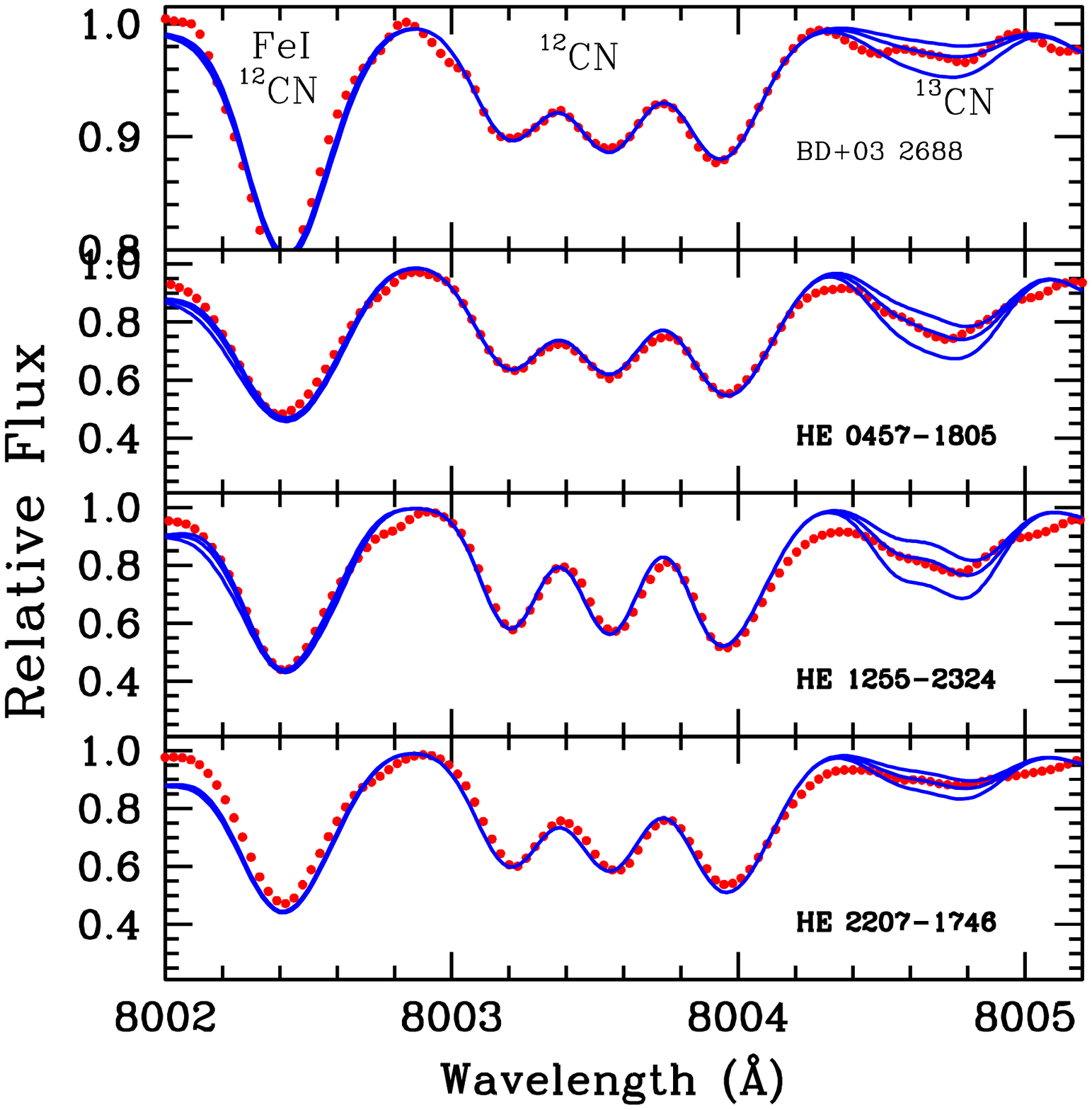}
    \caption{Observed (red dots) and synthetic (curves) spectra close to the region of the CN molecular band between 8\,002~\AA, and 8\,005~\AA\ for the stars BD$+03\degr$2688, HE~0457-1805, HE~1255-2324, and HE~2207-1746, calculated with different values for the $^{12}$C/$^{13}$C isotopic ratios, considering the previous CNO abundances determined for these stars. For BD$+03\degr$2688, we show the synthesis for $^{12}$C/$^{13}$C isotopic ratios of 10.0, 18.0 (adopted), and 30.0; for HE~0457-1805, the values are 30.0, 20.0 (adopted), and 10.0; for HE~1255-2324, the values are 30.0, 18.0 (adopted), and 10.0; for HE~2207-1746, the values are 20.0, 40.0 (adopted), and 90.0. For BD$+03\degr$2688, we enlarge the scale in order to better see the different values of $^{12}$C/$^{13}$C isotopic ratio, since it is a lower metallicity object.}
    \label{fig:syn_c12c13}
\end{figure}

\begin{figure*}
    \centering
    \includegraphics{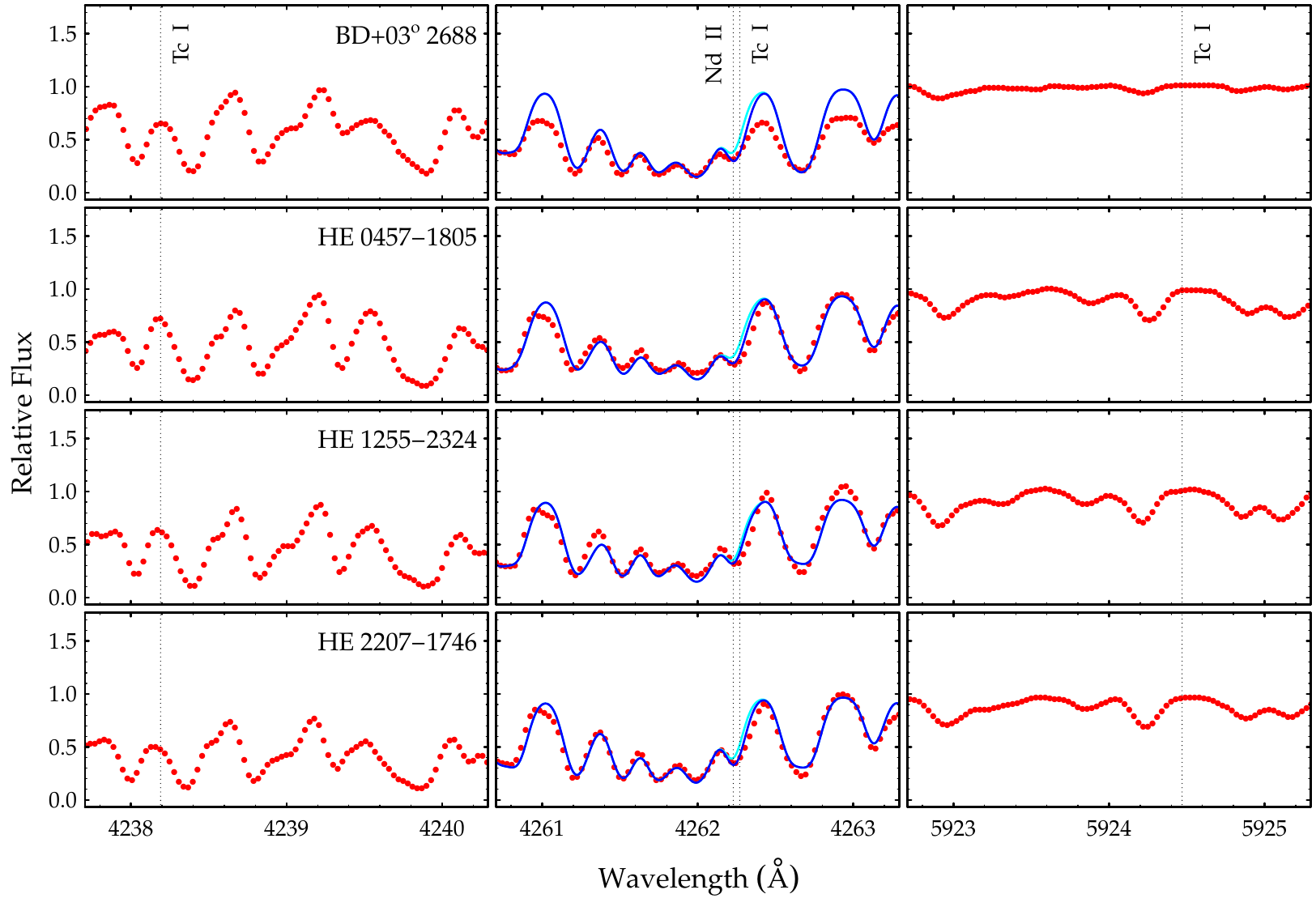}
    \caption{Observed (red dots) spectra close to the region of the Tc\,{\sc i} lines at 4\,238.19~\AA, 4\,262.27~\AA, and 5\,924.47~\AA\ (dashed black lines) for the stars BD$+03\degr$2688, HE~0457-1805, HE~1255-2324, and HE~2207-1746. The spectral feature at 4\,262.27~\AA\ is strongly affected by the Nd\,{\sc ii} line at 4\,262.23~\AA, also identified on the middle panels. We then performed spectral synthesis for this region without contribution from the technetium line (cyan curves) and for $\log \epsilon (\textrm{Tc})=1.0$ dex (blue curves).}
    \label{fig:technetium}
\end{figure*}

\begin{figure}
    \centering
    \includegraphics{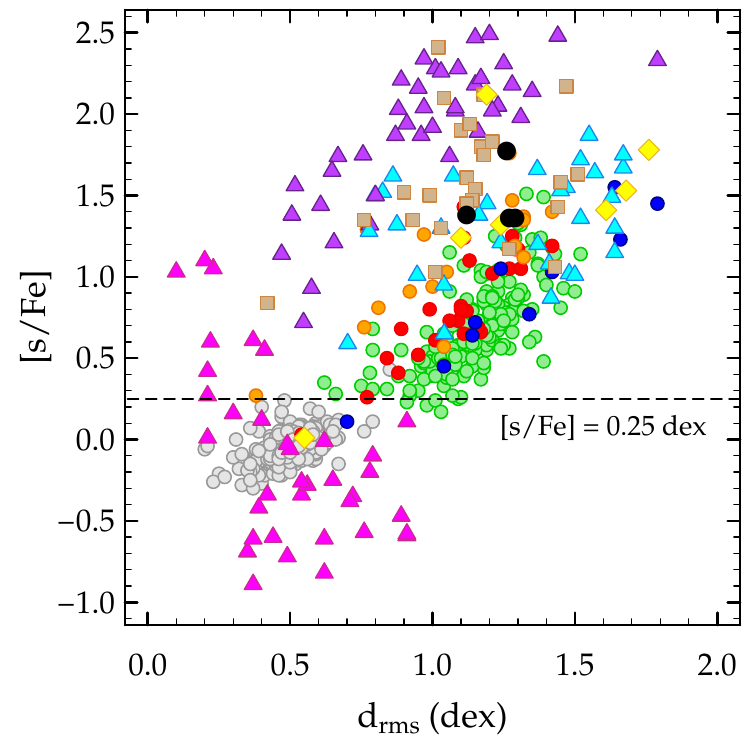}
    \caption{Diagram showing the [$s$/Fe] index against the $d_{\rm{rms}}$ distance for the program stars (black dots with error bars). Symbols have the same meaning as in Figures~\ref{fig:heavy_elements} and \ref{fig:la_vs_eu}. Data for field giant stars (grey dots) were taken from the samples of \citet{luck2007} and \citet{mishenina2007}. The dashed line in this figure represents the lower limit assumed by \citet{decastro2016} for classify an object as a barium star.}
    \label{fig:diagnostic}
\end{figure}

\onecolumn
\section{Abundance uncertainties}\label{app:uncertainties}

\begin{table*}
\centering
\caption{Abundance uncertainties for BD$+03\degr$2688. Second column gives $\sigma_{\rm ran}$, when three or more lines are used in the abundance derivation. Columns from 3 to 7 give the variation of the abundances caused by changes in $T_{\rm eff}$, $\log g$, $\xi$, [Fe/H], and equivalent width measurements ($W_\lambda$), respectively. The eight column gives the composed uncertainties, by quadratically combining the terms from the third to seventh columns. The last column gives the abundance dispersion observed among the lines for those elements with three or more lines available.}\label{tab:uncertainties_bd03}
    \begin{tabular}{lcccccccc}
    \toprule
        
    Species & $\sigma_{\rm ran}$ & $\Delta T_{\rm eff}$ & $\Delta\log g$ & $\Delta\xi$ & $\Delta$[Fe/H] & $\Delta W_{\lambda}$ & $ \sqrt{\sum \sigma^2}$ & $\sigma_{\rm obs}$ ($n$)\\
    & $_{\rule{0pt}{8pt}}$ & $+$40\,K & $+$0.3 & $+$0.2 km\,s$^{-1}$ & $-$0.1 & $+$3 m\AA &  \\
    \midrule
        
    Na\,{\sc i}    & 0.05 & $+$0.04 & $-$0.02 & $-$0.01 & $+$0.02 & $+$0.08 & 0.11  & 0.09  (04)\\ 
    Mg\,{\sc i}    & ---  & $+$0.02 & $-$0.01 & $-$0.04 & $+$0.01 & $+$0.08 & 0.09  & ---       \\ 
    Si\,{\sc i}    & 0.07 & $+$0.01 & $+$0.04 & $-$0.01 &    0.00 & $+$0.07 & 0.11  & 0.13  (04)\\ 
    Ca\,{\sc i}    & 0.03 & $+$0.06 & $-$0.02 & $-$0.08 & $+$0.03 & $+$0.07 & 0.13  & 0.10  (11)\\ 
    Ti\,{\sc i}    & 0.02 & $+$0.08 & $-$0.01 & $-$0.04 & $+$0.03 & $+$0.07 & 0.12  & 0.10  (20)\\ 
    Fe\,{\sc i}    & 0.02 & $+$0.05 & $-$0.01 & $-$0.09 & $+$0.01 & $+$0.06 & 0.12  & 0.13  (65)\\ 
    Fe\,{\sc ii}   & 0.03 & $-$0.02 & $+$0.14 & $-$0.04 & $-$0.02 & $+$0.07 & 0.17  & 0.09  (12)\\
    Cr\,{\sc i}    & 0.04 & $+$0.07 & $-$0.01 & $-$0.07 & $+$0.03 & $+$0.07 & 0.13  & 0.14  (10)\\ 
    Ni\,{\sc i}    & 0.02 & $+$0.05 & $+$0.02 & $-$0.04 & $+$0.01 & $+$0.07 & 0.10  & 0.11  (22)\\
    Sr\,{\sc i}    & 0.12 & $+$0.06 & $-$0.02 & $-$0.07 & $+$0.02 & $+$0.08 & 0.17  & 0.21  (03)\\ 
    Y\,{\sc ii}    & 0.07 & $+$0.01 & $+$0.10 & $-$0.14 & $-$0.02 & $+$0.07 & 0.20  & 0.17  (06)\\ 
    Zr\,{\sc i}    & ---  & $+$0.09 & $-$0.01 & $-$0.02 & $-$0.01 & $+$0.08 & 0.12  & ---       \\ 
    La\,{\sc ii}   & 0.04 & $+$0.01 & $-$0.12 & $-$0.05 & $-$0.03 & $+$0.04 & 0.14  & 0.10  (05)\\ 
    Ce\,{\sc ii}   & 0.08 & $+$0.02 & $+$0.11 & $-$0.15 & $-$0.02 & $+$0.09 & 0.22  & 0.22  (08)\\ 
    Nd\,{\sc ii}   & 0.04 & $+$0.02 & $+$0.12 & $-$0.15 & $-$0.01 & $+$0.10 & 0.22  & 0.14  (16)\\ 
    Sm\,{\sc ii}   & 0.08 & $+$0.03 & $+$0.11 & $-$0.15 & $-$0.01 & $+$0.09 & 0.22  & 0.21  (07)\\
    Eu\,{\sc ii}   & ---  & $-$0.02 & $+$0.09 &    0.00 & $-$0.05 &   ---   & 0.10  & ---       \\
    Pb\,{\sc i}    & ---  & $+$0.10 & $-$0.20 &    0.00 &    0.00 &   ---   & 0.22  & ---       \\
    \bottomrule
    
    \end{tabular}
\end{table*}

\begin{table*}
\centering
\caption{Abundance uncertainties for carbon, nitrogen, and oxygen, for the star BD$+03\degr$2688.} \label{tab:uncertainties_cno_bd03} 
    \begin{tabular}{lcccccccc}
    \toprule
    
    Species & $\Delta T_{\rm eff}$ & $\Delta\log g$ & $\Delta\xi$ & $\Delta\log {\rm (C)}$
    & $\Delta\log{\rm (N)}$ & $\Delta\log{\rm (O)}$ & $\sigma_{\rm tot}$ \\ 
    &   $+$40~K  &   $+$0.3   & $+$0.2~km\,s$^{-1}$  & $+$0.20  & $+$0.20   & $+$0.20  &  \\
    \midrule
        
    C  &  $+$0.02  &  $+$0.01   &  0.00  &     ---    &   0.00   &  $+$0.07  &  0.07 \\
    N  &  $+$0.10  &     0.00   &  0.00  &  $-$0.27   &   ---    &  $+$0.14  &  0.32 \\
    O  &  $+$0.02  &  $+$0.13   &  0.00  &     0.00   &   0.00   &     ---   &  0.13 \\
    \bottomrule
        
    \end{tabular}
\end{table*}

\begin{table*}
\centering
\caption{Same as in Table~\ref{tab:uncertainties_bd03}, for the star HE~1255-2324.}\label{tab:uncertainties_he1255}
    \begin{tabular}{lcccccccc}
    \toprule
        
    Species & $\sigma_{\rm ran}$ & $\Delta T_{\rm eff}$ & $\Delta\log g$ & $\Delta\xi$ & $\Delta$[Fe/H] & $\Delta W_{\lambda}$ & $ \sqrt{\sum \sigma^2}$ & $\sigma_{\rm obs}$ ($n$)\\
    & $_{\rule{0pt}{8pt}}$ & $+$110\,K & $+$0.3 & $+$0.2 km\,s$^{-1}$ & $-$0.1 & $+$3 m\AA &  \\
    \midrule
        
    Na\,{\sc i}    & 0.11 & $+$0.09 & $-$0.04 & $-$0.06 & $+$0.01 & $+$0.04 & 0.16  & 0.21 (04) \\
    Mg\,{\sc i}    & 0.05 & $+$0.06 &    0.00 & $-$0.03 &    0.00 & $+$0.05 & 0.11  & 0.10 (04) \\
    Al\,{\sc i}    & 0.01 & $+$0.05 & $-$0.02 & $-$0.04 &    0.00 & $+$0.04 & 0.08  & 0.02 (04) \\
    Si\,{\sc i}    & 0.04 & $-$0.03 & $+$0.07 & $-$0.03 & $-$0.02 & $+$0.05 & 0.11  & 0.07 (03) \\ 
    Ca\,{\sc i}    & 0.02 & $+$0.11 & $-$0.04 & $-$0.08 & $+$0.01 & $+$0.05 & 0.15  & 0.05 (05) \\ 
    Ti\,{\sc i}    & 0.02 & $+$0.15 & $-$0.02 & $-$0.08 & $+$0.01 & $+$0.06 & 0.18  & 0.07 (10) \\ 
    Fe\,{\sc i}    & 0.03 & $+$0.09 & $+$0.03 & $-$0.08 &    0.00 & $+$0.06 & 0.14  & 0.16 (31) \\ 
    Fe\,{\sc ii}   & 0.07 & $-$0.08 & $+$0.18 & $-$0.07 & $-$0.04 & $+$0.06 & 0.23  & 0.15 (05) \\
    Cr\,{\sc i}    & 0.04 & $+$0.10 & $-$0.02 & $-$0.05 & $+$0.01 & $+$0.06 & 0.13  & 0.12 (08) \\
    Ni\,{\sc i}    & 0.02 & $+$0.06 & $+$0.05 & $-$0.07 & $-$0.01 & $+$0.06 & 0.12  & 0.07 (13) \\
    Rb\,{\sc i}    & ---  & $+$0.10 &    0.00 & $-$0.10 &    0.00 &   ---   & 0.14  & ---       \\
    Sr\,{\sc i}    & 0.14 & $+$0.13 & $-$0.04 & $-$0.08 & $+$0.01 & $+$0.06 & 0.22  & 0.24 (03) \\ 
    Y\,{\sc ii}    & 0.07 & $+$0.01 & $+$0.09 & $-$0.15 & $-$0.04 & $+$0.05 & 0.20  & 0.14 (04) \\ 
    Zr\,{\sc i}    & 0.03 & $+$0.18 & $-$0.01 & $-$0.09 & $+$0.02 & $+$0.07 & 0.22  & 0.10 (10) \\
    Nb\,{\sc i}    & 0.09 & $+$0.20 &    0.00 & $-$0.05 &    0.00 &   ---   & 0.22  & 0.15 (03) \\
    Mo\,{\sc i}    & 0.01 & $+$0.17 & $-$0.01 & $-$0.14 & $+$0.01 & $+$0.06 & 0.23  & 0.02 (03) \\
    Ru\,{\sc i}    & 0.02 & $+$0.15 &    0.00 & $-$0.03 &    0.00 & $+$0.06 & 0.17  & 0.03 (03) \\
    La\,{\sc ii}   & 0.06 & $+$0.04 & $-$0.15 & $-$0.09 & $-$0.03 & $+$0.05 & 0.20  & 0.12 (04) \\
    Ce\,{\sc ii}   & 0.06 & $+$0.03 & $+$0.11 & $-$0.17 & $-$0.04 & $+$0.07 & 0.23  & 0.16 (07) \\ 
    Nd\,{\sc ii}   & 0.03 & $+$0.04 & $+$0.12 & $-$0.17 & $-$0.03 & $+$0.07 & 0.23  & 0.15 (20) \\ 
    Sm\,{\sc ii}   & 0.06 & $+$0.03 & $+$0.12 & $-$0.17 & $-$0.04 & $+$0.07 & 0.23  & 0.11 (04) \\ 
    Eu\,{\sc ii}   & ---  & $+$0.01 & $+$0.15 & $-$0.02 & $-$0.03 &   ---   & 0.15  & ---       \\
    Er\,{\sc ii}   & 0.08 & $+$0.01 & $+$0.13 & $+$0.11 & $-$0.03 & $+$0.07 & 0.20  & 0.16 (04) \\ 
    W\,{\sc i}     & ---  & $+$0.15 &    0.00 & $-$0.10 &    0.00 &   ---   & 0.18  & ---       \\ 
    Tl\,{\sc i}    & ---  & $+$0.15 & $+$0.05 & $-$0.05 &    0.00 &   ---   & 0.17  & ---       \\ 
    Pb\,{\sc i}    & ---  & $+$0.40 & $+$0.10 &    0.00  &    0.00 &   ---   & 0.41 & ---       \\
    \bottomrule
        
    \end{tabular}
\end{table*}

\begin{table*}
\centering
\caption{Abundance uncertainties for carbon, nitrogen, and oxygen, for the star HE~1255-2324. The last line of this table provides the uncertainty estimate in the $^{12}$C/$^{13}$C isotopic ratio.} \label{tab:uncertainties_cno_he1255} 
    \begin{tabular}{lcccccccc}
    \toprule
    
    Species & $\Delta T_{\rm eff}$ & $\Delta\log g$ & $\Delta\xi$ & $\Delta\log {\rm (C)}$ & $\Delta\log{\rm (N)}$ & $\Delta\log{\rm (O)}$ & $\sigma_{\rm tot}$ \\ 
                       & $+$110~K  &   $+$0.3   & $+$0.2~km\,s$^{-1}$  & $+$0.2     & $+$0.2   & $+$0.2    &        \\
    \midrule
        
    C                  &  $+$0.04  &     0.00   &  0.00                &     ---    &  0.00    &  $+$0.08   & 0.09  \\
    N                  &  $+$0.30  &  $+$0.06   &  0.00                &   $-$0.29  &   ---    &  $+$0.30   & 0.52  \\
    O                  &  $+$0.05  &  $+$0.16   &  0.00                &   $-$0.15  & $-$0.10  &    ---     & 0.25  \\
    \midrule
    $^{12}$C/$^{13}$C  &  $+$4.0  &    0.0     &  0.0                 &    ---     &  ---     &    ---     &   4.0 \\
    \bottomrule
        
    \end{tabular}
\end{table*}


\bsp	
\label{lastpage}
\end{document}